\documentclass[journal,10pt]{IEEEtran}

\usepackage[utf8]{inputenc}

\usepackage{epsfig}
\usepackage{booktabs}
\usepackage{epstopdf}
\usepackage{mathtools}

\usepackage{array}

\usepackage{bm}
\usepackage{mathrsfs}
\usepackage{epsfig}
\usepackage{algorithmic}
\usepackage{algorithm}
\usepackage{mathrsfs}
\usepackage{subcaption}

\usepackage{amssymb}
\usepackage{amsmath}
\usepackage{cite}
\usepackage{url}
\usepackage[table,svgnames,dvipsnames]{xcolor}
\usepackage{tabularx}
\usepackage{makecell,cellspace}
\usepackage{cite,graphicx,amsmath,amssymb}
\usepackage{fancyhdr}
\usepackage{mdwmath}
\usepackage{caption}
\usepackage{amsthm}

\usepackage{tkz-euclide}
\usepackage{pgfplots}
\usetikzlibrary{positioning,calc,fadings}
\usetikzlibrary{backgrounds, fit}
\usepackage{pgfplots}
\usepackage{pgfplotstable}

\usepackage[T5,T1]{fontenc}

\usepackage{ulem}
	\definecolor{brightpink}{rgb}{1.0, 0.0, 0.5}
		\definecolor{bondiblue}{rgb}{0.0, 0.58, 0.71}

\begin{document}
\title{Non-terrestrial Communications Assisted by Reconfigurable Intelligent Surfaces}

\author{Jia~Ye,~\IEEEmembership{Student Member,~IEEE}, Jingping~Qiao,~\IEEEmembership{Member,~IEEE}, Abla~Kammoun,~\IEEEmembership{Member,~IEEE}, and  Mohamed-Slim Alouini,~\IEEEmembership{Fellow,~IEEE}}

\maketitle

\begin{abstract}
Non-terrestrial communications have emerged as a key enabler for seamless connectivity in the upcoming generation networks. This kind of network can support high data rate communications among aerial platforms (i.e., unmanned aerial vehicles (UAVs), high-altitude platforms (HAPs), and satellites) and cellular networks, achieving anywhere and anytime connections. However, there are many practical implementation limitations, especially overload power consumption, high probability of blockage, and dynamic propagation environment. Fortunately, the recent technology reconfigurable intelligent surface (RIS) is expected to be one of the most cost-efficient solutions to address such issues. RIS with low-cost elements can bypass blockages and create multiple line-of-sight (LoS) links, and provide controllable communication channels. In this paper, we present a comprehensive literature review on the RIS-assisted non-terrestrial networks (RANTNs). Firstly, the framework of the RANTNs is introduced with detailed discussion about distinct properties of RIS in NTNs and the two types of RIS, that is, terrestrial RISs (TRISs), and aerial RISs (ARISs), and the classification of RANTNs including RIS-assisted air-to-ground (A2G)/ground-to-air (G2A), ARIS-assisted ground-to-ground (G2G), and RIS-assisted air-to-air (A2A) communications. In combination with next-generation communication technologies, the advanced technologies in RANTNs are discussed. Then we overview the literature related to RANTNs from the perspectives of performance analysis and optimization, followed by the widely used methodologies. Finally, open challenges and future research direction in the context of the RANTNs are highlighted. 
\end{abstract}


\IEEEpeerreviewmaketitle

\section{Introduction}
As the fifth-generation (5G) wireless network deployed worldwide successfully, the state-of-the-art of the 5G beyond and the sixth-generation network architecture is being enthusiastically investigated by researchers from both academia and industry. Future wireless communication networks are expected to provide a much more satisfying service to people by building uninterrupted and ubiquitous connectivity to everyone, everything, and everywhere with ultra-high data rate, extremely high reliability and low latency. However, due to the high deployment cost of typical ground base stations (gBSs) and wired backhauling infrastructure, it is expensive and unprofitable to build worldwide connectivity via terrestrial networks, especially in remote areas with sparse users. Considering the limitations of the conventional terrestrial cellular networks, the non-terrestrial networks (NTNs) supported by the developed aerial platforms become a fundamental technology in future communication blueprint \cite{giordani2020non, rinaldi2020non}. This promising vision has been confirmed by the 3rd Generation Partnership Project (3GPP) and reported in Technical Report (TR) 38.811 \cite{TR38811}. NTNs are able to tackle several tough problems in recent networks, such as the surge in throughput demands, blind spots, coverage holes, and terrestrial networks failures \cite{ejaz2019unmanned}. Specifically, practical experiments and projects have been initiated to provide ubiquitous internet services, such as Google Loon \cite{katikala2014google} and Thales Stratobus. 

Compared to the traditional ground wireless communication networks, NTNs tend to provide better channel conditions than terrestrial fading channels because aerial platforms usually have a strong line of sight connection with ground nodes \cite{7486987}. It will be easier to predict the channel state information (CSI) in three-dimensional (3D) positions based on the location information of terrestrial devices and communication performance \cite{qi2020deep}. Intuitively, the aerial platforms play a dominant role in the NTNs with 3D-mobility, flexibility, and adaptable altitude, who are often regarded as aerial base stations (ABSs) \cite{enayati2019moving, wang2018modeling, wang2019optimal}, relay platforms \cite{michailidis2019optimal,baek2017optimal,9177315}, or user equipment \cite{song2019probabilistic,hmamouche2020uplink} in NTNs. Different types of platforms possess distinct operating features, like frequency bands or wavelength, the operating altitude, flight duration, and the size of platforms. They can be divided into three types as unmanned aerial vehicles (UAVs), high altitude platforms (HAPs), and satellites without loss of generality. 

{\bf{UAV:} } UAVs operate at low altitudes of a few hundred meters with a coverage radius around 2 km \cite{zhang2018fast}. The fully controllable maneuverability 3D of UAVs enables the height and horizontal position adjustment to fit in different circumstances, which provides flexibility in easy deployment \cite{lin2018sky}. Because UAVs are battery powered and given the limited onboard energy, their lifetime in communication networks is generally time-limited, ranging from a few minutes to a few hours \cite{qin2020performance,abeywickrama2018comprehensive}. UAVs are of great help in establishing a temporary or specific communication network because they can fly directly over the users, thereby significantly improving  LoS communication probability and communication stability. It is expected that UAVs will bring considerable benefits to many fields including, but not limited to, emergency communications, smart city construction, rapid network recovery, and dense communication user networks.

{\bf{HAP:}} HAPs are network nodes that operate in the stratosphere at an altitude of around 20 km with a coverage radius of about 50km. HAPs are primarily located in the stratosphere, which enables HAPs staying at a quasi-stationary position relative to the earth \cite{KurtHAP}. Solar power coupled with energy storage has been regarded as the primary means of providing energy for HAPs since they have large surfaces suitable to accommodate solar panel films. Since the required power to stabilize HAPs is greatly reduced due to the low speed of weak wind in the stratosphere, HAPs can support flight duration varying from several hours to a few years. Typically, HAPs used in communications enable fast internet access and computation offloading, as well as data analysis to millions of devices in urban, suburban, and remote areas. Moreover, they also perform as the large-scale intelligent relays to build fast, reliable, and efficient connections between the satellites \cite{handley2019using}. As HAPs are distributed between UAVs and satellites, they can also act as the distributed data centers for recording, monitoring UAVs' and satellites' actions. 

{\bf{Satellite:}} The communication satellites are divided into three primary types according to their orbital altitudes. Specifically, the low Earth orbit (LEO) satellites are typically distributed in a circular orbit about 160 to 2000 km, and the Medium Earth orbit (MEO) satellites are in orbit somewhere between 2,000 and 35,786 kilometers above the earth surface. Both of them move in relation to the surface of the earth with a coverage radius exceeding 500 km. The geostationary orbit (GEO) satellites are located in a fixed position in the sky and cover about one-third of the Earth’s surface, and they have the same orbital period as the rotation rate of the Earth in 35785 km from the earth surface. The power source of satellites is the energy collected from the solar panels, which are exposed to direct solar radiation or to indirect radiation from albedo. Batteries of satellites are installed alongside the solar panels to store energy which can then be used when the satellite regularly passes through the shadow of the Earth or provide sufficient power during periods of peak demand. Consequently, satellites have long operating lifetimes in the range of 10 to 20 years on average, and provide a complementary role in wireless
communication coverage. Satellites can backhaul the traffic
load from the edge of the network, broadcast the popular
content to the edge, and ensure direct connectivity in remote areas where terrestrial infrastructure is difficult or impossible to implement. Moreover, they also can support 5G service onboard moving platforms, such as aircraft, vessels, and trains. 

The composed NTN systems, through its various aerial platforms, can enhance the network reliability by ensuring service continuity, guarantee the service ubiquity in un-served or under-served areas, as well as enable the 5G service scalability with efficient multi-casting or broadcasting. Although aerial platforms bring numerous advantages, they are not yet at their cutting edge of technology, with several vital problems waiting for handles. The specific challenges of NTNs can be summarized as follows:
\begin{itemize}
    \item {\textit{Weak Connectivity:} Down tilted antennas impair the connectivity at BSs or ground users. Since the lobes of gBS mainly point to the ground optimizing coverage, the aerial platforms flying over high altitude cannot build a reliable connection with the gBS. Moreover, the complex and uncontrollable wireless environment, especially a crowded area, makes the LoS links between ground users and aerial platforms more prone to be blocked, which can proliferate problems with coverage and connectivity. For example, the human bodies, buildings, and trees severely attenuate the signal strength, which also propose a new design challenge for the development of aerial platforms in the current environment. In most cases, it is hard to find a position where the ABSs are able to provide LoS connections to all ground devices.}
    \item {\textit{Interference:} However, once the BS antenna's main beam is directed toward the intended aerial platforms to concentrate the transmitted power and provide high signal gain, strong LoS propagation conditions in the aerial area cause only small signal losses, which impose unintentional but serious interference to aerial users in the same direction. This is a fatal problem, especially in cases requiring a critical data transmission quality. The cooperation work of multiple aerial devices and various aerial platform types also brings co-channel interference and thereby causing air-ground transmission performance degradation. Therefore, severe interference in the aerial area is a major hindrance to the coverage extension of the cellular system to the sky. }
    \item {\textit{Severe Path Loss:} In addition, the long-distance related large path loss in the 3D plane is a problem that cannot be ignored in NTNs. The long transmission distance and latency between satellites and terrestrial devices may impose some degradation on the quality of the received signals or the delivered data rate. Also, future communication networks are exploring the use of higher frequencies with wider spectrum bandwidths to tackle the increasing throughput demands and the spectrum scarcity problem. A major challenge hindering their widespread use is their short communication distance and vulnerability to blockages in the propagation environment. }
    \item {\textit{Information Leakage:} More eavesdropping threats are brought by the dominating LoS channels built by aerial platforms. Due to the broadcast nature of wireless transmissions and the dominating LoS channels in NTNs, the legitimate receivers are more vulnerable to malicious eavesdroppers, which greatly increases the potential risk of security in NTNs.  }
    
    \item {\textit{Power Constraint:} Moreover, the current size and weight of all kinds of aerial platforms are influenced by the carried power supply component. For instance, UAVs are generally time-limited due to the limited onboard energy. The energy-hungry issue seriously affects the promotion and popularization as the flying process consumes much more power than the hovering process. Due to the size and computation capability of UAVs, it is also impractical to perform power optimization with high complexity. Besides, under 6G communication in a high-frequency band, the system requires sophisticated radio frequency transceiver units and complex signal processing to achieve high-performance communication, which requires the system to have sufficient energy. }
    
    \item {\textit{Hardware Limitations:} It is well-known that mounting multiple antennas at wireless transceivers can boost the communication system performance significantly due to the potential to exploit multiplexing gains offered by spatial degrees of freedom. However, the size, weight, and power constraints of aerial platforms hinder the deployment of advanced multiple-input multiple-output (MIMO) techniques for mitigating the detrimental fading effects.}
\end{itemize}
Therefore, it is necessary to carry out innovative research to facilitate the realization of NTNs. In this context, reconfigurable intelligent surface (RIS) is envisioned as a cost-effective and energy-efficient reliable alternative to be integrated with aerial platforms. The RIS, also called intelligent reflecting surface, is a planar array comprising of a large number of low-cost and nearly passive reflecting elements. RISs can modify the wireless communication environment by introducing some changes in the phase, amplitude, frequency, or polarization of the incident electromagnetic wave. The destructive effect of multipath fading can be counteracted through RIS in a controllable fashion in that they reflect, refract, and scatter radio signals. These features can be leveraged to transform the propagation environment into a smart space that can be programmable for the benefit of the communication application. 

The real-time reconfiguration RISs are developed based on the invention of advanced micro-electrical-mechanical systems and metamaterials \cite{cui2014coding}, which enables the fixed phase shifters on surfaces to adapt the phase modification in time-varying wireless propagation environments. Owing to the important research effort is being devoted to developing hardware architectures based on this technology, RIS now can be implemented by various materials, including reflect arrays \cite{tan2018enabling,hum2013reconfigurable}, liquid crystal metasurfaces \cite{foo2017liquid}, ferroelectric films, or even metasurfaces\cite{liaskos2018using}. Compared to existing related technologies, such as active intelligent surface-based massive MIMO \cite{hu2018beyond}, multi-antenna relay \cite{sainath2012generalizing}, and backscatter communication\cite{yang2015multi}, RIS can
reflect ambient radio frequency (RF) signals in a passive way, without incurring additional energy cost for signal processing operations or transmitters, and without noise amplification. On the other hand, the lightweight and conformal geometry of RIS can enable their installment onto the facades of almost any object, which provides superior compatibility and high flexibility for practical deployment \cite{subrt2012intelligent}. Also, integrating RIS into the existing networks is transparent without any change in the hardware and software of the existing devices. Therefore, RIS stands out among existing advanced technologies with advantages like overcoming unfavorable propagation conditions, enriching the channel with more multi-paths, increasing the coverage area, improving the received signal power, avoiding interference, enhancing security/privacy, and consuming very low energy.

Motivated by the numerous aforementioned RIS's attendant benefits and the advanced features of aerial platforms, researchers have envisioned that RIS usage in NTNs will offer great agility, flexibility, rapid deployment and support for the wireless network in a cost-effective manner. RIS enables digitally tuned reflections of incident signals to improve service for isolated users in NTNs. In this way, integrating RIS into NTNs allows several benefits \cite{you2021enabling}. Firstly, RISs can be rapidly deployed to increase operators' revenue per user by filling coverage holes and meeting users' high-speed broadband needs. Secondly, compared to conventional relay-assisted NTNs, the data transmission between devices will experience less intermediate delays and additional interference to relay the information due to its full-duplex relaying mode. Thirdly, RIS greatly reduces the power consumption for communications since directing signals can be realized in a nearly-passive way without an RF source or power amplifier, and only a minimum amount of power is required for the RIS control unit. Additionally, thanks to the thin and lightweight materials of RISs, the aerial platforms’ payload is light even if a large number of reflecting elements are deployed. The greatly reduced energy consumption and payload at aerial platforms yield extended flight duration and reduced aerial platform deployment costs. Lastly, RISs consist of low-cost electronics, which does not put a heavy burden on hardware cost. 

\begin{figure*}
\centering
\scriptsize
\begin{tikzpicture}[tight background]
    \tikzstyle{sec} = [align=center, inner sep=0pt, rectangle, minimum width=3cm, minimum height=0.6cm, text centered, draw=black, fill=red!30]
    \tikzstyle{subsec} = [align=center, inner sep=0pt, rectangle, minimum width=3.2cm, minimum height=0.6cm, text centered, draw=black, fill=orange!30]
    \tikzstyle{subsubsec} = [align=center, inner sep=0pt, rectangle, minimum width=2 cm, minimum height=0.6cm, text centered, draw=black, fill=gray!30]
    
    \node[sec] (intro) {I. Introduction};
    
    \node[sec, below = 0.5cm of intro, minimum height=1cm] (framework) {II. Frameworks of \\ RANTNs};
    \draw[thick,->] (intro) -- (framework);
    
    \node[subsec, right = 1cm of intro, minimum height=0.8cm] (ristypes) {Reconfigurable \\ Intelligent surface};
    \node[subsec, below = 0.8cm of ristypes, minimum height=0.8cm] (risframeworks) {System models of \\ RANTNs};

    \coordinate[right = 0.5cm of framework] (framework_r);
    
    \draw[thick, ->] (framework) -| (framework_r) |- (ristypes);
    \draw[thick, ->] (framework) -| (framework_r) |- (risframeworks);

    \node[subsubsec, right = 1cm of ristypes] (terrestrial) {Terrestrial RIS};
    \node[subsubsec, below = 0.1cm of terrestrial] (aerial) {Aerial RIS};
    \node[subsubsec, below = 0.1cm of aerial] (comparisons) {Comparisons};

    \coordinate[right = 0.5cm of ristypes] (ristypes_r);
    
    \draw[thick, ->] (ristypes) -| (ristypes_r) |- (terrestrial);
    \draw[thick, ->] (ristypes) -| (ristypes_r) |- (aerial);
    \draw[thick, ->] (ristypes) -| (ristypes_r) |- (comparisons);
    
    \node[subsubsec, below = 0.17cm of comparisons, minimum height=1cm] (A2G) {RIS-assisted \\ A2G/G2A \\ Communications};
    \node[subsubsec, below = 0.1cm of A2G, minimum height=0.8cm] (G2G) {ARIS-assisted G2G \\ Communication};
    \node[subsubsec, below = 0.1cm of G2G, minimum height=0.8cm] (A2A) {RIS-assisted A2A \\ Communication};

    \coordinate[right = 0.5cm of risframeworks] (risframeworks_r);
    
    \draw[thick, ->] (risframeworks) -| (risframeworks_r) |- (A2G);
    \draw[thick, ->] (risframeworks) -| (risframeworks_r) |- (G2G);
    \draw[thick, ->] (risframeworks) -| (risframeworks_r) |- (A2A);
    
    \node[sec, below = 0.5cm of framework, minimum height=1cm] (risassisted) {III. RIS-Assisted \\ NTNs with Other \\ Technologies};
    \draw[thick,->] (framework) -- (risassisted);

    \node[subsec, left = 1cm of intro, minimum height=1.0cm] (optical) {Optical Wireless \\ Communication \\ System};
    \node[subsec, below = 0.1cm of optical] (mmwave) {mmWave/THz};
    \node[subsec, below = 0.1cm of mmwave] (wireless) {Wireless Power Transfer};
    \node[subsec, below = 0.1cm of wireless] (MAC) {Multiple Access Schemes};
    \node[subsec, below = 0.1cm of MAC] (SS) {Spectral Sharing};
    
    \coordinate[left = 0.5cm of risassisted] (risassisted_r);
    
    \draw[thick, ->] (risassisted) -| (risassisted_r) |- (optical);
    \draw[thick, ->] (risassisted_r) |- (mmwave);
    \draw[thick, ->] (risassisted_r) |- (wireless);
    \draw[thick, ->] (risassisted_r) |- (MAC);
    \draw[thick, ->] (risassisted_r) |- (SS);
    
    \node[sec, below = 0.5cm of risassisted, minimum height=0.8cm] (perfor) {IV. Performance Analysis};
    \draw[thick,->] (risassisted) -- (perfor);
    
    \node[sec, below = 0.5cm of perfor, minimum height=0.8cm] (analysis) {V. Optimization \\ of RANTNs};
    \draw[thick,->] (perfor) -- (analysis);
    
    \node[subsec, below = 0.6cm of risframeworks, minimum height=0.8cm] (SNR) {Received Power \\ /SNR};
    \node[subsec, below = 0.1cm of SNR] (interfe) {Interference};
    \node[subsec, below = 0.1cm of interfe, minimum height=0.8cm] (rate) {Achievable Rate\\ Maximization};
    \node[subsec, below = 0.1cm of rate] (energy) {Energy Consumption};  
    \node[subsec, below = 0.1cm of energy] (EE) {Energy Efficiency};
    \node[subsec, below = 0.1cm of EE] (PLS) {Physical Layer Security};    
    \node[subsec, below = 0.1cm of PLS] (time) {Time Consumption};       
    \node[subsec, below = 0.1cm of time] (probability) {Error Probability};  
    \node[subsec, below = 0.1cm of probability] (node) {Number of Served Devices};  
     
    \coordinate[right = 0.5cm of analysis] (analysis_r);
    
    \draw[thick, ->] (analysis) -| (analysis_r) |- (SNR);
    \draw[thick, ->] (analysis) -| (analysis_r) |- (interfe);
    \draw[thick, ->] (analysis) |- (analysis_r) |- (rate);
    \draw[thick, ->] (analysis) -| (analysis_r) |- (energy);
    \draw[thick, ->] (analysis) |- (analysis_r) |- (EE);    
    \draw[thick, ->] (analysis) -| (analysis_r) |- (PLS);
    \draw[thick, ->] (analysis) |- (analysis_r) |- (time);   
    \draw[thick, ->] (analysis) |- (analysis_r) |- (probability);
    \draw[thick, ->] (analysis) |- (analysis_r) |- (node); 
    
    \node[sec, below = 0.5cm of analysis] (math) {VI. Methodology};
    \draw[thick,->] (analysis) -- (math);
    
    \node[subsec, below = 0.6cm of SS, minimum height=0.8cm] (AO) {Alternating \\ Optimization Algorithm};
    \node[subsec, below = 0.1cm of AO, minimum height=0.8cm] (ML) {Machine Learning-based \\ Algorithm};
     
    \coordinate[left = 0.5cm of math] (math_r);
    \draw[thick, ->] (math) -| (math_r) |- (AO);
    \draw[thick, ->] (math) -| (math_r) |- (ML);

    \node[sec, below = 0.5cm of math, minimum height=1cm] (challenge) {VII. Challenges and \\ Future Research \\ Directions};
    \draw[thick,->] (math) -- (challenge);
    
    \node[subsec, below = 1 cm of ML] (channel) {Practical Channel Modeling};
    \node[subsec, below = 0.1cm of channel] (estimation) {Channel Estimation};
    \node[subsec, below = 0.1cm of estimation] (track) {Tracking};  
    \node[subsec, below = 0.1cm of track] (hardware) {Hardware Limitations};
    \node[subsec, below = 0.1cm of hardware] (control) {Backhual Control};    
    \node[subsec, below = 0.1cm of control, minimum height=0.8cm] (design) {Efficient Design \\ and Optimiztion};       
    \node[subsec, below = 0.1cm of design] (spectral) {Spectral Sharing};  
     
    \coordinate[left = 0.5cm of challenge] (challenge_r);
    
    \draw[thick, ->] (challenge) -| (challenge_r) |- (channel);
    \draw[thick, ->] (challenge) -| (challenge_r)  |- (estimation);
    \draw[thick, ->] (challenge) -| (challenge_r)  |- (track);
    \draw[thick, ->] (challenge) -| (challenge_r)  |- (hardware);    
    \draw[thick, ->] (challenge) -| (challenge_r) |- (control);
    \draw[thick, ->] (challenge) -| (challenge_r)  |- (design);   
    \draw[thick, ->] (challenge) -| (challenge_r)  |- (spectral); 
    
    \node[sec, below = 0.5cm of challenge] (con) {VIII. Conclusion};
    \draw[thick,->] (challenge) -- (con);
    
\end{tikzpicture}
\caption{Survey framework and outline of the main topics}
\label{orgnization}
\end{figure*}
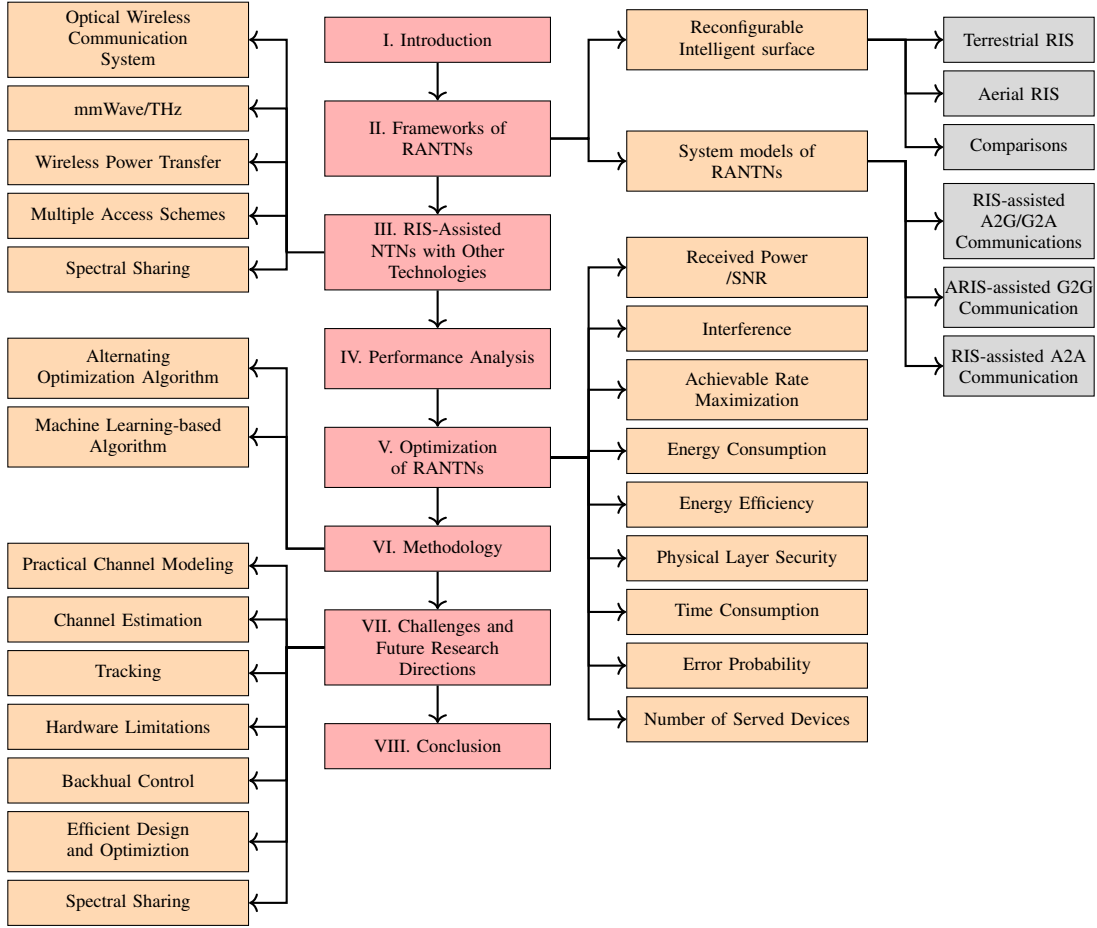

There are a few review papers that mentioned the integration of RIS and aerial platforms. In particular, the review paper in \cite{liu2020reconfigurable} covers an overview of the RIS-assisted UAV-enabled wireless networks. The tutorial in \cite{wu2021intelligent} introduces the concept of the RIS mounted on UAV to realize over-the-air intelligent reflection. The authors in \cite{almohamad2020smart} discuss the feasibility of using UAV carried RIS for enhancing physical layer security. The survey focusing on the RIS technology in \cite{gong2020toward} firstly concludes the RIS-assisted UAV communication in two different ways. One is utilizing the on-building RIS to enhance communication quality from UAV to the ground user, while the other one is to deploy aerial RIS to enjoy full-angle reflection. Motivated by the RIS' appealing properties, the authors in \cite{wu20205g} also discussed the methodologies of using RIS in UAV communication through the mentioned two ways and proposed several interesting and important problems remaining to be handled. There are few surveys \cite{saeed2020point,dao2021survey} mentioned the application potential of RIS in aerial networks with UAVs, HAPs, and satellites. Compared to the existing works, our survey provides a comprehensive study on the RANTN systems. A systematic organization of this paper is provided in Fig. \ref{orgnization}, and the main contributions can be summarized as follows:
\begin{itemize}
    \item We start from the review of the distinct characteristics of RIS in NTNs and the classification of the RIS types, that is, the terrestrial RIS (TRIS) coated on terrestrial objects, and the aerial RIS (ARIS) flying over the air. Then, we introduce three main kinds of network types studied in current literature involved with aerial platforms including RIS-assisted air-to-ground (A2G)/ground-to-air (G2A), ARIS-assisted ground-to-ground (G2G), and RIS-assisted air-to-air (A2A) communications. 
    \item We provide an extensive review on the combination of existing promising technologies with RANTNs, which are not covered by the existing review papers. Then, the stochastic performance analysis, the system optimization based on different performance indices, and the widely adopted methodologies are also summarized, which provide useful insights into practical implementation.
    \item Some limitations and impractical assumptions of the current literature are discussed and highlighted for future exploration. For example, the estimation of the time-varying channel due to the mobile aerial platforms is usually omitted, which is a critical problem in practice. We also notice that the joint design on phase shifts, platform deployment leads to a significant increase in hardware cost and design complexity. The derived conclusion under traditional channel models without jointly considering RIS capabilities and aerial platform properties might be useless in practical deployment. 
\end{itemize}

The rest of this paper is organized as follows. Section II introduces and compares the two types of RIS studied in current NTN systems. After this, this paper divides the RANTNs into three categories depending on the positions of transmitters and receivers. Section III discusses the integration of other promising technologies into RANTNs, which is followed by the performance analysis and optimization of RANTNs in Section IV and V, respectively. The methodology and tools used to enhance the system performance are concluded in Section VI. Section VII summarizes some practical challenges and future research directions, and Section VIII concludes the paper. A list of abbreviations used in this survey is given in Table \ref{Abbre}. 

\begin{table}[!htbp]
\centering
\small
\caption{List of Abbreviations}
\label{Abbre}
\begin{tabular}{| p{.11\textwidth}| p{.3\textwidth}| }
\hline
\rowcolor{gray!20}
Abbreviation &  Description  \\
\hline
\rowcolor{yellow!20}
2D/3D & 2/3-dimensional \\
\hline
\rowcolor{yellow!20}
3GPP & 3rd Generation Partnership Project \\
\hline
\rowcolor{yellow!20}
5G/6G & Fifth/sixth-generation \\
\hline
\rowcolor{yellow!20}
A2A/A2G/ G2A/G2G & Air-to-air/Air-to-ground /Ground-to-air/Ground-to-ground \\
\hline
\rowcolor{yellow!20}
ABS & Aerial base station\\
\hline
 \rowcolor{yellow!20}
 AoI & Age-of-information\\ 
\hline
 \rowcolor{yellow!20}
 BER/SER & Bit error rate/Symbol error rate \\ 
 \hline
\rowcolor{yellow!20}
CSI & Channel state information \\
\hline
\rowcolor{yellow!20}
DF & Decode-and-forward \\
\hline
\rowcolor{yellow!20}
DNN/DRL/ DQN &  Deep neural network / Deep reinforcement/Deep Q-network learning\\
\hline
\rowcolor{yellow!20}
DoF & degree of freedom\\
\hline
\rowcolor{yellow!20}
EH/ID & Energy harvesting/Information decoding\\
\hline 
\rowcolor{yellow!20}
EC/EE & Ergodic capacity/Energy efficiency\\
\hline 
\rowcolor{yellow!20}
FSO & Free space optics\\
\hline
\rowcolor{yellow!20}
gBS/gU&Ground base station/Ground user \\
\hline 
\rowcolor{yellow!20}
GEO/LEO/ MEO & Geostationary/Low Earth/Medium Earth orbit \\
\hline
\rowcolor{yellow!20}
GRU &Gated Recurrent unit \\
\hline 
\rowcolor{yellow!20}
HAPS & High Altitude platform stations\\
\hline
\rowcolor{yellow!20}
IoTD &  Internet of things device\\
\hline
\rowcolor{yellow!20}
LoS/NLoS &  Line-of-sight/Non LoS  \\
\hline
\rowcolor{yellow!20}
MIMO/MISO/ SISO & Multiple-input multiple-output / multiple-input single-output/single-input single-output\\
\hline
\rowcolor{yellow!20}
mmWave/THz &  Millimeter wave/ Terahertz\\
\hline
\rowcolor{yellow!20}
MU &  Multi-user\\
\hline
\rowcolor{yellow!20}
NMSE& Normalized mean square error \\
\hline
\rowcolor{yellow!20}
NOMA/OMA & Non-orthogonal/Orthogonal multiple access \\
\hline
\rowcolor{yellow!20}
NTN &  Non-terrestrial network\\
\hline
\rowcolor{yellow!20}
OFDMA/ FDMA/TDMA &  Orthogonal frequency/Frequency/time division multiple access\\
\hline
\rowcolor{yellow!20}
QoS & Quality of service\\
\hline
\rowcolor{yellow!20}
OP & Outage probability\\
\hline
\rowcolor{yellow!20}
OWC &  optical wireless communication \\
\hline
\rowcolor{yellow!20}
PDB & Power and data beacon \\
\hline
\rowcolor{yellow!20}
PPO & Proximal policy optimization \\
\hline
\rowcolor{yellow!20}
RF &  Radio frequency \\
\hline
\rowcolor{yellow!20}
RIS/ ARIS/TRIS/ & Reconfigurable intelligent surface/Aerial RIS/Terrestrial RIS\\
\hline
\rowcolor{yellow!20}
RL & Reinforcement learning \\
\hline
\rowcolor{yellow!20}
SWIPT & Simultaneous wireless information and power transfer\\
\hline
\rowcolor{yellow!20}
 SCA & Successive Convex Approximation \\
\hline
\rowcolor{yellow!20}
 SDP & Semidefinite Programming \\ 
 \hline
\rowcolor{yellow!20}
SINR/SNR & Signal-to-interference-plus-Noise Ratio/Signal-to-noise ratio\\
\hline
\rowcolor{yellow!20}
UAV & Unmanned aerial vehicle\\
\hline
\rowcolor{yellow!20}
URLLC & Ultra-reliable and lower latency communications\\
\hline
\rowcolor{yellow!20}
VLC & Visible light communication\\
\hline
\rowcolor{yellow!20}
ULA/UPA & Uniform linear/planar array\\
\hline
\end{tabular}
\end{table}

\section{Frameworks of RANTNs}
\begin{figure*}[t]
\centering
\includegraphics[width=1\textwidth]{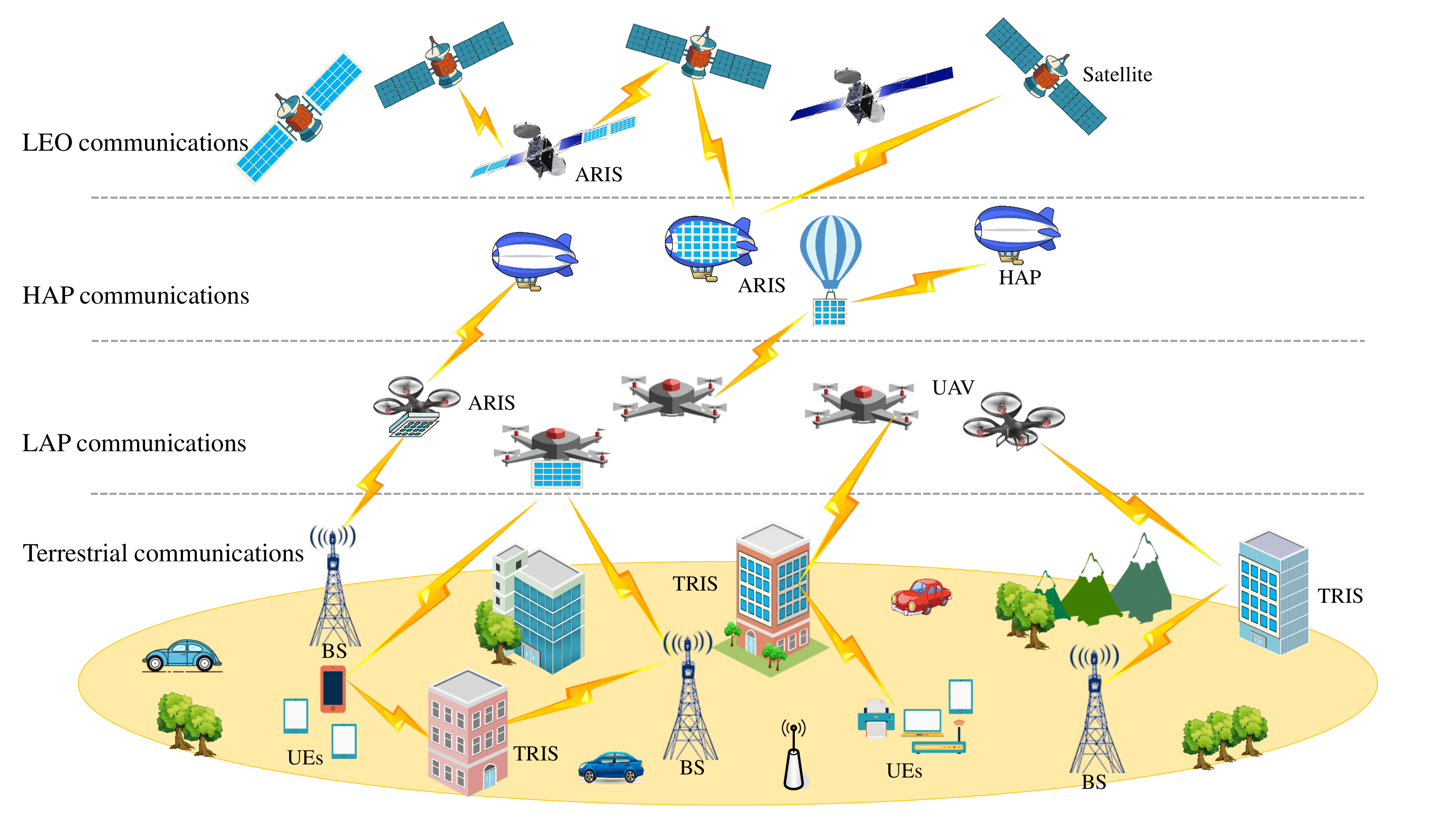}
\caption{RIS-assisted Non-terrestrial Network.}
\label{NTNs}
\end{figure*}

Under the assistance of RISs, the innovative NTN systems composed of various aerial platforms are shown in Fig. \ref{NTNs}. The RIS can contribute alongside aerial platforms to introduce rich scattering of LoS links for ground users (gUs) by optimizing the phase shifts of the reflecting elements. It is a promising technology regarding performance enhancements, especially when the target receiver is far from the serving transmitter or the direct link between the source and the destinations are blocked. In the absence of channel knowledge, RIS can be used for spatial diversity to combat channel impairments (fading and path loss). The deployment of the RIS at the edge of the base station coverage area greatly extends the reach of the incident signal signal-to-noise ratio (SNR). The optimization of the positions and phase shift vectors of RISs for reaching the desired end device locations, whether static or mobile, effectively extends the original cell coverage in the desired direction. Before introducing the various NTN systems, we firstly review the distinct functionality and characteristics of RIS in NTNs. 

\subsection{Reconfiguration Intelligent Surface}
As a key component in RANTNs, RISs can constructively combine the reflected waves at the destination or compensate blocked communication links to produce an enhanced transmission and reduce the channel attenuation. The jointly beamforming at transmitters and RISs provides the sharp directivity from the transmitter to receivers and suppresses the diffusion of transmitted signals and the transmission power. On the other hand, RISs provide an efficient solution to the problem of interference in the open area by controlling the direction in which radio waves are reflected to prevent them from penetrating the other users. Additionally, RIS is leveraged to enhance the communication quality of the legitimate links and weaken that of the eavesdropping links. 

Owing to a vast amount of low-cost reflection elements, RIS allows for more benefits than conventional relays based on sending and receiving data between devices through forwarding. 
\begin{itemize}
    \item {\textit{Less Power Consumption:} RIS reduces the used power for communication functions, since directing signals can be realized in a nearly-passive way, without an RF source or power amplifier, and only a minimum amount of power is required for the RIS control unit. }
    
    \item {\textit{Shorter Transmission Delay:} The data transmission through RIS will experience less intermediate delays to relay the information compared to conventional relays, which commonly adopt half-duplex mode. This is because, in a decode and forward half-duplex relaying mode, the transmission is executed over two time slots, while RIS requires only one time slot since the RIS operates in a full-duplex relaying mode. }
    
    \item {\textit{Simpler Hardware:} The passive nature of the RIS makes its relaying strategy overcome any antenna noise amplification and self-interference, which translates into less computation and lower power consumption than for active full-duplex relays. Additionally, RIS provides a promising but inexpensive solution to mimic the massive MIMO gain. As a result, RISs enable NTNs with single-antenna aerial platforms to enjoy a high passive beamforming gain without the need to deploy multiple antennas on aerial platforms.}
    
    \item {\textit{Longer Duration and Lower Cost:} The simplification in communication circuits and minimization of consumed energy indicates the energy-limited devices, no matter for ground devices or aerial platforms can reserve more energy for information transmission, which yield extended duration. }

\end{itemize}

Despite many advantages mentioned above, the realization of RANTNs faces several new challenges resulting from the controllability of reflecting elements at RIS and the aerial platforms. As shown in Fig. \ref{TRIS_buildling} and Fig. \ref{ARIS}, the additional control unit is required to deploy aerial platforms and adjusting the passive beamforming at RIS. Generally, the control can be managed at two levels: Ground control stations and aerial platforms.

\textbf{Ground Control Station}: The ground controller consists mainly of a processing unit that analyzes the sensed data from the aerial platform, the users' localization conditions, and the information exchange with gBS and/or gateways. Given a set of global policies (e.g., flying regulations, power, etc.) and objectives (e.g., coverage, SNR etc.), the processing unit ensures the joint management of the aerial platform's flying and communication functions. The ground control station coordinates the two working modes  of RIS, namely the receiving mode for channel estimation and the reflection mode for data transmission. The ground control station continuously controls the altitude of the aerial platforms and the phase-shift of the reflecting elements to serve the target devices and maintain their required quality of service (QoS)\cite{samir2020optimizing}. Specifically, the authors in \cite{9308937} assumed there is a smart controller on the ground with a central processing unit monitoring the channel estimation and location prediction. All transmission parameters are known and adjusted by the central processing unit, including the locations of aerial platforms and users, as well as the CSI of the propagation channels.

\textbf{Onboard Aerial Platform Control}: 
The onboard aerial platform's controller mainly consists of the flight control unit. It receives motion commands from the ground control station to ensure the platform's stabilization. For example, an embedded micro-controller is assumed in \cite{lu2020aerial}, allowing the aerial platform to communicate with the source node through a separate reliable wireless control link, thus enabling the instantaneous control of the aerial platform.  Another possibility is to embed  an active antenna  onto the UAV to receive a control signal from the ground control station \cite{zhang2020distributional}. In most cases, a specific channel is assigned between the ground control station and aerial platforms for control signaling \cite{zhang2019reflections}. 

Typically, it is assumed that the required phase shifts are transferred to the memory of the RIS controller. However, such an approach demands a fully synchronized and reliable control link between the computing node and the RIS. This can be achieved in stationary use cases or in other cases where the control link can be easily provided \cite{abdalla2020uavs}. More interestingly, a sensor can also be integrated into the RIS system, which for example, collects environmental information such as temperature, humidity, illuminating light, etc., and sends it to a smart controller at the RIS via a wired link. Then, the controller transmits the collected information to the BS by adjusting the on/off state of the RIS \cite{hua2020uav}. In fact, this idea has been tested experimentally in \cite{ma2019smart}, where a motion-sensitive smart metasurface was integrated with a three-axis gyroscope on an aircraft to sense its direction of motion, allowing a smart controller to adaptively adjust the RIS phase shifts to maintain a beam pointed at the desired receiver. Each RIS reflecting element updated by the smart controller is able to induce an independent phase shift on the incident signal to change the signal propagation such that the desired and interfering signals can be added constructively or destructively to assist the communication system.

Besides the phase shift control, where to deploy RIS is an importatn question  in the design of NTNs. Since the lightweight and conformal geometry of RIS enables the installment onto the facades of almost all ground and aerial objects, the deployment of RIS can be divided into terrestrial RISb(TRIS) and aerial RIS (ARIS). 
\subsubsection{Terrestrial RIS}
\begin{figure*}
     \centering
          \begin{subfigure}[t]{0.3\textwidth}
         \centering
         \includegraphics[width=\textwidth]{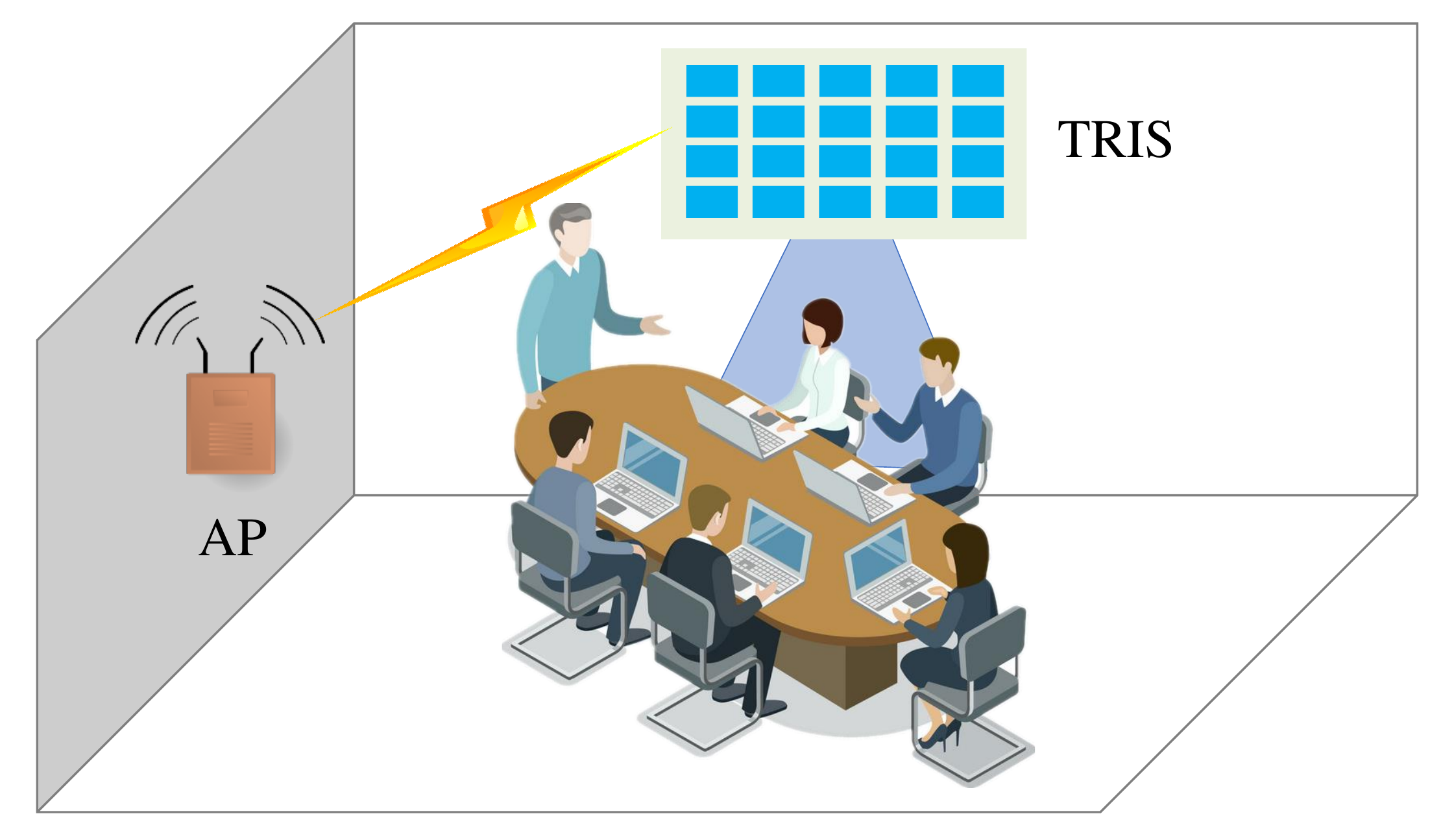}
         \caption{TRIS-aided indoor network}
          \label{TRIS_indoor}
     \end{subfigure}
     \hfill
     \begin{subfigure}[t]{0.3\textwidth}
         \centering
         \includegraphics[width=\textwidth]{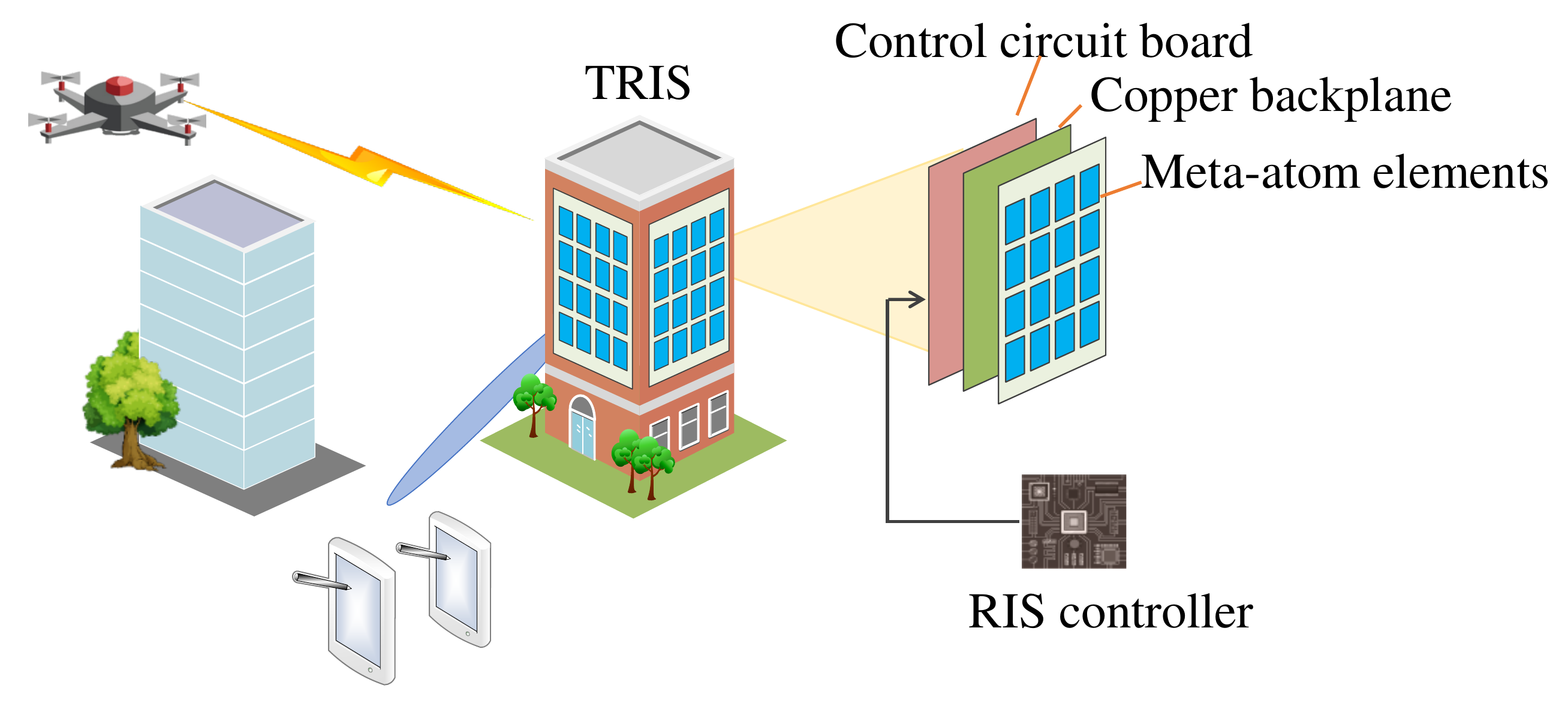}
         \caption{Architecture of TRIS}
         \label{TRIS_buildling}
     \end{subfigure}
      \hfill
     \begin{subfigure}[t]{0.3\textwidth}
         \centering
         \includegraphics[width=\textwidth]{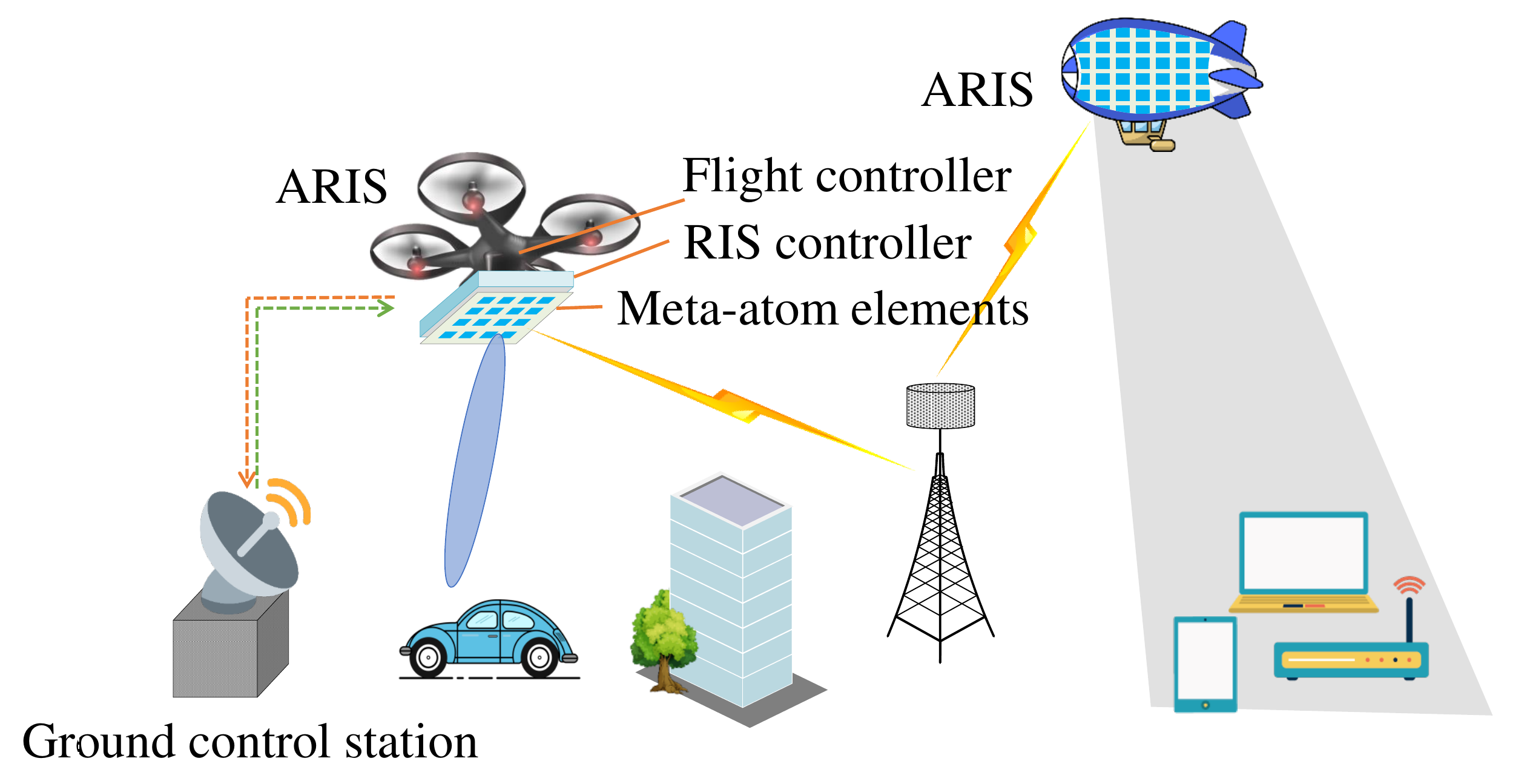}
         \caption{Architecture of ARIS}
          \label{ARIS}
     \end{subfigure}
        \caption{Architectures of TRIS and ARIS.}
        \label{TRIS}
\end{figure*}

As one of the key technologies to support air-ground communications in RANTNs, TRIS paves the way for the communication between air platforms and ground users with satisfactory service quality. The realization of TRIS into NTNs is completed by coating RIS to the facades of buildings for outdoor communications, or by installing it on the wall for indoor communications. As shown in Fig. \ref{TRIS}, TRIS is composed of meta-atom elements, copper backplane, and control circuit, and the phase shift of each reflection elements can be adjusted by the controller \cite{QingQu2020magazine}. The ground control station estimates the channel state information and angle of arrival between users and the aerial platform to determine the best TRIS configuration setup.

By integrating TRISs in NTNs, concatenated virtual LoS links between aerial platforms and mobile users can be formed via passively reflecting the incident signals, leading to extend the coverage and reduce the UAV's movement. 
 As a matter of fact, TRIS-assisted NTNs can effectively enhance the received signal power, extend the network coverage, and increase the link capacity. With the aid of TRIS, one can adjust the phase shift of the TRIS instead of controlling the movement of aerial platforms for forming concatenated virtual LoS propagation between the aerial platforms and the ground users. Therefore, the aerial platforms can maintain hovering status rather than deliberately altering their route and flying close to their users to establish strong communication links, which is usually time-and energy-consuming. The movement of aerial platforms happens only when concatenated virtual LoS links cannot be formed even with the aid of the TRIS, which in turn, leads to the reduction of aerial platforms' energy dissipation and maximizes the endurance of the aerial platforms.

\subsubsection{Aerial RIS}
With the deepening research on RIS, academia has proposed a new aerial architecture, the ARIS, to assist the information transmission, where RIS can be carefully mounted on an aerial platform to create an intermediate reflection layer between end devices. Besides the advantages mentioned above, ARISs enable lighter aerial platforms payload without signal processing circuits, even when a large number of reflectors is deployed. This comes from the fact that RIS is made of thin and lightweight materials. The decrease in the weight of satellites and HAPs enabled by RIS saves a huge amount of cost per operation due to the fact that the aerial platform launch cost is related to its weight. For instance, the cost of launching satellites into an orbit is related to the weight of the satellite system with \$6000 per kg. On the other hand, given the reduced communication components and payload, the required stabilization energy is minimized. 

Similar to TRIS, ARIS also contains meta-atom elements and RIS controller \cite{9356531}, as shown in Fig. \ref{ARIS}. Since AIRS is integrated on aerial platforms, an additional flight controller and a ground control station are needed to achieve its dynamical deployment and aerial platform's trajectory. Unlike TRIS, the onboard aerial platform is also equipped with the RIS control unit. The RIS control unit receives the optimized RIS configuration, translates it into a switch control ON/OFF activation map, and applies it to the metasurfaces layer for directing incident signals to targeted directions. 

Integrating RIS on aerial platforms can be done in several ways depending on the platform's shape. For instance, the RIS can coat the outer surface of the aerial platform, or it can be installed as a separate horizontal surface at the bottom of an aerial device, as can be seen in Fig. \ref{ARIS}. This enables intelligent reflection and MIMO transmissions from the sky, making aerial platforms act as aerial relay stations. In addition, a special case of RIS, known as intelligent omni-surface, has antenna elements on both sides of the meta-surface and can reflect incident signals coming from the opposite directions. An unobstructed intelligent omni-surface (IOS) reflects the incident signals on both sides of the sheet and can cover dead zones, providing massive $260$ degree coverage and higher spectral efficiency. The aerial platforms provide this capability by carrying the IOS underneath them and being able to fly at suitable heights to provide reflective RF surfaces where needed. The greatest strength point of the intelligent omni-surface is its ability to control the direction of the departure signal from the RIS to the potential receivers without any blind spots by adjusting the phase shift vectors on either side.  

\subsubsection{Comparisons}
It is obvious that whether the RIS is deployed on terrestrial objects or on aerial platforms, the integration of RIS can enhance the performance of NTN systems. Here, from the perspective of TRIS and ARIS features, we
briefly introduce the comparison between them.
\begin{itemize}
    \item {\textit{Deployment:} One of the most difficult tasks in the practical implementation of TRIS is finding the appropriate place. The TRIS installation process involves a lot of issues, such as, site rent, the impact of the urban landscape, and the willingness of owners to install large RIS on their properties. However, the deployment problem does not bother ARISs flying in the air, which enjoys higher deployment flexibility. }
    \item {\textit{Coverage:} TRIS deployed on the walls or facades of buildings can at most serve terminals located in half of the space, that is, both the source and destination nodes must lie on the same side of the RIS. Compared to the conventional TRIS, ARIS is able to achieve 360 degree panoramic full-angle reflecting, i.e., one ARIS can, in principle, manipulate signals between any pair of nodes located on the ground.  
    }
    \item {\textit{LoS Connection:} In a complex environment like urban areas, the radio signal coming from a source node has to be reflected many times before reaching the desired destination node, even with the presence of a sufficient number of TRISs. This  leads to significant signal attenuation since each reflection, even by TRIS, would cause signal scattering to undesired directions. The ARIS can be deployed more flexibly with an elevated position to cater to the real-time communication environment, where the LoS links between ARIS and the ground ends can be easily established with high probability, especially in the crowded urban scenario. The authors in \cite{zhang2020distributional} found that while TRIS yields an LoS probability of around 50\%, an  ARIS results in a probability of over 60\%. The placement of trajectory of aerial platforms can be more flexibly optimized to further improve the communication performance, thereby offering a new degree of freedom (DoF) for performance enhancement via 3D network design.  ARIS is usually able to achieve desired signal manipulation by one reflection only, even in the complex urban environment, thanks to its high likelihood of having LoS links with the ground nodes. This  greatly reduces the signal power loss due to multiple reflections in the case of TRIS. }
    
    \item {\textit{Channel Variation:} ARIS outperforms TRIS at the cost of experiencing higher fluctuating channels. Specifically, the position variation of the aerial platform leads to the channel variation in both hops in the ARIS-assisted system, while only the link between the aerial platform and TRIS will be impacted in the TRIS-assisted system. Due to the movement of aerial platforms, the signal misalignment happens in both hops of ARIS-assisted system, while it occurs in only one hop of TRIS-assisted system. Therefore, the ARIS passive beamforming is more ineffective than the TRIS passive beamforming. Moreover, the channel estimation/tracking to acquire the ARIS channels along their 3D trajectories,  is more challenging than in TRIS deployed at fixed locations. }
    \item {\textit{Lifetime:} The limited endurance of aerial platforms makes ARIS quite challenging to build a reliable network that is available for a long time. The power consumption minimization problem will be more important to ARIS to extend the service time. This problem is overcome by TRIS, which can be continuously powered by backhaul links. }
\end{itemize}

{\subsection{Frameworks of RIS-assisted NTNs}
In RANTNs, RISs are deployed to enhance the information transmission between aerial platforms and ground users, ground devices and ground devices, as well as aerial platforms and aerial platforms. In general, the RIS-assisted communications included in RANTNs can be divided into three types according to the types of end devices: RIS-assisted A2G/G2A communications, ARIS-assisted G2G communications, and RIS-assisted A2A communications. In the following, we will introduce the frameworks of these three communication types assisted by RIS.

\subsubsection{RIS-assisted A2G/G2A Communications}
The RIS-assisted A2G/G2A Communications have been widely investigated in the existing literature by treating the aerial platforms as one of the end terminals. Specifically, the aerial platforms can act as ABSs to serve ground users with the assistance of TRISs or ARISs, or they can also act as aerial users served by terrestrial base stations. Therefore, based on the functionality of the aerial platforms, the use cases of RIS-assisted A2G/G2A communications can be further divided into two cases. 

\noindent\textbf{Aerial Base station:} In this scene, ABSs replace ground base stations. In this way, the ground space is saved and the entire system is miniaturized.  As shown in Fig. \ref{ABS}, one common scenario involving aerial base stations is the one where  UAVs, HAPs, and satellites act as aerial BS to build uplink and downlink transmission links with a set of ground users, typically representing  internet of things devices (IoTDs).  
These aerial base stations can solely serve one user or  multiple ground users simultaneously by adopting suitable multiple access schemes, such as time-division multiple access (TDMA) \cite{cai2020resource,zhang2020data}, and frequency division multiple access (FDMA) \cite{9293155}.
To further improve the performance, 
an ARIS/TRIS can be used to provide an additional link between ABS and the ground user when the LoS link suffers from severe propagation deterioration. 
The use of ARIS/TRIS leads to several major benefits. First, it enables the aerial BS with limited capabilities and service duration to provide high data rate transmission with significantly improved coverage, capacity, and connectivity. Second, the size and weight of ABS can also be greatly reduced compared to the one adopting conventional MIMO technology, thus extending the battery life of the ground device because of reduced processing energy \cite{tekbiyik2021graph}.

In addition to information transmission, under the assistance of TRIS, these flying BSs can even provide power supply to ground IoTDs. In this context, ABSs serve in the scene of vast IoTDs by transferring all the necessary information and energy to all of the IoTDs with high efficiency and high security. The IoTDs distributed in every aspect of {our life} ranging from smart cities, home appliances, and transportation, can use the harvested energy to gather and disseminate information from the surrounding environment and upload it to ABSs. 

Moreover, few works proposed some interesting frameworks, which are worth more attention. Considering the communication scenarios with multiple ABS-gU pairs but with only limited RISs, the overall system capacity can be enhanced by dividing the RIS into sub-array partitions to serve different ABS-gU pairs \cite{cao2021reconfigurable}. Such a system aims not only to enhance the transmission of interest but also to cancel interference among multi-group A2G \cite{mu2021intelligent}.  
 To further enhance the communication quality, multiple available RISs are also considered to provide multiple reflections between the ABSs and the target ground users \cite{ge2020joint, wang2020joint}. The potential of RIS can be further exploited by dynamically allocating RISs to serve different users as the propagation environment varies \cite{cang2020optimal}. 
However, the consideration of multiple TRISs remain little explored in the context of A2G/G2A communication, while we are not aware of any works considering multiple ARISs. 

Overall, the A2G/G2A communications composed of ABSs, RISs, and ground terminals is expected to find a place in future networks. In practice, it will be useful for instance when the terrestrial infrastructures are destroyed due to natural disasters or had not been installed. It is also applicable in a crowded area for grand events, where the surrounding base stations are overloaded with heavy communication traffic. 

\begin{figure*}
     \centering
          \begin{subfigure}[t]{0.45\textwidth}
         \includegraphics[width=\textwidth]{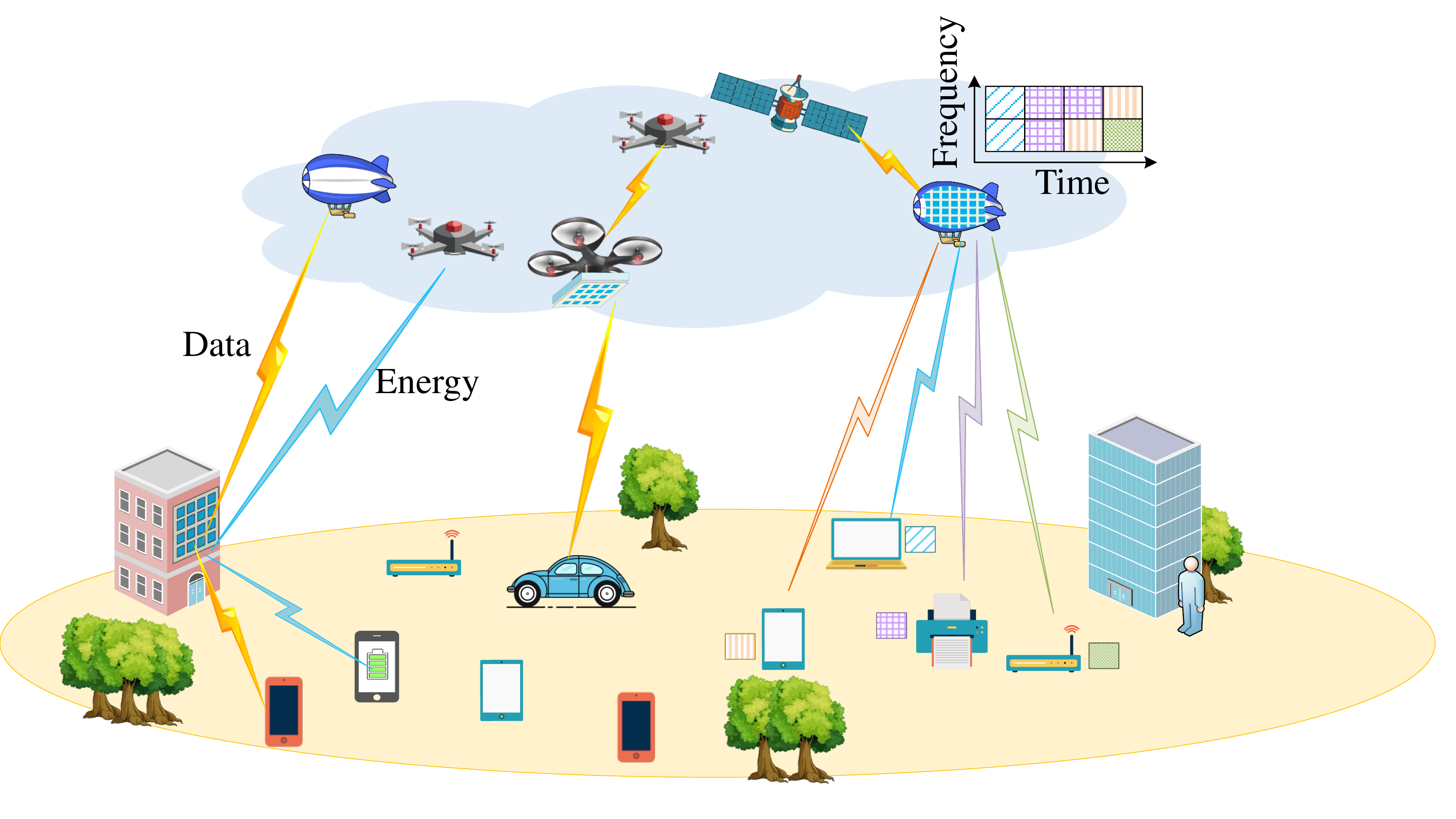}
         \caption{A2G/G2A communications with aerial base stations}
         \label{ABS}
     \end{subfigure}
     \hfill
      \begin{subfigure}[t]{0.5\textwidth}
         \includegraphics[width=\textwidth]{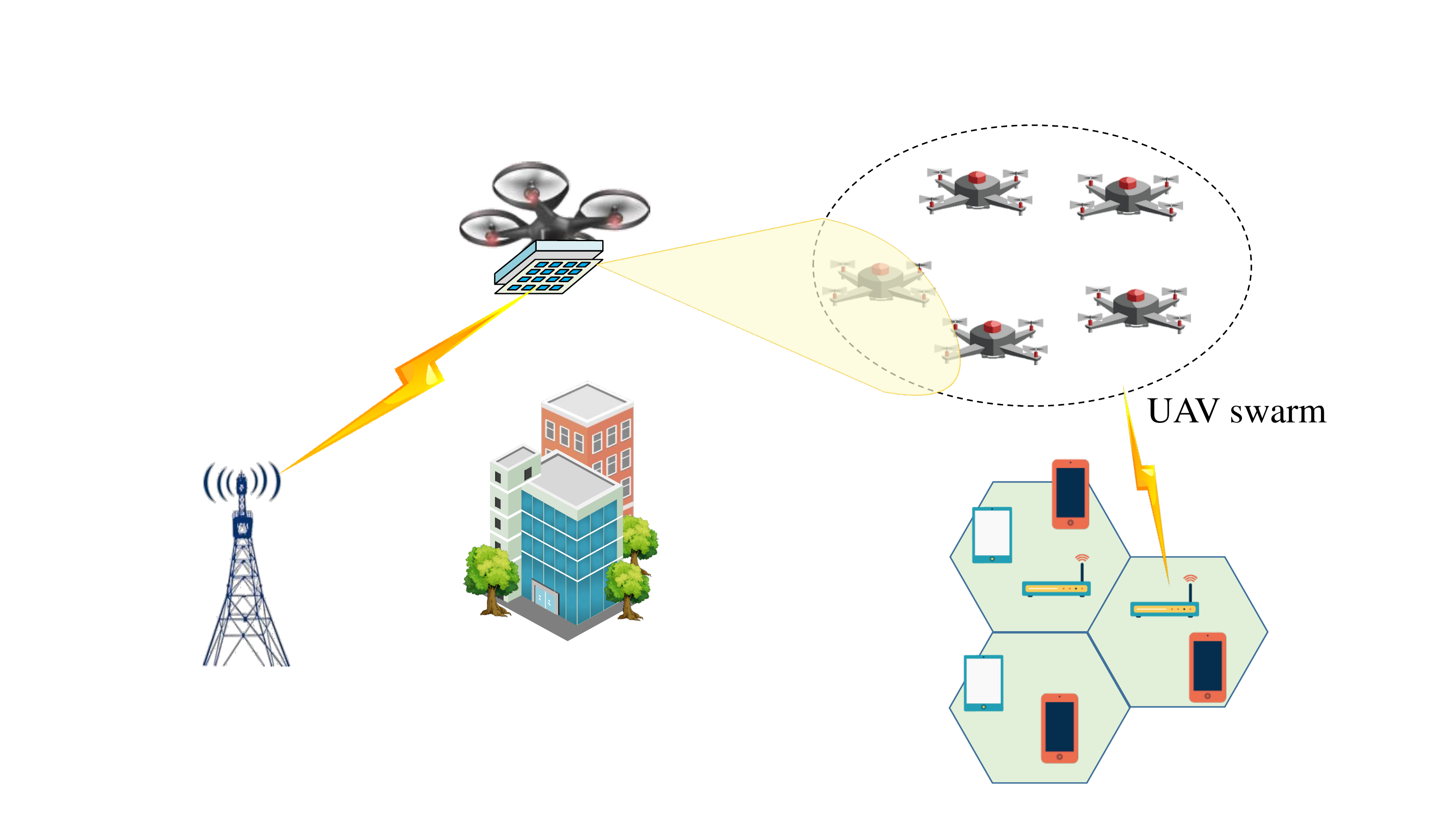}
         \caption{A2G/G2A communications used for backhauling}
         \label{Auser_B}
     \end{subfigure}
        \caption{A2G/G2A communications}
        \label{A2G}

\end{figure*}
\noindent\textbf{Aerial Devices:}
The A2G/G2A communications also include the use cases where aerial platforms act as aerial devices served by terrestrial base stations. This framework is similar to the one shown in Fig. (\ref{ABS}) but with aerial platforms acting as aerial users served by gBS. 
Similarly,  the RIS may be used to provide additional transmission links, combat the path loss, improve received channel gains, or mitigate the interference at other un-served aerial users caused by the omnidirectional antenna at BS and the open air communication environment \cite{hashida2020intelligent}. 
A typical scenario of this framework is represented by the case when these aerial devices act as IoTDs such as cameras and communication devices to collect environment information from the air and then send it to ground BSs with the help of RIS. Nevertheless, there are two less common yet interesting scenarios involving A2G/G2A communications. The first one corresponds to the case in which 
A2G/G2A networks  is  utilized for wireless traffic backhauling from remote area base stations. In this case, as shown in Fig. (\ref{Auser_B}),  signals transmitted from remote BSs are smartly reflected toward the gateway aerial stations connecting to the core network \cite{jeon2021energy,7486987}.
The second scenario is that of symbiotic communication in which the aerial platform plays the role of a primary transmitter assisted by a RIS that transmits information through the on/off states and performs passive beamforming at the same time.
A typical use case of symbiotic communication system is when   
 environmental information such as temperature, humidity, illuminating light, etc. is collected by a sensor deployed in the RIS and sent to a smart controller at the RIS via a wired link. Then, the controller transmits the collected information to the BS by adjusting the on/off state of the RIS. 
\subsubsection{RIS-assisted G2G Communication}
In the G2G NTNs, the aerial platforms assist data transmission between ground devices rather than acting as transmitters or receivers as in A2G/G2A NTNs. As shown in Fig. (\ref{G2GARIS}), the aerial platforms are equipped with RIS, which are called ARIS to refer to their being in the air. They serve in this case to  assist either the transmission between ground BS and served ground users \cite{zhang2020distributional,mahmoud2021intelligent,ma2020enhancing,jia2020ergodic,alfattani2020link}, or the transmission between ground communicating pairs \cite{shafique2020optimization}. The main role of ARIS in this kind of systems is to extend the coverage of the gBSs, enhance communication reliability, and improve spectral efficiency by the additional degree-of-freedom and the flexible deployment of ARIS. Such a functionality is typically beneficial to communication scenarios in which a sea of IoTDs are required to interact and exchange data with gUs but are often equipped with limited capabilities and cannot communicate over long distances in a reliable manner \cite{samir2020optimizing}. 

In some scenarios, the RIS can be deployed in the ground. As shown in Fig. (\ref{G2GTRIS}), the network in this case is 
composed of ground terminals, one TRIS, and one flying aerial platform. The transmitted signals from the ground device are forwarded at the aerial platform, then reflected at TRIS, prior to being received by the ground receiver. Typically, the TRIS can be employed to reflect the signals transmitted from a ground source to a UAV, and the UAV acts, in turn, as a relaying station employed for transmitting the signal to the destination by using a decode-and-forward (DF) protocol \cite{ranjha2020urllc, yang2020performance}. {In this way, such a network combines the reconfigurable channel abilities of RISs and the flexible deployment of UAVs in a novel way. Specifically, the TRIS enhances the connection between the transmitter and the UAV relay, while the UAV can adjust its position to avoid the signal reflected from the TRIS being blocked by obstacles.}

  Two other scenarios  have been considered within the framework of RIS-assisted G2G communications. 
  The first one is wireless traffic backhauling between  a remote gBS and a core network through ARIS. As compared to conventional backhauling, such a solution avoids the high cost of fiber optic infrastructure \cite{Tekbiyik}. The second one is that of cell-free massive MIMO systems \cite{9308937} which constitute the core architecture of next-generation mobile communication networks. In these systems, a number of users in a certain area are served by multiple APs simultaneously.  To overcome  the limitation of inter-cell interference encountered in traditional cellular mobile networks, these APs apply advanced beamforming technologies. In this context, the use of ARIS can help enhance communication links between APs and users with blocked direct links.  

\begin{figure*}
     \centering
          \begin{subfigure}[t]{0.45\textwidth}
         \includegraphics[width=\textwidth]{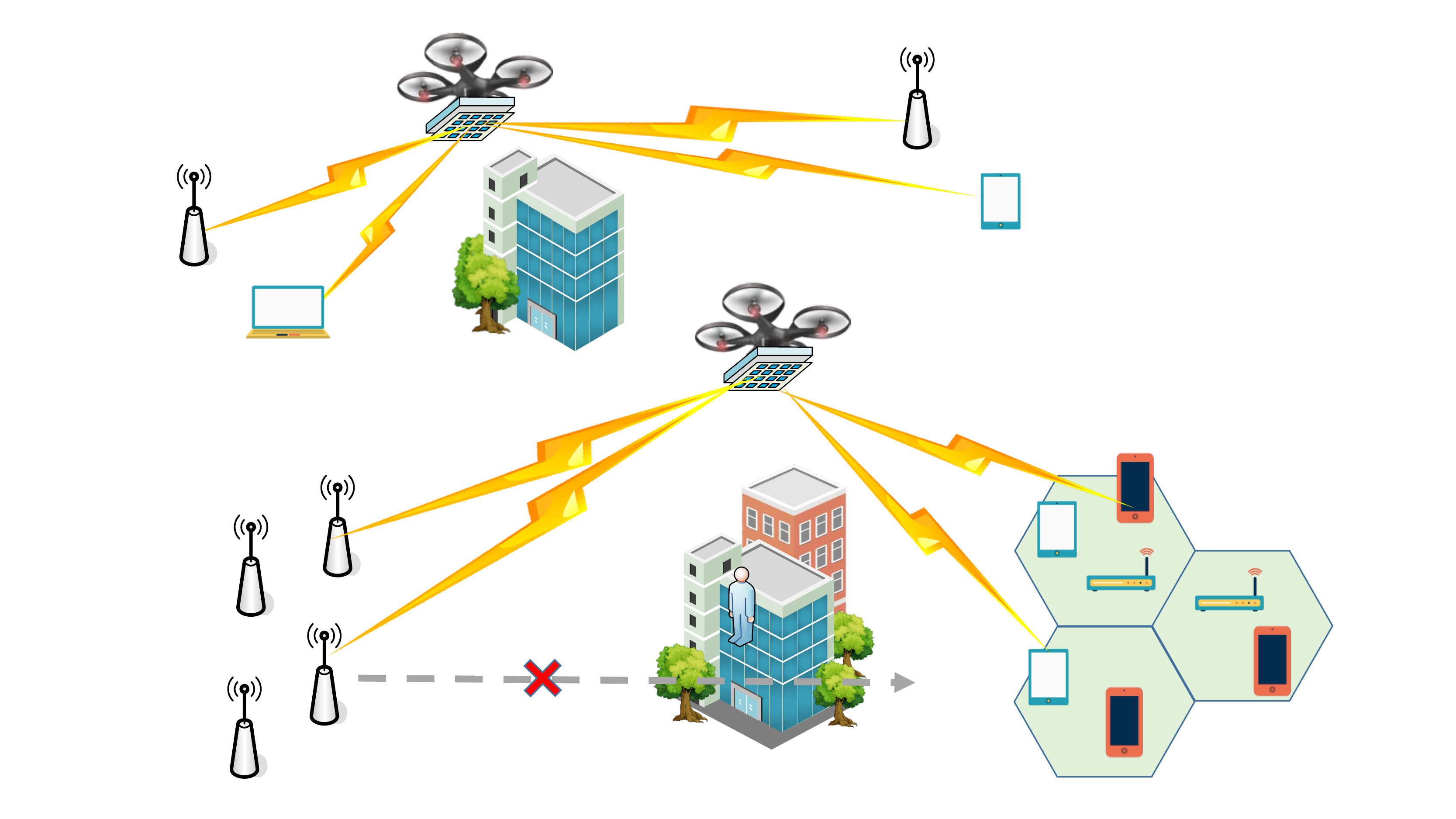}
         \caption{G2G communications with ARIS}
         \label{G2GARIS}
     \end{subfigure}
     \hfill
     \begin{subfigure}[t]{0.45\textwidth}
         \includegraphics[width=\textwidth]{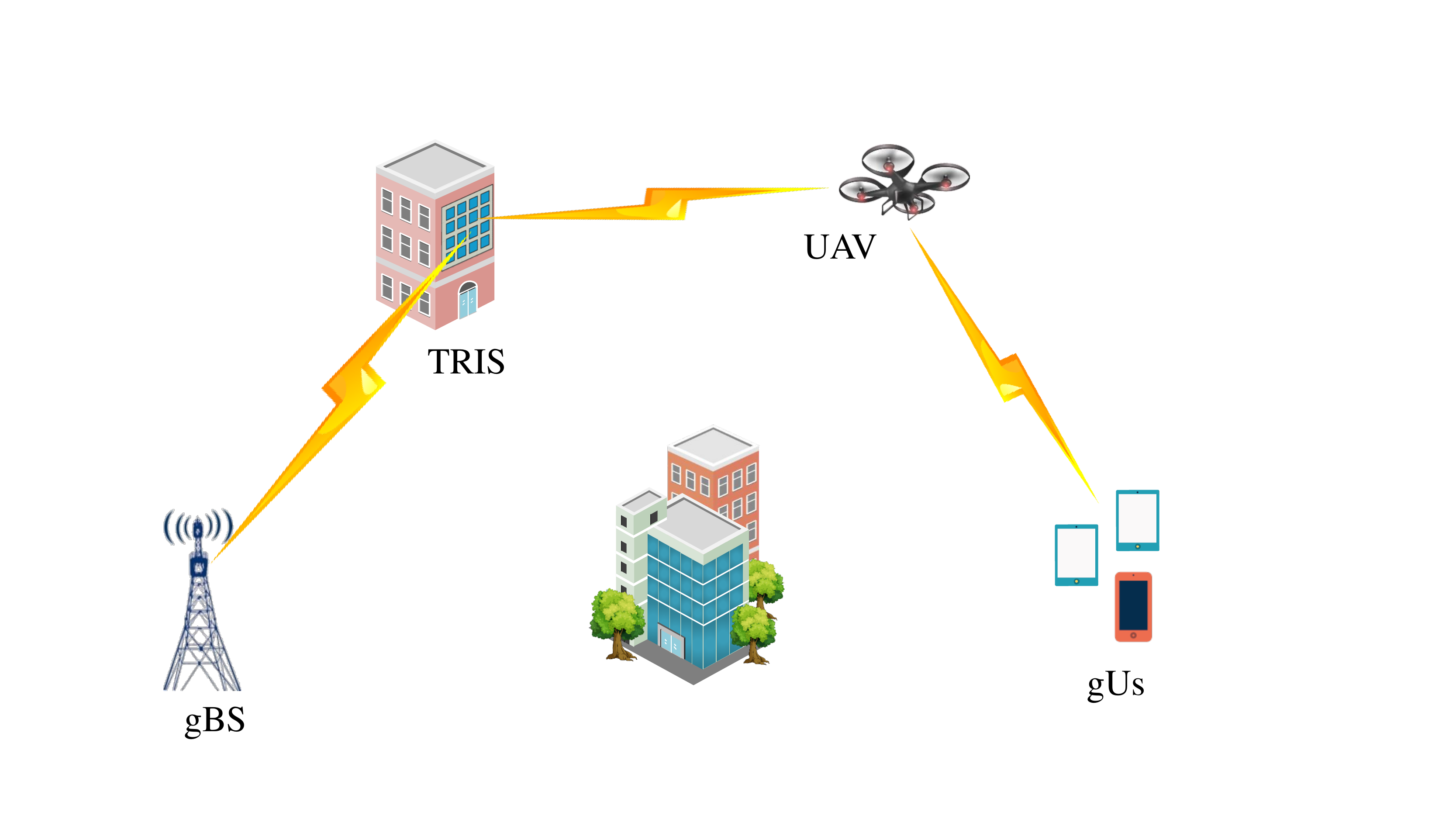}
         \caption{G2G communications with TRIS and UAV}
         \label{G2GTRIS}
     \end{subfigure}
        \caption{G2G communications}
        \label{G2G}
\end{figure*}


\subsubsection{RIS-assisted Air-to-Air Communication} Connections are also needed to be built between aerial platforms, as shown in Fig. \ref{A2A}. In this case, ARIS is considered to be the most effective way to assist the transmission since both end terminals are distributed in the air space. {In this context, the use of ARIS in LEO satellite communication systems to assist the inter-satellite link connection is a promising tool to address the high power consumption and low diversity order problem \cite{tekbiyik2020reconfigurable}.} The RIS carried by HAPs and LEO satellites provides seamless and ubiquitous connectivity for deep space networks. However, the studies on A2A communication assisted by RIS are still in their infancy. More valuable frameworks regarding the UAV-UAV, UAV-HAP, UAV-satellite, HAP-HAP, and HAP-satellite transmission are important topics that are worth deeper investigation. 

\begin{figure}[t]
\centering
\includegraphics[width= \linewidth]{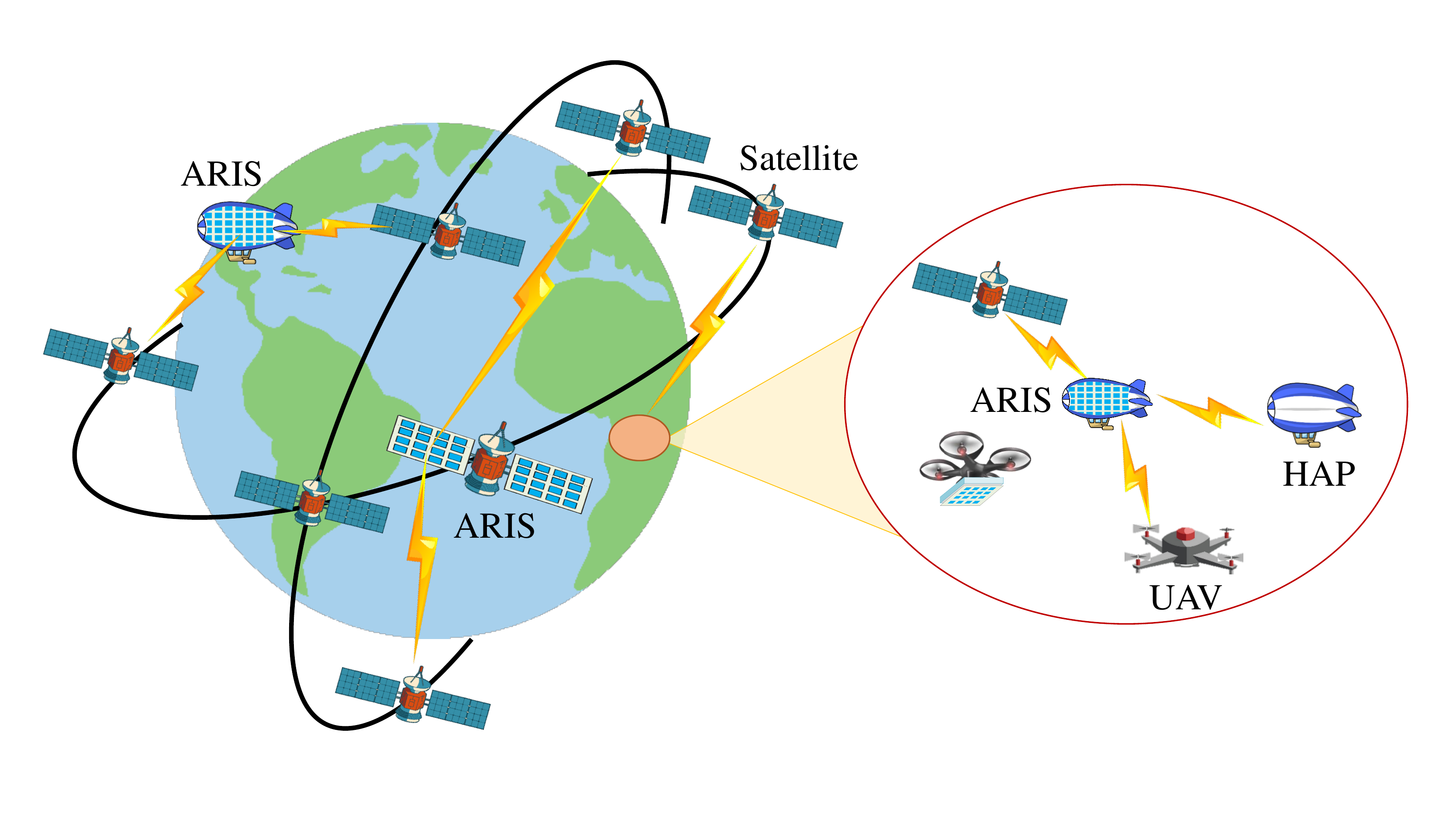}
\caption{A2A communications}
\label{A2A}
\end{figure}

\textit{Summary:} {In this section, we classified  RANTNs into three categories according to the types of end devices, namely RIS-assisted A2G/G2A communications, RIS-assisted G2G communications, and RIS-assisted A2A communications. We reviewed the novel frameworks proposed by existing works with great potential to be applied in the 5G beyond/6G network. However, additional research efforts are still in need  to build more powerful RANTNs to fully explore the potential of RISs.}

\section{RIS-Assisted NTNs with Other Technologies}
Combined with the advanced technologies in the next-generation wireless communication, including but are not limited to optical communication technologies, mmWave/THz communications, wireless power transfer, multiple access schemes, etc., the communication performances of RIS-assisted NTN can be further enhanced. Thus, we will review the current research work on RIS-assisted NTN with advanced technologies in this section. 

\subsection{Optical Wireless Communication System}
Conventional radio-frequency-based wireless communication is now seriously challenged by the overcrowding RF spectrum, leading to insufficient capacity to support the ever-increasing wireless data traffic. The idea of optical wireless communication (OWC) has been presented as a promising solution for obtaining larger bandwidth and offsetting the frequency spectrum crowding problem \cite{arnon2012advanced,elgala2011indoor}. 

 \noindent{\bf Free space optics.} Free space optics (FSO), which utilizes the coherent lightwave of the laser diode to transfer information in free space, is an attractive technology due to its numerous advantages such as high rate capability, license-free, wide spectrum available, long-distance, high energy efficiency (EE), reduced interference, directivity, inherent security and robustness \cite{khalighi2014survey}. These advantages make the FSO system a potential supplement candidate for new generation wireless communication. However, using a laser as a propagating medium requires FSO to have line-of-sight between the light source and the photo detector. This issue can be solved by using some relaying methods to satisfy the coverage requirement of FSO communication \cite{petkovic2017analytical,dabiri2018performance}.

 On the lookout for constructing  a flexible and energy-saving communication system, the combination of the described FSO communication into RANTNs remains an attractive option. 
In this context, the authors in \cite{jia2020ergodic} considered an FSO communication assisted by ARIS and showed that the system performance is influenced by the atmospheric conditions, the configurations of laser devices, and the steady ability of UAV. In particular, the ergodic capacity (EC) decreases as the index of refraction structure parameter increases, indicating a stronger atmospheric turbulence. On the other hand, EC decreases when the radius of the light source or the UAV vibration increases, as they cause pointing error losses. 
 Moreover, the EC will be reduced when the photoelectrical responsibility decreases, because the energy transformation efficiency drop leads to lower electrical SNR.  The authors also show that by changing the location of ARIS, the link distance variation will cause the area of beam footprint on photodetector to change, resulting in markedly different EC.

\noindent{\bf Visible light communication.} As far as optical communication is concerned, visible light communication (VLC) represents a potential  candidate that can be deployed in RIS-assisted NTN systems.  Based on utilizing the incoherent lightwave of light-emitting-diode to transfer information in the wireless channel, this technology has been presented as a popular solution for short-distance broadband wireless communication \cite{grobe2013high, wu2014visible}. It presents the advantages of incurring low deployment costs while providing ultra-high data rates, and operating in the unlicensed spectrum without health hazards \cite{pathak2015visible,karunatilaka2015led}.
 Besides, compared with traditional RF, VLC can provide communication and illumination simultaneously. However, there are several challenges towards VLC implementation, including the limited coverage range, the signal loss caused by little movement, the misalignment between the light-emitting diodes transmitter and photodetector receiver, and the need for LoS \cite{wu2014visible}. The RANTNs can help  overcome these drawbacks and improve the performance of VLC. 
 On the technological level, the RIS used in VLC may use a meta-lens or crystal liquid-based RIS to shape the environment of the incident light signals through dynamic artificial muscles and the refractive index \cite{ndjiongue2020towards}.  

A study about the TRIS-assisted VLC communication system was conducted in \cite{cang2020optimal}, where multiple ABSs are assumed to provide communication and illumination for ground users simultaneously. Each ABS is only under the assistance of one RIS at most, but can serve multiple users simultaneously. Targeting the minimization of the energy consumption of ABSs while meeting the transmission data rate and illumination requirements of users, the authors propose a joint optimization framework of the ABS deployment, the reflecting beamforming at RIS as well as the user and the RIS association. The obtained results show that the total transmit power at ABSs increases when more ground users are served. On the other hand, when the height of the UAV increases,  the total transmit power  decreases first and then increases later as the height of the UAV increases. This is in large part due to the fact  that
	the cosine of emission angles and the communication distance, both increasing when the ABSs fly from low to hight,  possess an opposite effect on the performance. While the increase in the cosine of emission angles enhances the system performance, the increase in the communication distance leads to a higher path-loss and thus degrades the performance.   
	Moreover, even when setting all phases equal to zero,  the total transmit power  can be reduced by 21.73\% on average.

\subsection{mmWave/THz}
One of the unique features of future 6G wireless networks is the use of frequency bands above 100GHz, which brings millimeter wave (mmWave) and terahertz (THz) communication to our life \cite{zhang2017investigation}. The large available bandwidth enables them to be the essential components to support the ultra-high data transmission for applications including but  not limited to virtual reality, high definition video broadcasting. However,  high-frequency bands experience tough environmental conditions due to molecular absorption, resulting in severe attenuation and   high path losses \cite{sulyman2014radio,jornet2011channel}. Furthermore, the small wavelength of mmWave and THz spectrum yields the high susceptibility to blockage caused by common objects, such as buildings and foliage, which seriously attenuate the reliability and availability of wireless communication services \cite{qiao2020secure}. In order to bypass obstacles and prolong the communication range, the proposed RANTNs have been considered as an energy-efficient solution for mmWave and THz communications. 

\noindent{\bf TRIS.} {RIS can establish line-of-sight links to combat signal blockage in mmWave/THz communication systems. The authors in \cite{wang2020joint,pan2020uav} adopted TRIS mounted on the exterior wall of the buildings to assist the downlink transmission of a UAV to serve multiple gUs  operating in mmWave and THz frequency, respectively.  As indicated in \cite{pan2020uav}, the THz frequencies experience severe pathlosses at some distance-dependent locations. 
	 As such, path loss peaks are going to appear when the communication distance varies according to the movement of the UAV. Consequently, terrestrial THz transmission approaches cannot be directly applied to  A2G communications. Considering the scenario of a UAV serving multiple users through a TRIS, the authors in  \cite{pan2020uav} proposed a solution based on dividing the THz band into sub-bands, each allocated to a specific UAV-user link. To avoid path-loss peaks, the sub-band allocation is updated for each UAV's position. }

\noindent{\bf ARIS.} Due to the blockage-prone nature of mmWave/THz signals, mobile ARISs are a better option  to further maintain LoS links and enhance communication than stationary RISs. ARISs can improve the reliability of mmWave/THz transmissions by optimizing their location intelligently. In this context, the works in \cite{zhang2020distributional} and \cite{zhang2019reflections} investigated a novel downlink framework using ARIS to assist mmWave BS for single user and multi-user communications, respectively. By leveraging the mobility of ARIS and adjusting its position, the direct NLoS links between BS and users can be replaced by two connected LOS links.   Targeting the optimization of the ARIS's location and its associated   reflection coefficients to optimize the total downlink transmissions, the works in \cite{zhang2020distributional} and \cite{zhang2019reflections} noted that compared to a system assisted by a static RIS, a significant improvement in LoS connection probability can be achieved. 
	Specifically, the authors in  \cite{zhang2020distributional} observed that the LoS downlink probability naturally increases with the altitude of the ARIS, and can be further improved by adopting a suitable trajectory design. However, as shown by the work in \cite{zhang2019reflections}, improving the probability of LoS links by increasing the altitude of the UAV does not result in better performances, due to the higher experienced pathloss. If the altitude of the UAV increases beyond a certain threshold, the gain brought by ARIS is lost, the performance becoming equivalent to that of TRIS.  
	However, with ARIS, it is possible to obtain a faster improvement of the downlink rate than with TRIS by increasing the transmit power. 
	 Furthermore, to ensure a better LoS link, the UAV is found to change its position more frequently in the case of a moving UE than in the case of a static UE. 
	   }

Besides ground-to-ground communications, ARIS operating over high-frequency bands can be used for air-to-ground communications. In this context, the authors in \cite{abuzainab2021deep} considered the setting of a multi-antenna base station serving a mobile aerial user in the THz band. Due to the movement of the aerial user, the LoS link between the BS and the aerial user may not be always maintained. In such a situation, the BS can serve the aerial user through ARIS. A design based on proactive hand-off and beam selection that predicts the best communication link (direct or RIS assisted) as well as the best beamforming vector was proposed in \cite{abuzainab2021deep}.

As far as THz communication is concerned, the use of THz band constitutes a promising solution to overcome the scarcity of spectrum resources and to provide high data rates for inter-satellite links \cite{suen2014global}. Although the path-loss due to molecular absorption is a non-issue for space-based applications of THz waves, a significant decrease of the received power can occur when the transmitter and receiver are misaligned. The impact of the misalignment fading for an RIS-empowered THz band in the inter-LEO satellite communications has been studied in \cite{tekbiyik2020reconfigurable}. In this system, the misalignment fading originates from the sharp beams of THz communication and the LEO satellites moving with high velocity. Under the circular beam assumption, the misalignment coefficient is related to the beam's radial distance and the receiver's jitter variance. The derived results in \cite{tekbiyik2020reconfigurable} showed that the impact of the misalignment loss could be re- reduced by using antennas with large bandwidths. On the technological level, implementing this solution is possible in the THz region using semi-conductors such as graphene and manufacturing techniques like micro-machining. To compensate for the high path loss experienced in high carrier frequencies, the work in \cite{tekbiyik2020reconfigurable} showed that increasing the number of RIS elements on each satellite is a cost-effective solution to reach the desired quality of service.

\subsection{Wireless Power Transfer}
Wireless power transfer has emerged as an effective solution to recharge low-power devices with limited lifetime  
\cite{xie2013wireless,zhang2018wireless}. Based on collecting energy from electromagnetic radiation carried by radio signals, it allows for charging devices wirelessly, thereby ensuring full operation while avoiding the cost of wired charging devices. It has been advocated as a promising technology to power IoTDs in IoT networks since it allows dispensing with wired charging circuits, paving the way for miniaturization. 

To enhance the performance in terms of collected energy, the authors in \cite{zhang2020data} considered the use of TRIS together with specific devices named data beacons (PDBs) that relay data and energy transmission between a UAV and vast ground IoTDs. The PDBs receive energy and data from UAV directly or reflected by RISs, before transmitting power and data to all IoTDs nearby.
Under this setting, the problem of determining the optimal flight trajectory to reach the minimum energy consumption or minimum time consumption is formulated and solved. 
It has been noted in \cite{zhang2020data} that using the UAV to charge IoTDs is not a good option for two major reasons. On the one hand, outfitting the UAV with an omnidirectional antenna to spread energy waves results in low received power, which, in turn, causes a long charging time. On the other hand, using a directional antenna at the UAV is not a feasible option since, in this case, the UAV has to know the exact positions of all IoTDs. Such a condition cannot be satisfied in vast IoTDs settings. As a solution, the work in \cite{zhang2020data} considered the use of PDBs. Since they have known positions, a directional antenna at the UAV can be thus used.  Based on a set of simulations, the work in \cite{zhang2020data}  illustrates the significant performance improvement in energy and time consumption brought by the deployment of RISs and the optimization of the UAV trajectory.

Simultaneous wireless information and power transfer (SWIPT) has emerged as an extension of conventional wireless transfer that enables the simultaneous transfer of data and power. Advanced UAV technologies were proposed in \cite{zhang2020data} to overcome the near-far problem occurring when the IoT devices are distributed far away from each other.  A significant improvement in the information decoding (ID) and energy harvesting (EH) in IoT networks based on SWIPT is thus achieved.  The use of ARIS and UAV offers further opportunities for improved performances by enabling passive beamforming beams alongside the original EH or ID beams.  A problem of interest under this setting is to ensure the joint optimization of many variables, including the  3D trajectory of the UAV, the phase shift vectors, ID and EH performances, the number of RIS elements, and the served ID and EH users. The 3D trajectory optimization aims to ensure the fairness of the EH and ID among the battery-limited ground nodes and
eliminate the side effects of the near-far problem while 
the beamforming optimization of the UAV targets to provide aligned energy and information beams to provide
more energy charging and information transfer capacity.


Although ARIS consumes a small amount of energy, it is advisable to avoid drawing its needs from aerial platforms because of their limited onboard energy. Instead, the RIS can self-power by harvesting energy from the unreflected fraction of the incident signals and converting it into electrical energy via the rectifier \cite{zhang2019reflections}. Structurally, the self-powered RIS is composed of an adjustable antenna and a rectifier. By varying the impedance of the antennas, the mismatch between the antenna structure with the carrier wave can reflect back a portion of the incident signal, while the rectifier can harvest the remaining part to power the reflector. Under this assumption, the amplitude of the reflection coefficient of each element on RIS is not unit anymore but is less than 1. The simulation results given in \cite{zhang2019reflections} showed that the energy harvested during transmission is sufficient for RIS to self-power without drawing any energy from the UAV when the BS transmit power is high enough.  


In NTN networks, smart and efficient management of the energy is necessary. High altitude platforms, like HAPs or satellites, can also be powered by solar energy in addition to energy harvested from RF signals \cite{9356531}. NTN networks are expected not only to improve terrestrial communication but also to enable deep space communication: 
They can act as a relay to exchange information between low-power sensor networks used in deep space networks, and they can power them by making use of RIS to transmit power from solar power satellites to the planetary surface \cite{nonterrestrial}. 

\subsection{Multiple Access Schemes}
Today's wireless networks allocate radio resources to users based on the orthogonal multiple access (OMA) schemes, such as TDMA, and FDMA. The TDMA scheme allows several users to share the same frequency channel by assigning their respective transmissions to different time slots, while FDMA allows multiple users to send data through a single communication channel by dividing the bandwidth of the channel into separate non-overlapping frequency sub-channels and allocating each sub-channel to a separate user \cite{al2021ris,jung1993advantages}. Among FDMA schemes, orthogonal frequency division multiple access (OFDMA) assigns information for each user to a subset of subcarriers \cite{kivanc2003computationally}. OFDMA has been widely adopted in practice to support multi-user communications owing to its flexibility in resource allocation design and the possibility of exploiting multi-user diversity \cite{morelli2007synchronization}.

As far as TRIS-assisted UAV multi-user communication is concerned, the TDMA scheme has been adopted in \cite{cai2020resource, zhang2020data} while the OFDMA scheme has been proposed in  \cite{9293155}. Targeting an RIS UAV-based OFDMA communication system serving multiple gUs, the work in \cite{9293155} focuses on the joint trajectory and resource allocation design that maximizes the system sum rate. 
Different from channels in narrow-band RIS-assisted communications, the composite channel gains from the UAV to gUs are frequency and spatial selective.
More specifically, for each gU, the channel gain presents a cosine pattern with respect to the subcarrier index. Moreover, on each subcarrier, the users' channel gains fluctuate with a cosine pattern depending on the differences between the differences in propagation distances of the UAV-to-gU
UAV-IRS-gU links. Based on this period cosine pattern, the authors in \cite{9293155} derived a lower bound for the formulated problem and solve it using alternating optimization techniques.
 {The simulation results showed that the proposed scheme adopting OFDMA outperforms the TDMA scheme, and the corresponding performance gain increases with the total transmit power. This is due to the fact that OFDMA enables a more efficient utilization of the power budget by exploiting the inherent multi-user diversity with flexible subcarrier allocation.}

\subsection{Spectral Sharing}
As the number of users increases, OMA-based approaches may not meet the stringent emerging requirements of very high spectral efficiency, very low latency, and massive device connectivity \cite{aldababsa2018tutorial}. 
Sparked by the concept of superimposing the signals of multiple associated users at different power levels, NOMA has been invoked for improving the spectrum efficiency, guaranteeing user fairness, and supporting massive connectivity of wireless networks \cite{saito2013non}. The key idea is to use the power domain for multiple access to exploit the spectrum more efficiently by opportunistically exploring the users' different channel conditions. In this way, it allows all  users to use the total communication resources, including frequency, space, and spreading code simultaneously. Central to NOMA, is the use successive-interference-cancellation (SIC)  at the strongest users to remove the interference caused by the weak users and decode the multiple data flows \cite{islam2016power}. 

The authors in \cite{jiao2020joint, liu2020machine} combined NOMA with RANTNs to further  enhance the achievable performance with higher spectrum efficiency. 
The authors in \cite{liu2020machine} integrated NOMA techniques into the dynamic air-to-ground communication scenario with multiple roaming ground users. In order to enhance the quality of wireless service and reduce the energy dissipation at UAV movement, TRIS is employed on the facade of a particular high-rise building to form concatenated virtual LoS propagation between UAV and users. 
Contrary to conventional NOMA, the authors showed that the  design of MISO-NOMA with dynamic environment has to fulfill the following requirements. First, the decoding order could not obtained from the order of the gU's channel gains, because a MISO-NOMA network is considered. Second, due to the dynamic nature of the scenario environment, the decoding order is required to be adapted at each time slot.  
Taking into consideration these aspects, the authors in \cite{liu2020machine} proposed a joint design of the movement of the UAVs, the phase shifts of the RIS, the power allocation policy from the UAV to MUs, as well as the dynamic decoding order, in order  to reduce the energy consumption. The proposed design was shown to be able to adapt to the dynamic environment ad guarantee successful SIC. 
  Compared with the OMA scheme, the authors in \cite{liu2020machine} showed that  NOMA is more efficient in terms of energy consumption and  spectrum efficiency. Moreover, fewer movement actions   the UAV are required to perform under the NOMA scheme because the transmit rate of gUs in NOMA networks is higher than that in OMA networks, which  indicates the data demand constraint is more likely to be satisfied in NOMA networks.

   More recently, the multi-group A2G network employing NOMA to serve multiple users in each group under the assistance of a single RIS was studied in \cite{mu2021intelligent}. The SIC decoding order among users in each groups are determined by the distance between users and the serving UAV, which is reasonable since the effective channel gains of users are dominated by the experienced path loss. This assumption facilitates the design as it avoids the need for instantaneous CSI. Compared to OMA-based networks and the interference-free network with orthogonal operating frequency bands and time slots, the authors  in \cite{mu2021intelligent} showed that the proposed NOMA-based scheme achieve the highest sum rate for limited transmit powers.
   The reasons are two fold. First, NOMA supports serving many users simultanesouly, leading to better spectral efficiency. Second, UAV and RIS introduce new-degrees of freedom brought by optimal UAV placement and RIS configuration that allows for efficient implementation of NOMA. 
  As a matter of fact, it was shown that the RIS gain is more pronounced in NOMA-based network than in other transmission schemes. However, as the transmit power increases, the achieveable sum rate saturates for NOMA-based and OMA-based network, while the interference-free solutions becomes more efficient. This is because of the existence of inter-group interference which makes the network for NOMA and OMA based networks becoming interference-limited for large transmit power. 

The ARIS-assisted downlink NOMA communication network with two user cases was studied in \cite{jiao2020joint}. The authors showed that the ARIS-assisted NOMA system outperforms the traditional NOMA system even with random phase shifts. Comparison with ARIS-assisted OFDMA has shown that ARIS OFDMA can only improve the performance of the strongest user, while the ARIS-assisted NOMA is capable of improving the performance of both users. 

The other efficient method to improve the spectral efficiency is the cognitive ratio (CR), where secondary users can opportunistically access the spectrum bands owned by the primary licensed users \cite{ye2019relay}. Driven by its efficient spectrum utilization, several researchers started looking at its combination with RANTNs. In this line, the work in \cite{xu2021intelligent} considered a primary network wherein a satellite communicates with a gU coexisting with a secondary network in which a gBS communicates with a gU. A TRIS is used in the secondary network to manage the secondary network's interference and prevent information leakage.

\textit{Summary:} This section opens up the vast possibility of using various promising technologies in RANTNs. The optical wireless communication enabled by free space optics and visible light communication, mmWave, and THz can be adopted to solve the spectrum scarcity problem. The wireless power transfer technology contribute towards extending the operation duration of aerial platforms, RISs, even IoTDs, whereas the multiple access schemes and spectral sharing strategies open opportunities to support massive users with limited channel resources. The summary of existing literature can be seen in Table \ref{Other_tec}. 

\begin{table*}[!htbp] 
\centering
\caption{Summary of existing literature about the integration of other promising technologies into RIS-assisted NTNs  }
\label{Other_tec}
\begin{tabular}{|p{0.08\textwidth}| p{.3\textwidth}|p{.08\textwidth}| p{.41\textwidth}| } 
\hline
\rowcolor{gray!20}
\textbf{Reference} & \textbf{Technologies} & \textbf{RIS types}& \textbf{System model}  \\
\hline
\rowcolor{yellow!20}%
\cite{zhang2020distributional}& mmWave 
& ARIS & G2G MU MISO downlink transmission\\
\hline
\rowcolor{yellow!20}%
\cite{zhang2019reflections}& mmWave, WPT
& ARIS & G2G P2P MISO downlink transmission\\
\hline
\rowcolor{yellow!20}%
 \cite{cai2020resource}&  Multiple Access technologies (TDMA)
& TRIS & A2G MU MISO downlink transmission\\
\hline
\rowcolor{yellow!20}%
\cite{zhang2020data}  &  Multiple Access technologies (TDMA), WPT
& TRIS & A2G MU SISO downlink transmission assisted by two RISs\\
\hline
\rowcolor{yellow!20}%
 \cite{9293155}&  Multiple Access technologies (OFDMA)
& TRIS & A2G MU SISO downlink transmission\\
\hline
\rowcolor{yellow!20} %
\cite{mu2021intelligent}&  Multiple Access technologies (NOMA) & TRIS & A2G multi-group SISO system\\
\hline
\rowcolor{yellow!20}%
 \cite{wang2020joint}& mmWave 
& TRIS & A2G MU SISO downlink transmission assisted by multiple TRISs\\
\hline
\rowcolor{yellow!20}\cite{cang2020optimal}
&OWC (VLC)
& TRIS & A2G MU SISO downlink transmission served by multiple ABSs\\
\hline
\rowcolor{yellow!20} \cite{jia2020ergodic}
& OWC (FSO)
& ARIS & G2G P2P SISO transmission\\
\hline
\rowcolor{yellow!20}%
\cite{tekbiyik2020reconfigurable}& THz 
& ARIS & A2A P2P inter-satellite transmission\\
\hline
\rowcolor{yellow!20}%
 \cite{pan2020uav}&  THz
& TRIS & A2G MU SISO downlink transmission\\
\hline
\rowcolor{yellow!20}%
\cite{abuzainab2021deep}&  THz 
& ARIS & G2A P2P SISO downlink transmission\\
\hline
\rowcolor{yellow!20}%
 \cite{jiao2020joint}&  Multiple Access technologies (NOMA)
& ARIS & G2G MU SISO downlink transmission\\
\hline
\rowcolor{yellow!20}%
\cite{ liu2020machine}&  Multiple Access technologies (NOMA) 
& TRIS & A2G MU MISO downlink transmission\\
\hline
\rowcolor{yellow!20} %
\cite{xu2021intelligent}&  Multiple Access technologies (CR) & TRIS & A2G CR system with primary satellite communication and secondary terrestrial network \\
\hline
\end{tabular}
\end{table*}

\section{Performance Analysis}
The various RANTN frameworks and the combination with other promising technologies lay a solid foundation for system design. However, how to evaluate the system performance becomes one of the leading research directions. Due to the randomness of the propagation environment between end devices, the quality of the received signal is also random, following particular distributions related to the properties of the wireless channel medium. Using tools such as probability theory, random matrix theory, or stochastic geometry, several essential metrics such as outage probability (OP), EC, and EE can be theoretically analyzed in closed-form expression depending on the system's parameters. The advantages of such theoretical studies are two-fold: they avoid the need for extensive numerical simulations and assist in finding optimal allocation resources. Recently, a significant research effort from both industry and academia has been made to assess the benefits of the RANTNs, by carrying out several performance analysis studies.      

\subsection{Terrestrial RIS}
The first theoretical study about TRIS-assisted NTNs is conducted by Yang {\textit {et al}} in \cite{yang2020performance} where a TRIS-assisted UAV is used to relay the communication between a source and a destination node, as illustrated in Fig. \ref{G2G}. 
Under the assumption of a Rician distribution for TRIS-UAV and UAV-gU channels, and Rayleigh distribution for the TRIS-gU channel, the authors provided an approximation of the cascaded channel distribution and derived the SNR of the RIS-assisted ground-air system. Considering the DF relaying protocol at UAV, the exact OP and BER, as well as the asymptotic approximation in high SNR regime are obtained in closed-form, while an upper-bound of the capacity is derived by applying Jensen's inequality. It has been shown that the OP performance are related  to the experienced path loss and the probability of LoS, both of which depends on the height of the UAV. More specifically, for small ranges of the UAV height, increasing the height increases the probability of LoS and thus improves the OP. However, by continuing to increase further the height, a larger path loss is experienced and thus the OP increases. This explains why the OP decreases first with  the height before increasing after a certain threshold.    

Curiously, it has also been shown that for many reflecting elements, deploying more of them improves only slightly the OP. Indeed, while increasing the number of elements improves the performance of the ground-to-air communication assisted by RIS, that of the second hop remains unchanged.  Since a DF protocol is employed, the end-to-end SNR depends on the minimum SNR of the two hops, and hence it becomes dominated by the performance of the second-hop. The same argument can be used to explain the behavior of the OP with respect to the horizontal distance between the transmitting gU and the UAV. 
If this distance is small, the performance of the end-to-end SNR is mainly determined by that of the second hop. Increasing the distance between the transmitting gU and the UAV of the first hop will make the UAV closer to the receiving gU, which leads to better OP performance. However, as the distance between the transmitting gU and the UAV increases, the performance of the first hop becomes dominant, hence the deterioration of the OP performance.

\subsection{Aerial RIS}
More theoretical studies were performed to explore the potential of ARIS. Targeting the ground-to-ground system, the authors in \cite{jia2020ergodic} considered the communication between a light source transmitter and a photodetector receiver through a UAV. The cascaded FSO channel is then the product of the atmospheric turbulence and the geometric pointing loss. The authors opted for a Gamma-Gamma distribution to model the atmospheric turbulence, which is known to present a good fit for a wide range of turbulence strengths. As for the geometric pointing error loss, it is a function of the movement of the aerial platform UAV and the misalignment vector between the centers of the photodetector and the beam footprint.  Its distribution is then derived by assuming that the misalignment distance follows a Hoyt distribution. Under this model, the asymptotic EC at high SNR was obtained, providing insights into the relationship between the capacity and the atmospheric conditions, the configurations of laser devices, as well as the steady ability of UAV.

Focusing again on ground-to-ground systems, the authors in \cite{shafique2020optimization}  considered an integrated UAV-RIS network in which a UAV carries a large array of reflecting elements to assist communication between two gUs. Comparison between three operating modes was investigated, namely the UAV-only mode, the RIS-only mode, and the UAV-RIS mode, wherein for the UAV-RIS mode, both the UAV and RIS forward the data, and the receiving gU combines the data through selection combining. For each of these modes, the authors derived approximations for the OP as well as upper bounds on the EC and the EE.     
The results provided in \cite{shafique2020optimization} directly showed that the integrated UAV-RIS mode outperforms the other two modes in terms of EC and OP, which is due to the opportunistic selection between UAV-only and RIS-only models. Compared to the UAV-only mode, the RIS-only mode presents better outage performance for a wide range of heights and for strong LOS, while it performs worse for weak Los and high UAV height. 
Targeting the same integrated UAV-RIS network, the authors in \cite{mahmoud2021intelligent} provided a closed-form for the PDF of a tight upper bound on the instantaneous SNR under the assumption of Rayleigh fading. The calculations hinge on the Cauchy-Schwarz-Buniakowsky inequality, which allows for obtaining the MGF of the SNR upper bound and its associated SER, EC, OP, and outage capacity. For the sake of comparison, the authors derived asymptotic analysis of these performance indexes based on central limit theorem and showed that while the CLT-based approach is in agreement with the simulation in the low SNR regime, it presents a gap for high SNR values.  


The authors in \cite{alfattani2020link} developed a complete link budget analysis by dividing the transmission into two different regimes: the specular reflection paradigm, and the scattering paradigm. 
These regimes are determined by the  geometrical size of the ARIS units, the communication frequency, and the link distances from ARIS to the transmitter and the receiver. In particular, when the ARIS is within relatively short distances from the transmitter and receiver or the ARIS units are electrically large,  e.g., their dimensions are 10 times larger than the wavelength, the path loss undergoes the specular reflection paradigm. In contrast, the path loss undergoes the scattering paradigm when both dimensions of the RIS units are 10 times greater than the wavelength, or the RIS dimensions are very small. 
 In other words, the scattering paradigm can be designated as the far-field, whereas the specular reflection  as the near-field. 
  The biggest difference between the two regimes in \cite{alfattani2020link} is that, while the number of reflecting elements plays the dominant role in the received power in the specular reflection paradigm,  
  the impact of the altitude of aerial platform, and wavelength becomes more important in the scattering paradigm. 
Under both paradigms, the authors derived the minimum feasible number of reflecting elements and the maximum number of them that can be deployed on a given aerial platform's surface. They also developped a link budget analysis for both reflection paradigms using the 3GPP and the log-distance channels models.  
Based on the link budget analysis for the log-distance channel model, the authors derived for both reflection paradigms the optimal horizontal location of the aerial platform at a given fixed altitude. It has been shown that regardless of the altitute of the aerial platform, the optimal horizontal placement of the aerial platform in the specular reflection regime is over   the perpendicular bisector of the segment from the transmitter to the receiver. However, in the scattering reflection regime, the optimal position becomes dependent on the height of the aerial platform.  
Particularizing the budget analysis to three different aerial platforms, namely UAV, HAP, and LEO satellites, the authors deduced that the specular reflection paradigm is not feasible for UAV and LEO satellites. The main reasons lie in the limited available surface on UAVs and the high altitude of satellites. However, the specular reflection paradigm can be realized using RIS-equipped HAPs, in which case they outperform TRIS. Under the same number of reflecting elements, and for both reflection regimes, the UAV achieves the best received-power performance due to its closeness to the terrestrial users. However, HAPs become the best one if each platform can be equipped with as many reflectors as it can support.  
Furthermore, the authors showed that UAVs are insensitive to frequency, as increasing the frequency, while leading to higher attenuation, enables the use of more reflecting elements. On the other hand, the performances of HAPs and LEO satellites are affected by the attenuation caused by water droplets or molecules at specific frequency ranges. Nevertheless, their  performances become insensitive to frequency over other spectrum regions, which can be then exploited to provide high-capacity communications.
The gains provided by ARIS-assisted systems require optimal phase compensation by reflecting elements. However, in practice, optimal phase compensation cannot be achieved because of the quantization error and the channel variation. As far as UAV systems are concerned,  channel variation occurs even when UAVs are hovering, making optimal phase compensation unrealistic. In \cite{al2020irs, al2021capacity}, the authors studied the effect of the imperfect phase compensation error on the performances of ARIS-assisted systems. Under the assumption of a Von Mises distributed phase error, the authors in   \cite{al2020irs} derived closed-form expressions for the symbol error rate (SER) and OP for less than three reflecting elements and obtained accurate approximations of them for a larger number of reflecting elements. The obtained results showed that a large number of reflecting elements confer robustness towards phase error. In contrast, the degradation caused by phase error may surpass the RIS gain when the number of reflecting elements is small. Under a similar setting, the authors in \cite{al2021capacity} analyzed the capacity performance for the constant channel fading model. The obtained results showed that the  capacity degradation due to phase errors is inversely proportional to the SNR, becoming negligible at high SNR. Moreover, increasing the number of elements enhances the capacity even for imperfect phase compensation scenarios.  

{The recent work in \cite{tekbiyik2021graph} derived the error probability expression for the direct communication between a satellite and ground IoTDs under  Rician channels and assuming RIS is deployed near the satellite. Taking the realistic discrete-phase RIS into account, the error probability decreases as the number of quantization bits increases. In particular, the 3-bit RIS design almost approaches that of the continuous-phase RIS, which indicates the 3-bit quantization level is cost- and energy-efficient for the satellite IoTDs communications. Another interesting experiment that has been conducted in this work is the error performance comparison when the phase shift at RIS belongs to discrete finite phases uniformly distributed over the intervals 
	 $\left[-\pi,\pi\right)$,  $\left(-\pi,-\frac{\pi}{2}\right] \cup\left[\frac{\pi}{2}, \pi\right)$, and $\left[-\frac{\pi}{2},\frac{\pi}{2}\right]$. The results showed that smaller phase intervals leads to obvious worse performance, because it cannot compensate for  the phase information of the cascaded channel. In \cite{tekbiyik2020reconfigurable}, the misalignment fading caused by the high relative velocity of LEO satellites and the sharp beam of THz antennas was taken into account when calculating the error probability of the inter-satellite communication assisted by ARIS, a system model introduced in Section II. B. Under the circular beam assumption, the misalignment coefficient is determined as a function of  the Rayleigh-distributed beam's radial distance  and the receiver's jitter variance. The channel amplitudes are assumed to follow the Rician distribution. 
	  Similar to \cite{shafique2020optimization}, the gain of the sum of the product channels was approximated by a Gaussian distribution based on the central limit theorem. Then, the closed-form expressions for the probability error rate under the assistance of multiple independent ARISs were derived.}

\textit{Summary:} Table \ref{performance_ana} summarized the existing performance analysis works reviewed above. The majority of works derived approximations of the underlying performance indexes due to the reflection at RIS introducing product channels that are less mathematically tractable. 
Moreover, important practical factors such as the near-field and the correlation between reflecting elements are hardly considered. The considered channel models are not realistic, with most of the studies relying on a Rician distributed channel. All this shows that there is still room for further analytic developments based on accurate channel models that fit with experimental data.  

\begin{table*}[!htbp] 
\centering
\caption{Summary of existing performance studies of RIS-assisted NTNs}
\label{performance_ana}
\begin{tabular}{|p{0.08\textwidth}| p{.42\textwidth}|p{.08\textwidth}|p{.08\textwidth}| p{.16\textwidth}| } 
\hline
\rowcolor{gray!20}
\textbf{Reference} & \textbf{Performance Metrics} & \textbf{RIS types}& \textbf{NTN types} & \textbf{Characterization} \\
\hline
\rowcolor{yellow!20}%
\cite{mahmoud2021intelligent} & OP, SER, and EC & ARIS &G2G & Upper bound \& Asymptotic \\
\hline
\rowcolor{yellow!20}\cite{jia2020ergodic}
&EC
& ARIS & G2G
& Asymptotic\\
\hline
\rowcolor{yellow!20}%
\cite{alfattani2020link} & Link budget & ARIS &G2G & Exact \\
\hline
\rowcolor{yellow!20}%
 \cite{shafique2020optimization}& OP, EC, and EE
& ARIS & G2G
& Approximation\\
\hline
\rowcolor{yellow!20} \cite{yang2020performance}
& OP, average BER, and average capacity
& TRIS & G2G
& Approximation\\
\hline
\rowcolor{yellow!20}%
\cite{tekbiyik2020reconfigurable}& BER
& ARIS & A2A & Approximation\\
\hline
\rowcolor{yellow!20}%
\cite{al2020irs}& OP and SER
& ARIS & G2A
& Exact \& Approximation \\
\hline
\rowcolor{yellow!20}%
\cite{al2021capacity}&Capacity
& ARIS & G2A
& Exact \& Approximation \\
\hline
\end{tabular}
\end{table*}

\section{Optimization of RANTNs}
Based on the performance analysis results, the controllability of RIS and flexibility of aerial platforms have generated excitement in the wireless community to fully explore the potential of RANTNs to optimize various performance metrics, such as SNR, data rate, EE, power/time consumption. All these works serve as a solid foundation and provide valuable insights for future research of RANTNs.

\subsection{Received Power/SNR}
\subsubsection{TRIS}
Several recent works considered the analysis of the performance gains brought by TRIS for G2A communications. In \cite{ma2020enhancing}, the authors studied the signal gain defined as the ratio of the received signal powers with TRIS and without TRIS under the practical path loss model for the urban macro scenario described in \cite{TR38901}.  As ground base
station antennas are down tilted,  UAVs
flying above gBS are only supported through side lobes resulting in the UAVs receiving low received powers. By deploying TRIS within the coverage of the main lobe of the BS, it is possible to gather more energy from the BS and reflect it to the UAV, allowing for a significant improvement of the received power. Similar to terrestrial networks, the signal gain was found to increase quadratically with the number of reflecting elements. However, it depends on the UAV height, becoming more significant when the UAV flies well above the gBS.  

The received power by a ground user served by ABS is investigated in \cite{ge2020joint}, where multiple TRISs are deployed to assist the transmission. 
Targeting the maximization of the average received power within multiple time slots, the authors proposed a joint optimization of the active beamforming at UAVs, the passive beamforming at TRISs and the UAV's trajectory. 
Interestingly, the results showed that the trajectory adaptation plays a more important role than the passive/active beamforming, which is reflected by the fact that the received power gain brought by trajectory optimization is much larger than that of passive beamforming and active beamforming. Moreover, because of the performance gain brought by TRISs, it was found that over the optimized trajectory,  ABS tries to fly as close as possible to the locations of TRISs to enhance the received signal strength, especially when the number of reflecting elements is large.

The recent work in \cite{matthiesen2020intelligent} focused on a TRIS-assisted system when the receiver position can be predictable. A typical system falling within this setting is that of uplink transmissions to a LEO satellite, since the positions of LEO satellites are known a priori. The authors developed a continuous-time for multipath channels and derived the optimal configurations of TRIS that optimize the received power, the doppler spread and the delay spread.  It has been shown that optimal received power can be achieved without incurring doppler spread, while the delay spread cannot be optimized without increasing the doppler spread and decreasing the received power. Moreover, simulations showed that RIS with planar elements can provide SNR gain and resilience to visibility outages, provided that it has a favorable orientation to the satellite. Rotating the RIS can achieve such a favorable orientation, yet at the cost of an increasing delay spread. 

\subsubsection{ARIS}
The preliminary contribution regarding the received SNR of ARIS-assisted G2G system appeared in \cite{lu2020enabling,lu2020aerial}, where an  ARIS is considered to assist the signal transmission between a source node and users located in a certain coverage area. Both works aim to design the transmit beamforming, the passive beamforming, and the ARIS location so as to maximize the worst received SNR over a coverage area of interest for a uniform linear array (ULA) at the ARIS in \cite{lu2020enabling} and a uniform planar array (UPA) at the RIS in \cite{lu2020aerial}. 
Considering a fixed location for the receiver,  the authors derived closed-form expressions for the optimal transmit and passive beamforming, as well as the optimal placement of the ARIS. Particularly, 
depending on the altitude of the ARIS, the authors showed that the optimal horizontal placement for a given location of the receiver could be either at the midpoint between the source and the destination or at one of two positions that are symmetric with respect to the midpoint. Consideration of the maximization of the minimum SNR over receiver's positions in a given region involves a non-convex problem for which a suboptimal yet efficient algorithm was proposed.   

In practice, the number of reflecting elements mounted on a UAV is limited by its payload and its battery capacity. According to \cite{shang2021uav}, one solution to overcome this limitation is to deploy multiple UAVs that form a UAV swarm to enable cooperation between UAVs. Such a system was designated in \cite{shang2021uav} as swarm-enabled ARIS-assisted G2G system. 
Contrary to TRIS, for which it is desirable to deploy RIS near the transmitter or the user so as to reduce the path loss of the cascaded channel, swarm-enabled ARIS systems may not present the same behavior.  
This is because although a high-altitude ARIS is more likely to establish a LoS connection, it undergoes  an increased signal attenuation due to the increase of the communication distance. On the other hand, when ARISs are close to gU, the path loss of LoS connections decreases, but the excessive path loss originating from NLoS connections becomes more severe. 
All this implies that there exists an optimal position for ARISs that strikes a favorable tradeoff between path loss and NLoS probability.

\subsection{Interference}
Besides enhancing the received power, RIS have been investigated to mitigate interference. In this line, the work in \cite{hashida2020intelligent} considered the scenario of a downlink ground-to-air communication system in which multiple BSs communicate with aerial users. To extend the communication coverage, TRISs are deployed in each cell boundary. Accordingly, the aerial area is
divided into two regions: BS coverage area and
TRIS coverage area. Aerial users in the BS coverage area are directly served by the BS, while BS and
TRIS collaborate to serve aerial users in the RIS coverage
area. The interference power in the
neighbor cells is determined by the elevation angle
of the interfering aerial user and the positions of the TRISs. Based on this observation, the work in \cite{hashida2020intelligent} proposes to find the optimal TRISs' positions to minimize the impact of interference to neighbor cells. This was shown to help prevent interference from spreading over a wider area. 

\subsection{Achievable Rate Maximization}
The other important performance index widely considered in the literature is the achievable rate. According to where the RIS is integrated, the related works concerning achievable rate maximization can be classified into two categories: TRIS-assisted performance optimization and ARIS-assisted performance optimization.

\subsubsection{TRIS}
The achievable rate of air-to-ground communication system assisted by TRIS is firstly studied in \cite{li2020reconfigurable}, where a rotary-wing UAV serves a gU within a finite time span composed of multiple time slots. Assuming the existence of a direct link between UAV and ground user, the work in \cite{li2020reconfigurable} proposed a joint design of the
UAV trajectory and the TRIS  beamforming that targets the maximization of the achievable rate. Comparison with benchmark algorithms revealed that the proposed design allows the gU to enjoy the most benefit of the channel gains from both the UAV and the RIS, and thereby obtaining the largest average achievable rate.

The above-reviewed literature analyzes and optimizes the achievable rate of point-to-point wireless communications by deploying TRIS in the propagation environment. The rate maximization problems can be naturally extended to multi-user scenarios. Consequently, the solutions will be more involved due to the interference and resource competition among different users. In this context,  the authors in \cite{li2020sum} considered an air-to-ground multi-user downlink system in which a UAV serves simultaneously single-antenna users through a TRIS within a limited time. Under this setting, the authors proposed a joint design of the UAV trajectory and the phase shifts of the TRIS with the aim of maximizing the achievable sum rate. It has been found that the behavior of the UAV trajectory differs according to the considered flight period. Particularly, for short flight periods, the UAV was found to  follow a relatively direct path to each user and the
TRIS and then remains stationary over those points as long as
possible.  Meanwhile,
the faster the UAV approaches the hovering points, the larger is the average sum rate. On the other hand, under a long flight period, the UAV bypasses the ground users and flies almost directly toward TRIS in an arc path for close connection.
The same system model is studied in \cite{pan2020uav} but in THz bands, where the minimum average achievable rate is maximized. It has been shown that the deployment of TRIS allows for shortening the trajectory of the UAV, which in turn implies less consumed energy.

Targeting multi-user communications, the work in \cite{9293155} investigated a TRIS-assisted UAV OFDMA system in which the phases of the TRIS are aligned with respect to one selected user in each time slot, while the remaining users are solely served by the UAV directly over other subcarriers. Under this setting, this work focused on the joint optimization of the UAV trajectory, the user scheduling and the resource allocation to maximize the system sum-rate over  quality-of-service (QoS) requirements for each user.
To set the phase shifts for the TRIS, the authors 
propose a design that maximizes the composite channel gain  of the LoS component. Such a procedure requires knowledge of the locations of the UAV and the gUs as well as the Rician factors of all the involved links. It presents the advantage of reducing  the required signaling overhead of CSI acquisition and enabling an offline design that is suitable to  a major range of the application scenarios of UAV communications dominated by a large Rician factor. Building on this phase shift design, the authors derived a lower bound of the sum-rate that is  optimized to jointly design the resource allocation and the UAV's trajectory. 
It was shown that the obtained optimal power allocation follows a multi-level water-filling principle while the  optimized user scheduling approach indicates that a user with a higher composite channel power gain or a more stringent data rate requirement has a higher chance to be scheduled as an RIS-assisted user. Furthermore, with a fixed data rate requirement, employing an RIS and appropriate power allocation scheme can scale down the transmit power of the UAV by the inverse of the number of relfecting elements, which is consistent with the power scaling law in \cite{wu2019intelligent}. 
As a major outcome of the authors' work, it has been shown that the use of the TRIS enables higher flexibility 
in designing UAV trajectories. Particularly, the size of the TRIS significantly affects the trajectory of the UAV. Assuming a not too large size of TRIS, the UAV directly flies towards centroid-formed users  for maximizing the system sum rate. This is because the minimum data rate constraint of the TRIS-assisted user can still be satisfied even if the UAV is far away from it. However, as the number of reflecting elements increases, the UAV in the proposed scheme would first detour to the TRIS and TRIS-assisted user at the beginning before flying to other users. In fact, equipping more reflecting elements allows the TRIS to reflect the radiated signal more efficiently, and thus approaching the TRIS and the TRIS-assisted user becomes more beneficial to the system sum-rate performance. Moreover, it has been made clear from the simulation results, that deploying TRIS at the boundary 
but close to the area with a high density of users is more beneficial for improving the system sum rate.

More recently, the achievable rate has been considered to optimize the trajectory and the phase shifts of TRIS in multi-UAV NOMA networks, \cite{mu2021intelligent}, wherein multiple UAVs are deployed to serve user groups through a TRIS.  The optimized trajectory has revealed that UAVs are likely to be deployed near from some of the served users to ensure the communication quality, while remaining considerably far away from the unserved users to reduce the inter-group interference. Moreover, they fly at a lower height, which may cause stronger interference to the unserved users but this appears to be more beneficial for enhancing the communication with target users. 
Comparisons with scenarios without TRIS have shown that the deployment of TRIS  enhances the channel power gains between UAVs and their served users, especially those located closed to TRIS, while  mitigating the interference between UAVs and their unserved users.

On the lookout for better performances, researchers started investigating the use of systems assisted by multiple RISs. 
In this context, the work in \cite{ wang2020joint} considered multiple RISs mounted on several buildings to enhance the communication quality between the downlink transmissions of one UAV to several gUs. Under this setting, the authors proposed to optimize the trajectory and the reflection coefficients over a set of discrete values for the phase shifts. Although the discretization causes a loss in the performances compared to optimization over continuous variables, it is often preferable for implementation due to the hardware limitations in practice. 

Rather than considering a single UAV, the authors in \cite{cao2021reconfigurable} maximized the overall  capacity of a system involving multiple UAV-gU pairs, where a single TRIS is divided into multiple groups to serve different pairs of UAVs and users. 
It was found that increasing the number of the TRIS's groups resulted in worse performances in terms of achievable rate and power consumption as this implies fewer available TRIS elements in each group. 
Moreover, assuming a fixed number of TRIS groups, the system throughput decreases as the number of UAV-gU pairs increases because more bandwidth is consumed with poor communication quality.

\subsubsection{ARIS}
A recent wave of research works contributed to investigate the data rate of ARIS-assisted networks.
In this context, the authors in \cite{zhang2019reflections} considered a UAV equipped with a ARIS to assist the communication between a gBS and a gU. To avoid the complexity of the design incurred by the  ARIS's movement, the authors assumed that the ARIS provides the downlink service toward the moving gU only during the hovering state. 
However, once the gU receives an SNR lower than the threshold yielding a blockage, the ARIS will move to a new place to rebuild an LoS link for downlink transmission. The authors proposed a method based on reinforcement learning that jointly optimizes the reflecting elements of the ARIS and its location to maximize the average sum rate. 
The results showed that the ARIS yields a significant improvement in the average data rate.  The obtained average rate increases faster  with the transmit power than for a static RIS, while it decreases as the altitude increases due to an increasing path loss.  
Comparisons with benchmark algorithms showed that the proposed method 
 yields a LoS probability over 90\%, while the static ARIS and the ARIS moving towards gUs present  LoS probabilities below 5\% and above 70\%, respectively. 

Researchers also put their eyes on the multi-user systems. 
In this line, the authors in \cite{jiao2020joint} considered a UAV equipped with a ARIS to assist downlink transmissions from a multi-antenna gBS to two gUs. Under this setting, the optimal location of the ARIS that minimizes the pathloss to the strongest user is obtained. This location coincides with the one derived in \cite{lu2020enabling,lu2020aerial}, and is shown to depend on the ratio of the height of ARIS and the distance between gBS and the strongest user.  With the ARIS placed at its optimal position, the transmit beamforming vector and the phase shifts of the ARIS are optimized to maximize the data rate at the strongest user while ensuring a minimum target rate for the weak user.  
Considering the same system model but with multiple users, the authors in \cite{zhang2020distributional} proposed a method based on alternating two processes named communication process and movement process. In the communication process,  the position of the ARIS is assumed to be fixed while the transmit beamforming and the ARIS phase shifts are optimized to maximize the sum rate. In case the average rate per gU becomes below a certain threshold, the communication stage ends, and a movement process  starts to optimize the location of the UAV-IR so as to maximize the
long-term downlink communication capacity.  
 As compared with the scheme without ARIS, the proposed ARIS  yields a performance gain of over two folds in the downlink data rate. Meanwhile, the deep RL (DRL)-enabled deployment used in the movement process to learn the relationship between the channel and the ARIS position was shown to improve the communication performance by over 25\% and 50\% compared to a non-learning ARIS moving towards the user by one meter once an outage occurs, and the static ARIS, respectively.
 Targeting cell-free massive MIMO systems, the work in \cite{9308937} considered the case of multiple APs serving through an ARIS a gU. In order to maximize the user’s achievable rate, a  joint optimization strategy to  maximize the user’s achievable rate is proposed. 

\subsection{Energy Consumption}
Energy consumption is another important issue that limits the superior performance of RANTNs.
Indeed, due to the fact that UAVs are battery-powered, energy consumption is one of the most important challenges in commercial and civilian applications. The limited endurance of batteries hampers the practical implementation of the UAVs. The advantages of UAV-enabled wireless systems could not be reaped without appopriate solutions of  
how to overcome the energy-hungry issue. 
As stated in \cite{liu2020machine}, the total energy dissipated by the UAV consists of two components, namely, the communication-related and the propulsion-related energy consumption. The first component is dissipated for radiation, signal processing, and hardware circuitry, while the other one is dissipated for supporting the hovering and mobility of the UAV. It is worth noting that the propulsion-related energy consumption accounted for the vast majority of the sum energy consumption (usually more than 95\%), which emphasizes the importance of considering the propulsion-related energy consumption model of UAVs when aiming for designing environment-friendly wireless networks. In other words, in order to further reduce the energy consumption of the UAV, it is better to keep the UAV aloft without consuming its own energy. {Based on this property, RISs with low-cost passive elements, which can extend the coverage while keeping UAVs aloft, are introduced to assist aerial communications and reduce energy consumption.}

\subsubsection{TRIS}
With the assistance of RIS coated on the surface of buildings, the authors in \cite{liu2020machine} proposed to control the phase shift of the TRIS instead of changing the position of the UAV to establish LoS wireless links between continuously roaming mobile users. By invoking this protocol, the UAV can maintain hovering status only when concatenated virtual LoS links cannot be formed even with the aid of the TRIS, which in consequence minimizes the total energy dissipation of the UAV and maximizes the endurance of the UAV. Intuitively, invoking more reflecting elements leads to the reduction of energy consumption, while serving more gUs results in an increase of energy consumption. Moreover, when the altitude of the UAV is high, it consumes more energy. This is because increasing the altitude of the UAV increases the LoS probability between the UAV and MUs but leads to a higher path loss due to the increasing distance. 

The optimization of the transmit power becomes more important when the VLC technology is used, as in this case, it is necessary to simultaneously provide wireless services as well as illumination for users. In this context, the authors in \cite{cang2020optimal} considered a  VLC-enabled UAV network consisting of a group of UAVs to  provide communication and illumination to ground users through multiple RISs. The problem of interest is to  minimize the transmit power of UAVs via adjusting four variables which are the UAV deployment, the phase shift of RISs, the user association and the RIS association under the constraints of data rate and illumination demand. For comparison, the authors compared between the cases where only one control variable is considered. In this case, the authors noted that the schemes  considering the user association outperforms all other schemes based on either optimizing UAVs deployment, the phase shifts of RISs or RIS association.
 This indicates that an appropriate user association can greatly reduce the total transmit power. Hence, it is reasonable to only optimize user association when an urgent deployment is required. Practical insights into the impact of the number of RISs and the association of users and RISs were drawn. Particularly, it has been shown that the increase of the number of RISs contributes to reducing the total transmit power since optimized RISs constructively improve the channel state between a UAV and its associated gUs. Moreover, the application of the proposed algorithms revealed that UAVs should be surrounded by their associated users and RISs. On the other hand, a small number of users distant from the majority of them can be served by one UAV, resulting in reduced transmit power. Furthermore, RIS should be associated with the closest UAV, as this reduces the path loss and thus enhance the channel gains between UAV and gUs. 

Referring to the work in \cite{zhang2020data}, described in section III-C where the authors considered the use of TRIS mounted on two buildings and PDBs to relay energy and data between a UAV and IoTDs, the authors noted that the optimal UAV position that minimizes the energy consumption depends on the considered scenario.  Particularly, if one PDB is considered, then the optimal UAV's hovering position coincides with that of the PDB.
However, if we account for the UAV's energy to reach the optimal position, the best  hovering position occurs before the UAV arrives at the same position as the PDB. This is because there is a tradeoff occurring between the UAV's energy flying, the charging energy, and the data transmission energy.

To investigate the role of TRIS in reducing the consumed energy, the authors in \cite{cai2020resource} considered the case in which a multi-antenna UAV serves multiple gUs through either a direct transmission mode or a reflecting mode using TRIS. Under this setting, 
the authors proposed a model for the consumed energy that accounts for the energy consumed by the TRIS whenever it is used to reflect the incoming signal. 
With this model at hand,  the problem of interest is to optimize the transmit beamforming, the phase shifts of TRIS, the UAV trajectory, and the transmission mode for each gU in order to minimize the transmit power under minimum rate constraints. 
The simulation results showed that not only the TRIS enables a reduction in the average total power consumption but also it accelerates the convergence of the proposed algorithm. The reason lies in that an effective solution can be easily reached in the time slots associated with TRIS activation. Moreover, it was noted that with the help of TRIS, the UAV is not required to hover at the positions of the served users, as in the case without TRIS. Indeed, with TRIS, it is possible to avoid requiring the UAV to approach the served users while satisfying the minimum rate constraints.  
In other words, TRIS can improve the flexibility of  designing the UAV’s trajectory, which can lead to a substantial reduction in power consumption. 
 
\subsubsection{ARIS}
The work in \cite{shafique2020optimization} considered the optimization of the power consumption of an ARIS assisted system in which a UAV equipped with RIS  assists communication between two users. A model for the total power consumption is provided, which accounts for the hovering power consumption as well as the power for data transmission and that consumed by hardware. 
The power consumption is then minimized subject to rate constraints by optimizing the number of active RIS elements. It is noteworthy that optimizing the power consumption with respect to the number of RIS elements in an ARIS system is crucial for two reasons. 
First,  the number of RIS elements that can be deployed on a UAV is limited by its size. Second, each RIS element consumes energy, which while being low, may lead to a significant power consumption depending on the considered phase resolution for each element. 

\subsection{Energy Efficiency}
It cannot be denied that designs based on minimizing the power consumption sacrifices the achievable data rate. To strike a tradeoff between the consumed power and the achieved data rate, maximization of the EE is more advisable. In short, EE advocates for smaller energy consumption in regards to performing higher.
In \cite{al2020irs}, the authors showed that the use of RIS can reduce the energy consumed for transmission of one bit, becoming smaller and smaller as the size of the RIS increases.
Referring to the work in \cite{shafique2020optimization} described in the previous section, the authors showed that the EE firstly increases and then decreases as the number of RIS elements increases. The reason is that the effect of capacity dominates that of power consumption for a small number of RIS elements. However, as the number of RIS elements increases, the power consumption prevails over the capacity. An optimal number of RIS elements that maximizes the EE exists. However, its value depends on the power consumed by each RIS element. 
Particularly, if the power consumed by each element is small, the EE continues to increase for a wide range of values of the number of elements, because the increase in the number of RIS elements does not significantly increase the power consumption, whereas the capacity keeps increasing. On the other hand, when a high resolution is used, the power consumed by each element increases. As such, the impact of power consumption of RIS elements dominates that of the gain brought by using more of them, leading to a decrease in the EE beyond a specific number of RIS elements.
It can be concluded that if the bit resolution power is very small, then using the maximum number of RIS elements is optimal, whereas when the bit resolution power is significantly large, then using a small number of RIS elements is optimal. 

The EE maximization problem of an ARIS-assisted multi-user downlink communication system is studied in \cite{mohamed2020leveraging}, in which the BS is assisted by a hovering UAV equipped with  ARIS  to  communicate with cell-edge gUs. The total dissipated power considered in this system consists of the transmit power at the BS, the hardware related power that is consumed in the BS equipment, the RIS power consumption depending on the resolution of the reflecting elements, and the UAV power consumption. 
Targeting the maximization of the total energy efficiency under minimum rate constraints for each cell-edge users, the authors proposed a  joint design of the beamforming vector at the base station and the
phase shifts matrix of the passive reflecting elements at the
UAV.
The results showed that the achieved EE increases as the transmit power at the BS increases. Moreover, the ARIS yields higher EE compared to the scenario with conventional UAV operating in AF mode. This is the consequence of the AF relaying being an active equipment and consuming power to amplify the incoming signal, thus having higher energy consumption compared to the RIS. On the BS side, increasing the number of antennas clearly yields a higher EE. On the other hand, an increase in the rate requirement reduces the achievable EE as the BS will need higher transmit power to meet the users' demands. Moreover, if RIS is closer to the users, higher EE can be achieved than when the RIS is close to the BS.  On the other hand,  the ARIS located between the BS and the UE achieves lower EE since the NLoS link becomes dominant and the reflected signal becomes weaker due to a higher path loss. 
Finally, it is important to note the paper in \cite{jeon2021energy}, which unlike all aforementioned research works focusing on multi-user downlink communication systems, considered the use of ARIS to allow for high data rate yet energy-efficient backhaul links. Particularly, the system model of interest is the downlink backhaul link between a gBS and multiple ABSs through the assistance of a HAP carrying an ARIS.  The considered problem is to determine the HAP's position and the ARIS's phase shifts to minimize the total transmit power under minimum rate constraints for each ABS. As a solution, the authors applied the global criterion method to find Pareto-optimal values for the considered optimization variables. A refinement of this algorithm is also proposed to handle the cases wherein the channel gains of some ABSs are below the half-power beamwidth of the beamforming gain. As a solution, the authors proposed partitioning ARIS into parts; each serves a group of ABSs with a channel gain above the half-power beamwidth of the beamforming gain. Comparisons with other benchmark schemes proved that the proposed approach achieves higher EE by ensuring a source-nearby ARIS placement and providing a high probability of selecting the full-array RIS structure. 


\subsection{Physical Layer Security}
The broadcast and shared nature of the wireless medium makes it challenging to guarantee the secrecy of wireless communication networks. Physical layer security has emerged as a new technique to significantly improve the communication security of wireless networks by exploiting the communication channel with unauthorized users. In theory, any achievable secrecy rate representing the quantity of information transmitted from the source to the destination that unauthorized users cannot decode could not exceed an information-theoretic limit known as the secrecy capacity. This latter depends on the channels between all communication nodes, and thus, if all nodes remain static,  the achievable secure rate remains below the secrecy capacity, regardless of the deployed techniques.  
To further boost the security performance, a potential solution is to leverage UAVs, which, owing to their high mobility, offer opportunities of creating stronger LoS links between legitimate users to boost the secrecy capacity. However, such LoS links can be maliciously leveraged by eavesdroppers to increase their SNR, posing a potential security risk in UAV communications.  
Recently, several works demonstrated that the use of intelligent reflecting surfaces is key to solving all these issues by opening up possibilities of reshaping the propagation channels to improve the secrecy rate of wireless communication systems.

\subsubsection{TRIS}
The authors in \cite{fang2020joint,wang2020intelligent} studied the secrecy rate of a TRIS-assisted UAV network, where an ABS transmits information to a legitimate receiver while a passive eavesdropper tries to intercept this information. In \cite{wang2020intelligent}, the problem of interest is to jointly optimize the RIS's phase shifts, the UAV location, and the transmit beamforming to maximize the secrecy rate under minimum rate and power constraints. 
 The secrecy rate obtained in \cite{wang2020intelligent} was found to increase with the maximum allowed transmit power. The effect of the distance between the receiver and the TRIS was  investigated. The authors concluded that when TRIS is far away from the receiver, the scheme without TRIS achieves a similar performance to that with TRIS. The gain of the TRIS becomes more significant when it is near the receiver, allowing for full exploitation of the passive beamforming to enhance the desired signal at the legitimate receiver. 
Considering a similar setting, the authors in \cite{fang2020joint} assumed  that the position of the eavesdropper is known but its small scale fading with respect to the TRIS could not be acquired. As such, only a lower bound of the secrecy rate can be acquired, which is then  optimized with respect to the UAV's trajectory, the TRIS's phase shifts and the power transmitted at each slot under total and peak power constraints, as well as given fixed start and end positions. The obtained results showed that for long flight durations, the trajectory of the UAV  is such that it flies to a hovering position before it flies again to the destination. If a TRIS is used, the UAV hovers over a position close to the TRIS to leverage the reflect beamforming gain, while it hovers over a position near the receiver in the case without TRIS. 
Both works in \cite{wang2020intelligent} and \cite{fang2020joint} considered either complete or partial knowledge of the channel with the eavesdropper. In practice, the eavesdropper  usually makes it difficult for the legitimate users to detect or track its location. Hence, a more plausible assumption is to assume  the estimated eavesdropping channels inaccurate.
In this context, the authors in \cite{sixian2020robust} proposed a deterministic uncertainty model, according to which the true  eavesdropper's channel  is within a ball centered at their associated estimates with a given uncertainty radius. Compared to the system model described in \cite{wang2020intelligent} and \cite{fang2020joint}, the work in \cite{sixian2020robust} assumed that each time slot is divided into downlink communication and uplink communication. Under this setting, the problem of interest is to jointly optimize the trajectory of the UAV, the phase shifts of the TRIS and the power of the UAV in both downlink and uplink so as to maximize an average of the worst weighted sum of the uplink and downlink achievable secrecy rate  over the eavesdropper's uncertainty region. 
Since in  uplink communications the achievable rate from the gU to the eavesdropper does not depend on the UAV trajectory, placing more priorty to uplink rate link in the objective would lead the UAV to fly so as to achieve the best communication quality between the UAV and the gU. On the other hand, as the weight associated with the downlink secrecy rate becomes more important, the algorithm tries to find a tradeoff between the links with the legitimate user and the eavesdropper.
Therefore, the UAV trajectories are in this case inclined to fly along relatively direct paths to the final location, so as to avoid information leakage and increase the worst achievable secrecy rate. 
Furthermore, the average worst-case secrecy rate decreases as the CSI uncertainty of the wiretap channels increases. This is because of the high uncertainty in the eavesdropper's  channel making it difficult to achieve a robust design. While too high uncertainty leads to the failure of RIS's passive beamforming, it presents a marginal effect on the trajectory or the transmit power optimization.
All the aforementioned works are based on leveraging the impinging signal on the RIS  from the aerial platform to degrade the quality link with the eavesdropper.  This approach can be impractical if the link with the RIS undergoes a high path loss.  A possible solution to address this issue is to rather use the reflections of signals from terrestrial networks to produce a strong interfering signal to the eavesdropper. Such a solution was adopted in \cite{xu2021intelligent} where a satellite was considered together with a terrestrial BS to serve a satellite user and a terrestrial user. The terrestrial BS is required to reflect its own signal on a TRIS so as to degrade the communication link between an eavesdropper and the satellite. In light of this, the authors define the problem of interest to jointly design the transmit beamforming and the TRIS's phase shifts to minimize the SNR at the eavesdropper under minimum SNR requirements on the terrestrial user and the satellite user. 
  Based on simulation results, the authors observed that increasing the number of reflecting elements or the transmit power of the terrestrial BS  yield lower SNR at the eavesdropper. However, increasing the number of antennas at the BS  does not necessarily decrease the SINR at the eavesdropper despite the increasing spatial degrees of freedom. This is because the spatial degrees of freedom are rather used to satisfy the constraints on the SNR of the terrestrial and satellite users.

\subsubsection{ARIS}
Besides  improving communication links, RIS-equipped UAVs can also be used to enhance secrecy of communication networks. In this context, the authors in  \cite{long2020reflections} considered an uplink wireless communication network in which legitimate users communicate through a ARIS-assisted UAV with a terrestrial base station in the presence of an eavesdropper. Under this setting, the authors focused on maximizing the secure energy efficiency of the system defined as the ratio between the secrecy rate and the total consumed power via jointly optimizing the UAV's trajectory, the RIS's phase shifts, the user association, and the power allocation under maximum per-user power constraints, fixed UAV's height and maximal speed, and a closed-loop trajectory for the UAV.  
Simulation results demonstrated that the designed system can enhance the secure EE by up to 38\% gains, as compared to the traditional schemes with an AF relay. As for the ARIS, it tends to establish communication with the users away from eavesdroppers to achieve high secrecy rates, while keeping silent to avoid information leakage when ARIS is close to the eavesdropper. 
As the considered ARIS is a totally passive reflecting structure and does not cost any specific energy, the secure EE increases as the number of reflecting elements becomes large. 
On the other hand, with the ARIS's getting at higher positions, the secure EE decreases because of the increase in the communication distance. 
Furthermore, the UAV follows a closed-loop trajectory over which it sequentially visits all users.   However if the maximal flying speed decreases, the UAV is found to fly closer to BS to secure confidential information transmission.  
Interestingly, the authors noted that secure EE  improves with the increase in the maximum transmit power but not always with the same pace. Indeed, for small power values, it increases fast then slowly before keeping stable as the maximum transmit power keeps increasing. 
This can be explained by the fact that for high transmit powers, the excess in transmit power is not used and as such would not result in a significant improvement in the secure EE. 

To further enhance the secure achievable data rate, collaboration between the transmit users can be envisioned. In this context, the work in \cite{nnamani2021joint} considered a ARIS-assisted multi-sensor IoT system in which several nodes  collaboratively beamform the transmitted signal to be reflected by a 
RIS-equipped UAV to a terrestrial BS in the presence of an eavesdropper. The problem of interest is to maximize the achievable secrecy rate  via jointly optimizing
the collaborative beamforming weights of the sensor nodes, the trajectory of the UAV and the ARIS's phase shifts. For that, the cases of availability and non-availability of the eavesdropper's channel are considered.
 Simulations showed that the UAV tries to be far from the eavesdropper while keeping reasonable distance with the receiving BS. Moreover, if the channels with the BS and the eavesdropper are strong, the beamforming design accounting for the eavesdropper's channel outperform the one ignoring it. In contrast, both schemes present similar performances, when these channels are weak. 
This suggests that, in noisy environments, it is a viable strategy to use the scheme discarding the channel with the eavesdropper, when the location of the eavesdropper could not be determined exactly. 
On the other hand, when the sensors become scattered over a larger area, the average secrecy rate improves. This can be explained by the fact that the channel matrix aggregating all channels with the sensors become better conditioned, offering larger degrees of freedom to protect the signal from the eavesdropper. 

\subsection{Time Consumption}
Several emerging smart city applications are based on the freshness of sensory data (i.e., status updates), which is being monitored and generated by a plethora of IoTDs. Examples of such applications include 
smart environment monitoring, industrial control systems, and intelligent transportation systems, which all require reliability and timeliness in delivering status-update information. Outdated updates may be inconsistent with the current status of the physical process being monitored and controlled, which may lead to erroneous decisions.
 Several works in the literature considered the time cost as the metric of interest. For instance, the work in \cite{zhang2020data}, described in section III-C aimed to design the UAV's trajectory so as to minimize the total cost time. 
Considering the scenario in which multiple IoTDs are scheduled using TDMA to transmit through a ARIS-equipped UAV their status-updates to a gBS, the authors used  the expected sum Age-of-Information (AoI) as their metric of interest, which represents the  elapsed time since the generated/sampled of the most  recently received status-update.  
Under this setting, a joint optimization of the
altitude of the UAV, the communication schedule, and the ARIS's phases-shift was proposed so as to minimize the AoI under  SNR and ARIS altitude constraints.  The use of ARIS is motivated by the fact that ARISs operate in a full-duplex relaying mode, which should lead to reducing the AoI. Comparisons with other schemes considering either UAVs following a random walk policy with random selection of IoTDs or hovering at a fixed position that satisfies reliability constraints for all IoTDs has shown that the proposed scheme presents a considerable gain in terms of AoI.

\subsection{Error Probability}
Ultra-reliable and lower latency communications (URLLC) is one of the three diverse features to be offered by upcoming 5G networks next to enhanced mobile broadband, and massive machine-type communication. URLLC is indispensable to support important use cases of mission-critical applications, including UAVs control information delivery, V2V communications, self-driving cars, intelligent transportation, tactile internet, E-health, industrial automation, etc. Such URLLC applications require deterministic communications with ultra-high reliability where the packet decoding error rates are $10^{-9}$ or even lower, depending on the considered mission-critical application. 

Therefore, the authors in \cite{ranjha2020urllc} studied and formulated the problem of ultra-high reliability in URLLC assisted by a mobile UAV and RIS in a short packet transmissions communication scenario. 
Particularly, the authors considered the setting in which a ground transmitter sends short packets to a UAV that reflect them through a TRIS to a ground receiver. 
Under this setting, the SNR between the ground transmitter and the UAV depends on the UAV position, whereas the SNR between the UAV and the ground receiver depends on the UAV position and the phase shifts of the TRIS. For each hop, the transmitted packet uses a given blocklength. Based on that, the authors derive an expression for the total decoding error rate that involve the packets' blocklengths and the SNR in each hop.  
Targeting the minimization of the decoding error rate, they proposed a joint design of the optimal  UAV's   position and blocklength in each hop under a total blocklength constraint. 
It is observed that when the number of passive reflecting  elements  increases, the decoding error rate decreases dramatically. Such a behavior results from the improvement in spatial diversity brought about by the increase in the number of TRIS's elements leading to a higher total channel gain with the ground receiver. 
Furthermore, optimizing the UAV's position is a crucial step that can pave the way ultra-high reliable transmission of short pakcets. 
Moreover, when increasing the blocklength, the probability of decoding error rate decreases, but this comes at the cost of decreasing the message transmission rate. 

{In addition to assisting transmissions of existing systems, RIS can be used to simultaneously enable passive beamforming and information transmission. To achieve this functionality known as symbiotic transmission, RIS use existing radio to carry information bits encoded by the on/off states of its reflecting elements while adjusting their associated phase shifts to enable passive beamforming. In this context, the authors in \cite{hua2020uav} proposed a UAV-assisted RIS symbiotic radio system in which a UAV played the role of a primary transmitter for multiple TRIS to help them transmit their signals to a gBS. 
 For the sake of simplicity, a wake-up communication scheduling approach is adopted, according to which the UAV can assist at most one TRIS at each time slot. The authors proposed to optimize two different objectives involving the BER with respect to the UAV trajectory, the reflection coefficient at RIS, and the RIS association, and under a minimum rate constraint for the primary communications between UAV and gBS.
  The first optimization problem aims  at maximizing the weight sum BER of all RIS, where a higher value weighting factor represents a higher priority over other RISs. In this case, the UAV tends to sequentially visit all RIS when all RIS have the same weight factors to obtain the maximal sum rate, while, for unequal priorty weights,  it is likely to fly by the RIS with lower weight factor than hovering over it.  
  The other objective aims at minimizing the maximum BER among all RISs over all the time slots, which results in the UAV flying with maximum speed to become closer to each RIS, and then remaining stationary above it for a certain amount of time. }

\subsection{Number of Served Devices}
In \cite{al2021ris}, the authors considered the scenario in which an ABS tries to gather data sent by a set of IoTDs scattered in an urban environment, using FDMA. Each device alternates between an active and passive
activation mode and has a period during
which its information should be collected before it becomes of no value. To enhance the communication link between the IoTDs and ABS, a TRIS is used. The problem of interest is to maximize the number of IoTDs that were served under limited available resources and trajectory constraints.  It has been observed that 
under the assistance of TRIS, ABS tends to have a complex routing to serve more IoTDs around TRIS, while the IoTDs located far away from the TRIS cannot be served due to the restricted speed at ABS and their poor indirect LoS. The performance in terms of percentage of served IoTDs degrade with  adding more IoTDs because of the limited available resources. It also degrades when increasing the transmitted data size, due to the fact that smaller data can be uploaded within shorter periods and as such the ABS can have more free resources to serve more devices. However, both limitations can be tackled by   by increasing the number of reflecting elements at the TRIS. Comparisons with schemes using non-optimized trajectories adopting random trajectory, random TRIS configuration, and stationary ABS revealed the importance of trajectory planning and  TRIS optimization to adapt to more complex scenarios. A major important fact is that for short data sizes, the trajectory optimized scheme using random phases at the TRIS  is more efficient than the same scheme but without TRIS.   

\section{Methodology}
The above-reviewed optimization works on RANTNs revealed that the system performance optimization is always one of the key issues in RANTNs. Not only because of the reconfigurable reflecting elements and the controllable communication environment, but also due to the flexible and dynamic deployment of aerial platforms, which is different from the traditional terrestrial network. The NTNs introduce an important adjustable parameter, that is, the position/trajectory of aerial platforms. The main research effort that has been carried out is to develop efficient algorithms to jointly design the position/trajectory of aerial platforms, RIS configuration, and other system parameters such as transmit beamforming, resource allocation, user scheduling and so on. In fact, the formulated optimization problems in RANTNs are mostly involved with non-linear and non-convex objective functions and constraints, and the involved multiple optimization variables are often intricately coupled with each other, which makes them hard to solve. As a solution, many efficient methodologies are proposed which can be mainly divided into two types, i.e., alternating optimization algorithms and machine learning-based algorithms. 

\subsection{Alternating Optimization Algorithm}
To obtain a favorable performance with an efficient algorithm, the alternating optimization algorithm is adopted. It consists in decomposing the original problem into several subproblems regarding different optimization variables and solving them iteratively. Specifically, in each subproblem, one or two variables are optimized while the other variables are assumed to be fixed. 
This procedure is repeated in turn until a specific convergence criterion is reached. 
For instance, the authors in \cite{ge2020joint} aimed at maximizing the received power at the ground user in a TRIS-assisted UAV communication system by jointly optimizing the active beamforming at ABS, the passive beamforming at the TRIS, and the trajectory of ABS. With the alternating optimization framework, the formulated optimization problem is decomposed into three subproblems, and solutions of all variables are derived iteratively. Specifically, the optimal beamforming at ABS is solved by fixing the phase shift matrix and the ABS trajectory, then the optimal reflecting elements of passive beamforming are derived based on the fixed trajectory and the given optimal active beamforming, and finally the ABS trajectory is optimized based on the given active and passive beamforming. 

In recent works using alternative optimization algorithms, some of the  subproblems admit a closed-form solution. For example, the optimal solution of the involved active beamforming design at the transmitter is verified to be the maximum ratio transmission precoding \cite{9308937, mohamed2020leveraging, tan2018enabling, ge2020joint, lu2020aerial, nnamani2021joint} once other variables are determined. Also, in most of the subproblems, the power allocations problems \cite{wang2020intelligent,9293155,sixian2020robust, fang2020joint,jiao2020joint,9308937, nnamani2021joint, jeon2021energy}, RIS configurations \cite{fang2020joint,lu2020aerial,ge2020joint,lu2020enabling,9308937,pan2020uav,ranjha2020urllc,li2020reconfigurable,hua2020uav, nnamani2021joint}, and aerial platforms positions \cite{jiao2020joint,lu2020aerial,lu2020enabling} have closed-form solutions. Optimization variables in decoupled convex subproblems without closed-form solutions can also be solved by existing algorithms or tools efficiently, such as, the Lagrange dual method, CVX, interior-point algorithm, and linear search. 

Sometimes, the reformulated subproblems are still non-convex and cannot be solved by traditional convex optimization methods. Depending on the considered  objectives and their associated constraints, researchers propose to adopt different algorithms for solving optimization problems with particular forms.

\begin{itemize}
    \item \textbf{Successive convex approximation} The successive convex approximation algorithm acts as one of the most popular methods for solving non-convex optimization problem such as, power control \cite{long2020reflections,mu2021intelligent},  RIS configuration \cite{sixian2020robust,jiao2020joint, mu2021intelligent}, position/trajectory optimization of aerial platforms \cite{cang2020optimal,li2020sum,wang2020intelligent,9293155,sixian2020robust,fang2020joint,ge2020joint,al2020irs,pan2020uav,li2020reconfigurable, mu2021intelligent}. Specifically, at each iteration, the original non-convex functions are approximated by some convex upper-bounds of them with the same first order behavior. Then, the approximate solution of the original problems can be obtained by iteratively solving these convex problems. Once the convex upper-bounds are properly selected, SCA has been shown to achieve a favorable convergence behavior.
    \item \textbf{Semidefinite relaxation} The semidefinite relaxation technique constitutes an efficient tool to solve nonconvex quadratically constrained quadratic programs (QCQPs) in an almost mechanical fashion \cite{luo2010semidefinite}. It is generally used in RANTNs to transform the problems into a standard semidefinite programming (SDP) by relaxing the rank-one constraint \cite{cang2020optimal,jiao2020joint,sixian2020robust, xu2021intelligent}. The relaxed problem is convex and can be easily solved by existing toolboxes.  
    \item \textbf{Riemannian manifold} The unit modulus constraints usually required for phase shift optimization is one of the main obstacle towards solving the optimization problem. As a solution, the authors in \cite{yu2019miso} show that it can be efficiently solved by manifold optimization, as the unit modulus constraints define a Riemannian manifold. Therefore, the Riemannian conjugate gradient algorithm is adopted in \cite{li2020sum}, thereby enabling an efficient optimization for the RIS configuration. 
    \item \textbf{Fractional programming} For the problems whose objective function in a fractional program is a ratio of two functions that are in general nonlinear, the fractional programming method is preferred to be invoked. This method is especially useful in the scenarios concerning about the sum rate \cite{zhang2020distributional}, secrecy rate \cite{wang2020intelligent}, power minimization \cite{cang2020optimal}, and EE \cite{long2020reflections}. 
    \item \textbf{Difference-of-convex} Additionally, the difference-of-convex programming and its difference-of-convex algorithm also adopted in \cite{wang2020intelligent,mohamed2020leveraging} to address the problem of optimizing a objective composed by the difference of functions. It is able to quite often gave global solutions to a lot of different and various non-differentiable nonconvex optimization problems based on local optimality conditions and difference-of-convex duality. This method is proved to be robust and efficient, especially in the large scale setting. 
    \item \textbf{Quadratic transform} There are also some objectives in the form of ratios of concave and convex functions with reference to the same variable. Fortunately, due to the structure of the problem, the globally optimal solution can be obtained by applying quadratic transform proposed. The quadratic transform converts the ratio of concave and convex function to the convex form by introducing an auxiliary variable \cite{shafique2020optimization}.
    \item \textbf{Binary constraint relaxation} Consideration of multiple users or multiple RIS lead to user/RIS scheduling problems with binary constraints. 
    The non-convexity of this kind of problems can be handled by relaxing the binary constraint to a continuous one. \cite{cai2020resource, hua2020uav}.
    \item \textbf{$\mathcal{S}$-procedure} The authors in \cite{sixian2020robust} also adopted a Lagrange relaxation technique called $\mathcal{S}$-procedure to transfer the quadratic constraints into linear matrix inequalities to ease the optimization process.
    \item \textbf{Penalty method} This method is widely used for solving constrained optimization problems through replacing a constrained optimization problem by a series of unconstrained problems whose solutions ideally converge to the solution of the original constrained problem \cite{mu2021intelligent}. The unconstrained problems are formed by adding a penalty function to the objective function. The penalty function consists of a penalty parameter multiplied with a measure of violation of the constraints. The measure of violation is nonzero when the constraints are violated, while it is zero in the region where constraints are not violated.  
\end{itemize}
 
In addition to these existing methods, several works also proposed other approaches to solve some specific optimization problems. Specifically, to ensure all users in the particular area lie within the main lobe of the ARIS and connect to the transmitter successfully, the reflecting elements were divided into sub-arrays in \cite{lu2020enabling}. The use of sub-arrays produce larger main lobes but comes at the cost of smaller peak gains.  The complete and improved ARIS design regarding the sub-arrays division based on a novel 3D beam broadening and flattening technique is provided in its extension work \cite{lu2020aerial}. The phase shifts of the sub-arrays are designed to form a flattened beam pattern with adjustable beam-width, catering to the size of the coverage area. 

\textit{Summary: }The alternating optimization is the most general method to get a favorable solutions for the jointly coupled optimization variables in different system settings. Table. \ref{AO} summarizes all the literature adopting the alternating optimization method. As reviewed above, although the original problems can be divided into several subproblems, it is necessary to choose an efficient method for solving each subproblem whose objectives and constraints have certain properties. 

\begin{table*}[!htbp] 
\centering
\caption{Summary of existing literature adopting alternating algorithm}
\label{AO}
\begin{tabular}{|p{0.07\textwidth}| p{.28\textwidth}|p{.25\textwidth}|p{.25\textwidth}| } 
\hline
\rowcolor{gray!20}
\textbf{Reference} & \textbf{Optimization variables} & \textbf{Algorithms}& \textbf{System model}  \\
\hline
\rowcolor{yellow!20}%
 \cite{9308937}& APs' power allocation and beamforming, ARIS's phase shifts and placement
& MRT
& G2G MU downlink transmission served by multiple APs\\
\hline
\rowcolor{yellow!20}%
\cite{lu2020aerial}& ARIS's position and phase shifts, transmit beamforming
& 3D beam flattening
& G2G MU downlink system\\
\hline
\rowcolor{yellow!20}%
\cite{zhang2020distributional}& ARIS's phase shifts, transmitter beamforming
& fraction programming, dual decomposition& G2G MU MISO downlink system\\
\hline
\rowcolor{yellow!20}%
 \cite{hua2020uav}& UAV's trajectory, TRIS' phase shifts, RIS scheduling
& SCA, binary constraint relaxation & A2G P2P system\\
\hline
\rowcolor{yellow!20}%
\cite{cai2020resource}& UAV's trajectory and velocity, TRIS' phase shifts, user scheduling, power allocation
& SCA, dual decomposition, binary constraint relaxation
& A2G MU downlink system\\
\hline
\rowcolor{yellow!20}%
\cite{9293155}& UAV's trajectory, TRIS scheduling, subcarrier allocation, power allocation 
& SCA, dual decomposition, water-filling &A2G MU downlink system\\
\hline
\rowcolor{yellow!20} \cite{mu2021intelligent}& TRIS's phase shift, UAV's location, power allocation, and NOMA decoding orders, &penalty method, SCA& A2G multi-group SISO downlink system\\
\hline
\rowcolor{yellow!20}%
 \cite{ge2020joint}& UAV's trajectory, TRISs' phase shifts, transmit beamforming
& SCA & A2G P2P downlink system\\
\hline
\rowcolor{yellow!20}%
\cite{cang2020optimal}& UAVs' position, TRISs' phase shifts, user scheduling, RIS scheduling 
& SCA, SDR, fraction programming & A2G multicell downlink system\\
\hline
\rowcolor{yellow!20} \cite{jeon2021energy}& ARIS's phase shift and array partition, HAP's position, power allocation, & convex sub-problems& G2A MISO gBS-multiple ABS backhual system\\
\hline
\rowcolor{yellow!20}%
\cite{shafique2020optimization}& ARIS's height
& Quadratic transform
& G2G P2P system\\
\hline
\rowcolor{yellow!20}%
 \cite{pan2020uav}& UAV's trajectory, TRIS's phase shifts, THz sub-bands allocation, power allocation
& SCA, dual-based method & A2G MU downlink system \\
\hline
\rowcolor{yellow!20}%
\cite{jiao2020joint}& ARIS's position and phase shifts, transmit beamforming
& SCA, SDR & ground-to-ground NOMA downlink system\\
\hline
\rowcolor{yellow!20} \cite{xu2021intelligent}& TRIS's phase shift, transmit precoding, &SCR & A2G CR system with primary satellite communication and secondary terrestrial network.\\
\hline
\rowcolor{yellow!20}%
\cite{lu2020enabling}& ARIS's position and phase shifts, transmit beamforming
& ARIS sub-array division
& G2G MU downlink system\\
\hline
\rowcolor{yellow!20}%
 \cite{li2020reconfigurable}&  UAV's trajectory, TRIS's phase shifts
& SCA & A2G P2P downlink system\\
\hline
\rowcolor{yellow!20} \cite{li2020sum}
& UAV's trajectory, TRIS's phase shift
& Riemannian manifold, SCA
& A2G MU downlink system\\
\hline
\rowcolor{yellow!20}%
 \cite{mohamed2020leveraging}& transmit beamfoming, ARIS's phase shifts, 
& DC  & G2G MU downlink system\\
\hline
\rowcolor{yellow!20}%
\cite{fang2020joint}& UAV's trajectory and transmit power, TRISs' phase shifts
& SCA
& A2G P2P downlink secure system\\
\hline
\rowcolor{yellow!20}\cite{wang2020intelligent}
& UAV's transmit power and location, TRIS's phase shifts, 
& SCA, fractional progamming, DC
& A2G P2P downlink secure system \\
\hline
\rowcolor{yellow!20}%
\cite{sixian2020robust}& UAV's trajectory, TRIS's phase shift, and gU's transmit power
& SCA, S-produce, SDR 
& A2G P2P downlink and uplink secure system\\
\hline
\rowcolor{yellow!20}%
\cite{long2020reflections}  &  UAV's trajectory, TRIS's phase shifts, user scheduling, power allocation
& SCA, fraction programming & A2G MU downlink secure system\\
\hline
\rowcolor{yellow!20} \cite{nnamani2021joint}& ARIS's phase shift and location, transmit precoding, &fractional progamming, linear search& G2G MU SIMO uplink secure system\\
\hline
\end{tabular}
\end{table*}

\subsection{Machine Learning-based Algorithm} 
 Alternating algorithms are able to obtain near-optimal performance, but most of them are highly complex and time-consuming. On the other hand, the time-varying and highly dynamic nature of wireless NTNs requires the proposed solutions to be easily implemented with low complexity. 
 Such a goal could not be achieved by conventional optimization methods. A recent research activity attempts to investigate the use of machine learning techniques, known for their suitability to tackle non-convex and sophisticated optimization problems with nearly-optimal solutions. As far as RANTNs are considered, machine learning stands as an efficient tool to handle highly dynamic wireless environment. 

{\bf Reinforcement learning.} Reinforcement learning gathers a group of methods aimed to make decisions in a given environment. Such an environment is typically modeled through a Markov decision process which involves 4-tuple 
 $\mathcal{S}, \mathcal{A}, \mathcal{P}_a, r(s,a)$, where $\mathcal{S}$ is the state space grouping all possible states, $\mathcal{A}$ is the action space containing all possible actions, 
 $\mathcal{P}_a(s,s')$ is the probability that action $a$ in state $s$ at time $t$ will lead to state $s'$ at time $t+1$, and the reward $r(s,a)$ is the immediate reward (or expected immediate reward) received  due to action $a$ while being at state $s$. 
 A reinforcement problem interacts with the MDP through three elements:
 \begin{itemize}
 	\item {\bf Agent.}  An entity that takes actions and receive accordingly a reward.
 	\item {\bf Policy.} It is the mapping of each state $s$ to an action $a$ and is denoted usually by $\pi$
 	\item {\bf Value functions.}  These include the state-value function returning the total expected future discounted rewards starting from a given state, and the state-action value function, known also as Q-value function, which returns these rewards starting from a given state and a given action and following a policy $\pi$ or considering optimal policy.  
 \end{itemize}
  Reinforcement learning  involves different approaches which can be decomposed into three main categories:  model-based techniques, value-based techniques and policy-based techniques. In model-based techniques, the agent alternates between two processes: estimating the model of the underlying environment and determining the optimal decision from the estimated model. In value-based techniques, the agent estimates the value function and finds the policy that optimizes it, while in policy-based techniques, the agent directly targets the estimation of the optimal policy without estimating the value function. In the sequel, we will briefly describe  popular algorithms from the categories of value-based and policy based techniques and explain how they are applied to solve trajectory planning problems in RANTNs.

\noindent{\bf Q learning.} Q.learning is one of the most popular algorithms in RL that belongs to the category of value-based techniques. It is based on  learning a Q-value function that can be exploited to find the optimal action while being at a given state.  
The Q-value function is a quantity denoted by $Q^\star(s,a)$ which   measures the total expected value of the cumulative discounted reward of choosing action $a$ when being at state $s$ and then following the optimal policy. Formally, $Q^\star(s,a)$ writes as:
$$
Q^\star(s,a)=\mathbb{E}_{s'\sim P_a}\left[r(s,a)+\gamma \max_{a'} Q^\star(s',a')\right]
$$
where $\gamma$ is a discount factor. In practice, the agent perform either an exploration step or an exploitation step. In the exploration step, the agent  tests new actions to update the Q-table storing the Q-value estimates for every state and every possible action. In this step, the Q-values estimates are updated based on the following rule: 
\begin{equation}
Q^\star(s,a)\leftarrow (1-\alpha) Q^\star(s,a) + \alpha\left[r(s,a)+\gamma \max_a Q^\star(s',a)\right]
\label{eq:q_value}
\end{equation}
where $\alpha$ is the learning rate and $s'$ is the new state after performing action $a$. 
In the exploitation step, the agent exploits the Q-table and performs the action that presents the highest Q-value for the current state. To find a good balance between exploration and exploitation steps, one frequently used method is the $\epsilon$-greedy exploration method in which the exploration step is performed with a decaying probability $\epsilon$. In this way, the agent will start by exploring the environment. As it acquired more accurate estimates for the Q-values, it can start exploiting them to perform optimal actions. 

\noindent{\bf Deep Q. learning.} When the number of states is too high, Q-learning faces two major problems. First, the amount of memory required to update new states can be prohibitively high. Second, the exploration step to create the required Q-table would require a lot of time, making it impractical. One solution to this problem is to use a neural network to approximate the Q-value function for any possible state and action. For that, the value function is parametrized by a parameter $\theta$, hence denoted by $Q^\star(s,a;\theta)$, where $\theta$ are the weights of the used neural network. 
The aim of the neural network is to minimize the following loss function:
\begin{equation}
L(\theta)=\mathbb{E}\left[\left({\rm target}(s,a)-Q^\star(s,a;\theta)\right)^2\right]
\label{eq:loss}
\end{equation}
where ${\rm target}(s,a)$ is the target Q-value which represents a refined estimate for the expected future reward from taking an action $a$ while being in state $s$ and is given by
$$
{\rm target}(s,a)=r(s,a)+\gamma \max_{a'}Q^\star(s',a')
$$
The main difference with classical deep learning is that here the target changes constantly during the process, which may lead to unstability issues in practice. One solution to this problem is to employ two neural networks, where the first one is trained to optimize the loss in \eqref{eq:loss}   and the second one termed ``target network''  is a copy of an old version of the former used to provide the unknown target and is updated less frequently. Such a method is known as Deep Q-network (DQN) and applies only when the set of actions is discrete. 

\noindent{\bf  Policy gradient methods.} To handle the case of continuous  state and action spaces, policy gradient methods have been proposed. Central to these methods is to represent the policy by a parametrized function of the parameter vector $\theta$ and optimize it via gradient ascent to maximize the expected return. Such a policy  may be either deterministic $(a=\pi_\theta(s))$ or stochastic, representing the probability distribution of the action given a state $s$ $(a\sim \pi(\theta|s))$.
One issue that policy gradient methods should address is how to estimate the gradient and particularly the Q-value function that arises in its expression. The most natural way to estimate this function is by Monte Carlo averaging where states and actions are sampled at each episode according to the previous policy  to update the Q-value function and the policy parameter vector $\theta$ for each sampled state and action.  
 Such an approach is known as the REINFORCE algorithm and while being simple, suffers from high variance gradient estimates and slow learning rate. 
As a solution to the limitations of the REINFORCE algorithm,  hybrid architectures combining policy and value based techniques have been proposed. They are often termed actor-critic architectures, and  as their name suggests, they involve an actor to update by stochastic gradient the parameter $\theta$ parametrizing the policy, and a critic to estimate the Q-value function and feed it to the actor. Several reinforcement algorithms adopt the actor critic architecture, of which we distinguish the deep deterministic policy gradient (DDPG), the trust region policy optimization (TRPO) and the proximal policy optimization (PPO). While DDPG can be thought of as an extension of Deep Q-learning for continuous action spaces, PPO and TRPO aim to prevent performance collapse due to large policy updates. To do this, PPO uses a surrogate objective function which guarantees monotonic policy improvement. Compared to TRPO,  PPO is a first-order approximation of TRPO that is simpler to implement and tune and was shown to achieve better empirical performance.

\noindent{\bf Application to RANTNs.} Both policy gradient and Deep Q. learning methods have been used in RANTNs to assist the design of the UAV's trajectory. In this context, targeting IoT networks, the authors in \cite{samir2020optimizing, al2021ris} applied the PPO algorithm to perform joint IoT scheduling and UAV's trajectory optimization. The authors in \cite{wang2020joint} considered the design of the UAV's trajectory to maximize the weighted fairness and data rate for a network composed of  a UAV communicating with the best  rate user at each time slot in the presence of multiple TRISs. Under this setting, the authors modeled the system as an environment in which the UAV's coordinate and the energy represent the set of states and the UAV's flying direction and its traveled distance the set of actions. Both cases of continuous and discrete actions were considered. For the discrete action case, a solution based on DQN is proposed. To handle the continuous action case, the use of DDPG algorithm making use of two target networks associated with the actor and the critic networks has been proposed. Focusing on energy minimization for NOMA-UAV-RIS-enhanced wireless networks in which a UAV is deployed to provide service for multiple users, the authors in \cite{liu2020machine} proposed a DQN to jointly optimize the trajectory and the RIS's phase shifts under per-user minimum rate constraints. Contrary to the above mentioned works, the RIS's phase shifts are part of  the actions and are optimized by the RL algorithm. In order to attain a tradeoff between
 training speed and convergence to the local optimal, the authors proposed to employ a decaying learning rate, starting from a large value to accelerate the training and decreasing its value with the training episodes to help the algorithm converge to a local optimum.  The Q-learning approach can be also useful to cope with a dynamic environment in which both the UAV and the served user are moving. Such a setting was considered in \cite{zhang2019reflections} wherein a UAV equipped with RIS relays a signal transmitted from a BS to a moving outdoor gU. To cope with the dynamic channel variation, a solution based on Deep Q-learning was proposed, wherein a deep neural network with a long-short term memory architecture is used to approximate the Q-value function. 

\noindent {\bf Distributional reinforcement learning.} 
As seen above, RL algorithms like Q-learning involve estimating the Q-value function through the update rule in \eqref{eq:q_value}. Such a value represent the expectation of the cumulative  discounted rewards. A different alternative to conventional RL that has shown impressive performance improvements is to rather estimate the distribution of the sum of the discounted rewards instead of their expectation. So, instead of updating the Q-value function as in \eqref{eq:q_value},  
the distribution of the cumulative discounted rewards is modeled by a parametrized distribution, the parameters of which are estimated so as to minimize a certain distance to the estimated target distribution. Based on the estimated parametrized distribution, the Q-value function is then computed and the action associated with the maximum Q-value function is taken. 
Building on this approach, distributional versions of Q-learning and DQN have been proposed and were shown to outperform their conventional counterparts.   As far as RANTNs are considered, the distributional reinforcement learning algorithm was applied in \cite{zhang2020distributional} where a RIS-equipped UAV is employed to enhance the downlink communication between the BS and multiple moving gUs. Targeting the maximization of the transmitted data volume under a limited energy constraint, the authors proposed an alternating optimization algorithm to optimize the transmit beamforming and the ARIS's phase shifts while a distributional RL is used to update the location of the UAV once a blockage occurs.   

\noindent{\bf Deep learning.} 
Deep learning has been proved to be a powerful tool for the development of data-driven algorithms. As it does not require knowledge of the data model, it has  naturally been applied to predict information from data  that cannot be easily modeled like photos or audio recordings. 
Although in wireless communication the propagation channel model is easy to model and the transmitted signal is man-made, deep learning can still play an important role for the development of future communication systems. 
As shown in \cite{9186132}, deep learning can be used for problems where an efficient algorithm is known but suffers from a prohibitively high complexity, making it impratical for real-time implementation.
In this case, a deep neural network can be used to approximate the output of the high-complexity  algorithm based on an offline training phase, and then used instead for real-time implementation. Doing so, the complexity of the online computation is moved to offline training, which  reduces the computational complexity and facilitates the implementation. 
%
 In practice, the training dataset can be obtained by running the high complexity algorithm on a multitude of settings. 
 In this context, 
 the authors in \cite{cao2021reconfigurable} applied this approach to the scenario in which multiple UAVs serve  through a multi-group RIS multiple users on orthogonal sub-carriers. Of interest is the problem of optimizing the RIS group allocation to each UAV-user pair as well as the RIS phase shifts to optimize the total throughput.  As a first solution, the authors proposed an algorithm based on  exhaustive search, the real implementation of which is not possible as it possesses an exponential complexity in  the number of the UAV-user pairs. To solve this issue, the authors proposed a solution based on deep learning to approximate the outputs of the exhaustive search-based algorithm. The proposed solution develops a two-task learning model that performs a classification task to predict an integer allocation vector describing the RIS group allocation and a regression task to predict a real-valued vector representing the RIS phase shifts. In order to train the associated deep learning model, the exhaustive search algorithm is run offline for different settings to yield a training data set formed by the system parameters such as the number of UAV-user pairs, the RIS size and the channel conditions and their associated labels represented by the solutions of the exhaustive-search algorithm. 
 Another application of  deep learning in wireless communication is when a theoretical model describing the relationship between  physical quantities like the propagation environment and some information of interest could not be obtained. In this case, deep learning can help provide such a model. 
 One example of such an application of deep learning to RANTNs is represented by the work in \cite{abuzainab2021deep} wherein the authors considered a THz drone network in which a mobile drone user is served by a
 base station and a flying RIS. Of interest is the problem of proactively predicting the best beamforming vector at the BS and the best communication link between  direct and RIS-assisted links. To account for the drone mobility, at each time, the prediction task is required to learn the temporal correlation in the learning sequence, so that it can predict the values of the best communication link and the BS beamforming vectors from their past values. The authors have thus opted for a recurrent neural
 network based on gated recurrent units, which has been proven to be
 effective in learning sequence dependency, especially long
 sequences. 
 This architecture is trained over a training data set built using the DeepMIMO generation framework in \cite{alkhateeb2019deepmimo}.

Another neural network architecture that has been used in RANTNs is the graph neural network which, unlike conventional DL methods that exhibit a grid-like structure, can produce state-of-the-art solutions for problems involving data in irregular domains. Such a property can be useful for channel estimation tasks since the observed data frequently changes because of the dynamic nature of the propagation channel. In this context,  K. Tekbıyık et al. first used GAT  to estimate the channels of RIS-equipped HAP assisted full-duplex communications \cite{Tekbiyik}, and the channels in the LEO-satellite enabled IoT communications with RIS deployed near the satellite \cite{tekbiyik2021graph}. The numerical
results show that for the full-duplex channel estimation, the
performance of the GAT estimator is better than the least
squares. Contrary to the previous studied method, GAT has
the ability to estimate the concatenated channel coefficients at
each node separately. As such, there is no need to use the
time division duplex mode during the pilot signaling in the
full-duplex communication. Moreover, the numerical results
also show that, the GAT estimator is robust to hardware impairments
and small-scale fading characteristics changes, event when the training data does not include these changes.

\textit{Summary:} Further enhancements are achievable for RANTNs when leveraging machine learning algorithms to empower the integrated technologies with RIS and aerial platforms. The RIS/UAV will provide autonomous decision-making, knowledge extraction and prediction, and near-optimal optimization performance. Machine learning algorithms can be used for enhancing the channel estimation, spectral efficiency, and balancing different tradeoffs by automatically learning from the collected data, the propagation environment and their past experience. All the machine learning-based algorithms share the general property that more extra iterations are needed for convergence when the number of reflecting elements increases. The adopted machine learning algorithms in the existing literature aiming to find the optimal solutions for different control variables are summarized in Table. \ref{ML}. 
Despite the important number of tools that have been thus far applied, some existing tools have not been explored in the context of RANTNs, such as semi-supervised learning or unsupervised learning without labeled data. 

\begin{table*}[!htbp] 
\centering
\caption{Summary of existing literature adopting alternating algorithm}
\label{ML}
\begin{tabular}{|p{0.07\textwidth}| p{.28\textwidth}|p{.22\textwidth}|p{.28\textwidth}| } 
\hline
\rowcolor{gray!20}
\textbf{Reference} & \textbf{Optimization variables} & \textbf{Machine learning algorithms}& \textbf{System model}  \\
\hline
\rowcolor{yellow!20}%
\cite{samir2020optimizing}& UAV's altitudes, user scheduling
& DRL, PPO, MDP 
& G2G MU uplink system\\
\hline
\rowcolor{yellow!20}%
 \cite{zhang2020distributional}& CSI, ARIS's position
& DRL, MDP & G2G MU downlink system \\
\hline
\rowcolor{yellow!20}%
 \cite{zhang2019reflections}& CSI, ARIS's position
& Q-learning, MDP
& G2G P2P downlink transmission\\
\hline
\rowcolor{yellow!20}%
\cite{tekbiyik2021graph}& CSI 
& GAT & A2G/GA2 satellite IoT SISO downlink/uplink communications  \\
\hline
\rowcolor{yellow!20} \cite{cao2021reconfigurable}& TRIS's phase shift and location, TRIS's elements allocation & Multi-task learning& G2A multiple UAV-gU pairs SISO system\\
\hline
\rowcolor{yellow!20} \cite{wang2020joint}
& UAV's trajectory
& DQN
& A2G MU downlink system assisted by multiple TRISs\\
\hline
\rowcolor{yellow!20}%
\cite{Tekbiyik}& CSI 
& GAT & G2G P2P full-duplex system\\
\hline
\rowcolor{yellow!20}%
\cite{abuzainab2021deep}& communication link, transmission beam
& recurrent neural network, GRU
& G2A P2P downlink system\\
\hline
\rowcolor{yellow!20} \cite{al2021ris}& UAV's trajectory, user scheduling &MDP, PPO& G2A MU SISO uplink system\\
\hline
\rowcolor{yellow!20}\cite{liu2020machine}
& UAV's trajectory, TRIS's phase shifts
& D-DQN, MDP
& A2G MU downlink system \\
\hline
\end{tabular}
\end{table*}

\section{Challenges and Future Research Directions}
Although the existing literature we reviewed has made a great contribution to RANTNs, there are some critical unsolved problems undermining the full potential of this cutting-edge technology in practical implementation. Since the studies on the integration of RISs into NTNs are still at initial stages, we briefly illustrate some potential but significant research directions in RANTNs.  

\subsection{Practical Channel Modeling}
Channel modeling is vital to enable fundamental and applied research. Theoretical channel modeling on RANTN is beneficial to assist the advanced system design by paving the way for theoretical performance evaluation and optimization by means of simulations.  
It is therefore of paramount importance to build accurate channel models for NTNs as well as for RIS, as the current models adopted in most existing works are inaccurate. 
As a matter of fact,  most measurement studies on NTNs focused on ideal rural, suburban and open fields environments and are not suitable to model dense environments characterized by
the existence of dense buildings, various weather conditions, streets, trees and lake water. Additionally, it is often that, for the sake of simplicity, they rely on channel models applied to low frequency bands and rarely considered to model the channel in high frequency bands, which suffer from severe attenuation, and high probability to be blocked. Furthermore, in RANTNs, a doppler frequency shift is generally experienced due to the mobility of the aerial platforms. While a large doppler spread can be experienced when different signal paths are associated with largely different Doppler frequencies, their effect is in general ignored. Besides doppler spread, mobility of users  make it necessary to evolve current channel models based on static ground users to  non-stationaly models accounting for the users' mobility.  
A future important research direction is to propose accurate channel models to account for wave propagation at high frequencies, doppler spread and mobility of users. 

For the channel modeling on RIS, most of the existing works are based on far-field channel models, which allows one to assume  the distances between an end terminal and all reflecting elements are the same. However, the near-field channel model is more accurate for some communication scenarios. Specifically, when the deployment of RIS is in the vicinity of end-terminals, or the RIS has large size capturing a large number of incident electromagnetic waves, the distances from the end terminal to all the reflecting elements is different. In addition to near-field communications, accurate models should account for other important effects such as 
 reflection loss, correlation between RIS units, and atmospheric attenuation. 

These unsolved problems further complicate the channel modeling for RANTNs, since there is a necessity to investigate new aerial models that consider RIS capabilities together with aerial platforms properties. Specifically, the non-stationary channels with excessive spatial and temporal variations are general in NTNs systems, which are caused by the rapid dynamic movements, rotations and aerial shadowing of aerial platforms. The RIS adds complexity to defining appropriate channel models because of its passive and reflective behavior and the near-field propagation that needs to be taken into account. The above two main components together make the channel modeling sophisticated and challenging, whose unique characteristics are still unknown. Moreover, the realistic channel modeling should be derived from real-world implementations and experiments, which involves accurate data to confirm the existing theoretical results.

\subsection{Channel Estimation}
The configuration of RISs in RANTNs require solving an optimization problem depending on the CSI. Such a step is key to achieving optimal beamforming and control of the radio channel. Since the gain brought by RIS depends on how much accurate is the CSI, channel acquisition in RANTNs is an important research topic that deserve particular attention. It poses several challenges    
 due to the passive nature of the RIS and the mobility of aerial platforms.

\noindent {\bf Channel estimation with passive RISs.} From the RIS perspective, the passive nature of transceiver chains makes it more difficult to estimate the channel, since only the cascaded channels are observed. The number of channel coefficients to be estimated proportionally increases with the number of elements, resulting in huge estimation overhead, enormous energy consumption at transceivers, as well as heavy signal processing burdens and long delay at the RIS's controller that has limited computational capabilities. 
Most channel estimation methods are no longer applicable for RIS-assisted full-duplex communications. For instance, the assumption of channel reciprocity used in typical time-division duplex (TDD) does not likely hold. As a matter of fact, several experimental results showed that the reflection coefficient at RIS is sensitive to the  arrival angle of the impinging wave. As this arrival angle in the uplink  differs from that in the downlink, reciprocity of channels could not be met.    
 The frequency division duplex-based estimation protocol is also unrealistic as it involves high feedback overhead due to  the large-dimensional channel matrices.   
 On top of that, the mutual coupling between RIS elements adds an additional layer of complexity to an already complicated problem. 

Several complexity-reduced methods are proposed by grouping the adjacent RIS elements with the same configuration but come at the cost of a performance loss. Another approach is to alter the passive nature of the RIS by incorporating few low-power active sensors, which enables sensing and channel estimation directly at the RIS. However, the use of active elements is against the attractive characteristic of RISs that is to control the channel using passive elements. Moreover, even with active elements, a control loop is still required to jointly  
adjust the RIS configuration and the beamforming at the transmitter/receiver. 

\noindent{\bf Impact of mobility on channel estimation.} In high frequency communications, the propagation channel can experience an important variation due to only a few millimeters of movement. This is the reason why in RANTNs,  real-time channel estimation is more challenging due to mobile aerial platforms and becomes even more difficult when ARISs instead of TRIS are considered.  
Achieving real-time reconfigurability in such various mobility conditions is still an open question that requires development of appropriate tools. 
In this respect, machine learning techniques, such as supervised learning and RL \cite{zhang2019reflections, zhang2020distributional}, might be a promising way to facilitate the estimation process \cite{Tekbiyik}. 

\noindent{\bf Potential research directions.} Beyond the channel estimation accuracy of a given method, it is important that the energy cost and the long delay it entails be taken into consideration. 
Overall, acquiring the channel with low latency, affordable energy consumption, low signaling overhead, and reduced computational complexity in dynamic wireless environments is still an open problem due to the mobility of aerial platforms  the passive reflection of RIS, and the massive number of reflecting elements. 
A potential method to reduce the latency of channel estimation procedures is 
 is to utilize the slowly varying long-term CSI depending on both angular and location information, which can be estimated by the direction of arrival/departure estimation. It dramatically reduces the burden on channel estimation and control signaling transmission, because the phase shifts designed based on the slowly-varying, which yields lower computation complexity and lower requirement for the control links. However, there are still not enough works in the system design based on long-term CSI. 

\subsection{Tracking}
Tracking users/aerial platforming is also challenging in RANTNs, 
since RISs cannot send pilot signals to enable tracking of their movement, especially when the direct links between the end devices are blocked. This problem is very important in future wireless communication systems, which are required to sense their surrounding environment and perform accurate localization. The utilization of higher frequency bands to support applications also greatly reduces the accuracy of sensing and localization. The reason is that the number of propagation paths in high frequency is reduced, which is mainly due to large penetration losses, high values of path loss, and low scattering. Adding to that, the conventional sensing and localization models are based on far-field  models and as such may not apply to the case of near-field  models . A potential research direction is to develop near-field sensing and localization models that exploit the information in the wavefront curvature \cite{wymeersch2020radio,abu2021near}. 

\subsection{Hardware Limitations}
Existing analysis algorithms for RANTNs have been developed based on the assumption of perfect hardware. Yet, the various types of hardware limitations existing in practical RANTNs must be accounted for.

Most of existing research works are based on the unrealistic assumption that RIS can perfect manipulate the impinging electromagnetic waves and reflect it with the optimized ideal phase shifts. Such an assumption could not be met in practice. It is thus necessary to build practical models accounting for properties of the used physical materials, the manufacturing processes, and  the RIS's reliability and configuration capabilities for  different communication frequencies and diverse numbers and sizes of RIS units.
More specifically, a widely considered assumption is to assume that the RIS's phase shifts are represented by an infinite bit resolution, allowing them to take any continuous value. Obviously, such an assumption  iss unrealistc and could not be supported by the controller or the control link.  
 An important research direction is to perform the RIS control under  the assumption of  finite resolution phase shifts, supporting only a finite number of phases. In this respect, the 1-bit resolution with only two possible phases is appealing as it is more suitable for a high number of reflecting elements.  
 The performance loss caused by using practical low-resolution RISs is also an interesting question that deserves investigation. 
 
Another aspect that is worth studying is to consider a realistic RIS reflection model  capturing the fact that part of the incident wave by each reflecting element is consumed at the resistance of the reflecting circuit,   which induces  a loss that depend on the phase shifts of the reflecting element. 
 \cite{jung2021optimality}. 
  As a consequence,   the amplitude of the reflection coefficients less than or equal
 to one depending  on the phase shifts of the incident wave. 


As far as aerial platforms are considered, designs based on optimizing their trajectories and positions  may not perform as expected \cite{xu2020multiuser}, due to the finite precision of electronic circuits and imperfect manufacturing of mechanical components. The imperfect {transmitting}/receiving hardware modules (such as power amplifier non-linearity, non-linear phase noise, frequency and phase offsets, in-phase and quadrature imbalance, etc.), and quantization noise jointly degrade communication quality. Hence, it is still unclear whether the favorable performance can be guaranteed considering these hardware limitations.

\subsection{Backhaul Control}
Because of the limited signal processing and computing capabilities of the RIS, the configuration of RIS is performed through a smart controller that communicates with a computing node to achieve real-time adaptive beamforming. 
However, such an approach requires a fully synchronized and reliable control link between the computing node and the RIS. A reliable control can be achieved in static propagation environments, which is not the case of RANTNs. 
For instance, the control link between the computing node and the ARIS is likely to experience  time-varying channel fading conditions with shadowing, thus affecting the process of uploading the phase shift modifications in real-time. In addition, the number of elements in the RIS ranging from a few to hundreds or more generate a tremendous amount of signaling overhead. More importantly, the energy consumption at powerful controller degrades the system performance, especially for ARISs, which are energy-limited devices. In advanced system involving multiple end nodes and aerial platforms, the centralized controller is required to collect all the complex-valued channel matrices, compute all passive beamforming, before sending them back to the corresponding nodes over the network. This results in a  heavy feedback overhead and a prohibitively high computational complexity that impede the RIS configuration. 

Therefore, new solutions are needed to provide  stable control links with low latency and limit the control signaling and processing overhead without compromising the performance. The control protocol via a separate wireless link or via dedicated time slots is a possible option that has not been investigated  in the literature. Other options include using distributed algorithms which possess appealing advantages over centralized algorithms such as low information exchange overhead, reduced computational complexity and increased scalability. 
These algorithms can be enabled by using for instance UAV swarms that can offer distributed computing and communications capabilities.

\subsection{Efficient Design and Optimization}
\noindent{\bf Trajectory optimization.} The development of robust and efficient algorithms for enhancing the RANTNs is still a practically challenging task, because of the mobility of aerial platforms, and in particular the controllable highly mobile UAVs. 
Most important topics in NTNs revolve around the dynamic 3D location adjustment to provide an additional DoF for improving the communication performance. 
They involve solving a trajectory optimization problem under two types of constraints, namely constraints on the aerial platforms' flight such as minimum/maximum flying altitude/speed, initial/final locations, maximum acceleration, obstacle and collision avoidance, and no-fly zone, and communication-related constraints including transmit power, serving time, and frequency resource limitations. 
Despite the important research works on trajectory optimization of aerial platforms,  most contributions are based on that aerial platforms are deployed in free space,  ignoring the scenario that low-altitude aerial platforms such as, UAVs can be deployed in closed and roofed environments. The impact of the propagation environment including the speed and direction of wind, the propulsive efficiency, and weather conditions is always overlooked. In general, the trajectory optimization problem is usually non-convex, thus accounting for all these constraints would make it even more difficult to solve. 

\noindent{\bf RIS configuration with moving aerial platforms.} 
In RANTNs, the configuration of RISs poses several challenges. From an optimization perspective, the aerial platform's trajectory is coupled with that of the RIS, leading to a more involved problem. 
From an implementation perspective, the jittering of aerial platforms and the uncertaintly of user location increases channel fluctuations \cite{xu2020multiuser}, while the movement of aerial platforms introduces signal alignment. All this can lead to the RIS passive beamforming becoming ineffective. 
	 It is noticeable that all the performance improvement in RANTNs is obtained at the price of solving a complex optimization problem with high hardware cost and design complexity. Hence, how to strike an optimal performance-complexity/cost tradeoff remains an open problem. To address this issue, the RIS grouping scheme was proposed \cite{zheng2020intelligent,yang2020intelligent,mu2021intelligent}, where adjacent RIS reflecting elements with high channel correlation are grouped into a {sub-surface} and are assumed to have the same reflection elements. Moreover, several complexity reduced-algorithm were developped under the unrealistic trajectory discretization assumption, where the UAV is considered to be at a fixed sample location during each interval. However, in reality, the UAV is continually moving. 

\noindent{\bf Distributed devices.} 
To enable data collection applications, advanced networks are 
 supported by a large number of ground devices, aerial platforms and RISs distributed in a certain area. This thus induces a higher design complexity because the computational resources are distributed everywhere across heterogeneous aerial and ground nodes with distinct communication and computation capabilities. How to match the time- and spatial-varying communication/computation/prediction demands with distributed communication/computation/data supplies in such highly dynamic 3D networks is a challenging task. The inter-aerial platform cooperation among aerial platform swarms constitutes a promising solution \cite{zeng2019accessing, wu20205g}, but has not been studied in RANTNs. It is based on dynamically selecting a cluster head whereas other aerial platforms and devices first collect data in a small area and then transmit them to the cluster head for further processing. Such an architecture has powerful computing capabilities, but it also has the drawback of requiring  a transmission delay to coordinate between clusters, as well as a possibly heavy loading on the cluster head, especially when the number of devices becomes large. 
 
 \noindent{\bf Integration of RISs: challenges and possible solutions.} As far as RISs are concerned, the large-scale deployment of RISs may bring inter-RIS interference to the network, degrading its performance. Specifically, transmissions coming from one RIS can leak to others, which causes strong inter-RIS interference. Such an impairment could be mitigated by letting RISs with high interference levels work in a cooperative mode rather than operating independently. However, the coordination of multiple RISs can trigger an explosion in signaling and processing overhead, and potentially results in high communication latency. 
 Apart from inteference, safeguarding wireless communication with such massive aerial platforms, RISs, and ground devices, is another vital issue that is worth studying. In this context, the authors in \cite{shang2021uav} presented several applications of swarm-enabled ARIS including assisting G2G/A2G/G2A networks to tackle the physical layer security problem, and mentioned the key challenges in the ARIS swarm network, such as beamforming design, channel estimation,  deployment and trajectory design. Although some potential solutions have been suggested to rise to these challenges, their implementation in practical scenarios have not been thoroughly assessed. 
 More research is required to examine the most effective network typology and communication protocol that allows for both efficient sensing and high data rate communication. 
 
 More importantly, the potential of RIS needs to be further dug. For instance, the RIS's full-duplex characteristics should be further incorporated with the existing networks by optimizing the fraction overlap between the uplink and downlink, which yields new design problems in terms of power allocation and beamforming. Additionally, most existing studies focus on using RIS to assist communication between devices. 
 With the emergence of the  new paradigm of symbiotic communication, RIS can be envisioned to have the additional role of transmitting information.   In such settings, RISs would be able to simultaneously enhance the communication quality of the primary link, and transmit their own information that may include control signals to acknowledge their current status, and some environmental parameters. The most simplest way to achieve this is to equip each RIS with a dedicated transmitter, at the cost of an extra power consumption. The other appealing approach to achieve this is to modulate the RIS information onto the reflected signals to establish passive information transfer.

Since different types of aerial platforms, including satellites, HAPs, and UAVs, have their own advantages and disadvantages such as cost, latency, persistence, and mobility, a hot research topic is the design and optimization of integrated networks consisting of various types of aerial platforms and various RISs. For the integrated network, new research topics emerge that concern the design of  efficient and fault-tolerant network mechanisms,  reliable transmission protocols,  seamless information exchange among different layers,  dynamic network operation control mechanisms, all of which  aim to provide continuous service, expanded coverage, rapid mission-response, and reliable transmission.

\subsection{Spectral Sharing}
A critical issue in RANTNs is the limited spectrum available for simultaneous ground-to-ground, ground-to-air, and air-to-air communications. Unlike terrestrial communication systems where ground BSs are connected to a data hub via high-speed fixed-line backhaul links, e.g., optical fibres, the implementation of RANTNs usually relies on dedicated wireless communication channels. In particular, RANTNs have a stringent demand for system resources due to the required support of high data rate backhauling and exchange of time-critical control signals of aerial platforms. Furthermore, multiple RISs and aerial platforms tend to be deployed simultaneously in next-generation networks, which puts a significant burden on the need for available spectrum.

In this context, it is generally believed that traditional multiple access schemes based on orthogonal spectrum partition cannot even support moderate numbers of UAVs and ground users due to the rapid exhaustion of available resources. Actually, the scarce wireless spectrum is already congested by existing communication systems. Thus, efficient multiple access schemes based on non-orthogonal spectrum utilizations have to be developed for enabling RANTN systems.

As reviewed in the previous section, NOMA is a promising technology for multiplexing in RANTNs. However, since RISs are dynamically optimized based on the instantaneous realizations of combined channels and dependent on the current set of meta-atoms, the users' effective channel gains become coupled of the RIS coefficients. This results in a highly complex problem to determine the optimal user ordering in a multi-user NOMA scenario, especially when the number of users and the number of reflecting elements increase,  posing major challenges for the practical integration of NOMA into RANTNs. Therefore, the environment-dependent NOMA scheme with limited gain facilitates researchers to introduce rate-splitting multiple access (RSMA), which has better spectral efficiency for an integrated network. However, this scheme is just utilized in the context of integrated spatial-terrestrial networks, where RSMA can be distributively or centrally controlled to be employed horizontally at one of the layers or vertically at each layer. {Moreover, RISs can be enablers of  cognitive radio systems to save spectral resources, where RISs are used to ensure quality satisfying communication in targeted primary or secondary networks.} Nevertheless, the investigation of the mentioned and other efficient spectrum sharing schemes in RANTNs is missing in the literature and needs the researchers’ attention.

\section{Conclusion}
RIS provides efficient and cost-saving solutions to the existing challenges in NTNs, such as, endurance, blockage, information leakage, and so on. This paper has presented a comprehensive survey on the RANTNs from perspectives of framework, methodology, design, and applications. Firstly, we have introduced key properties of RIS in NTNs and the two kinds of RISs utilized in NTNs, which followed by the overview of the structure of RANTNs including A2G/G2A communications, G2G communications, and A2A communications. Then, we have focused on the integration of other promising technologies to further enhance  RANTNs. Then, we have provided detailed reviews on the insights brought by the performance analysis and optimization of RANTNs for different performance indices, such as SNR, achievable data rate, energy and time consumption, EE, etc. Afterwards, we have concluded the methodology and tools utilized in the existing literature, which provides readers with efficient methods to pursuing solutions for various optimization frameworks in RANTNs. Finally, we have outlined key challenges and future research directions. 

Meaningful insights are revealed in the literature and summarized in this survey. The joint design of the deployment of ARIS/aerial platforms and RIS configurations is the key factor in reaping the full benefits of RANTNs. The AO framework and ML algorithms are widely used in the current literature to obtain feasible solutions. The well-designed RANTNs tends to a key component of future networks, which can be applied in smart city construction, worldwide coverage realization, emergency network etc. However, the studies on RANTNs have not been supported by realistic data due to the lack of practical experiments, especially on the practical channel modeling. The majority of recent works ignored the practical limitations such as the RIS's finite phase resolution, signal misalignment, imperfect CSI, hardware impairments, and limited control resources. There is still limited literature on the advanced networks with multiple users, RISs, and aerial platforms, as well as the combination with other promising technologies, such as communications in high-frequency bands, spectral sharing, and wireless power supply. This work opens a significantly larger space for the RANTNs to explore than the current research scope. 

\bibliographystyle{IEEEtran}
\bibliography{reference}

\end{document}